\pdfoutput=1
\documentclass[PAPER, atlasdraft=false, texlive=2023, UKenglish, cernpreprint, texmf, orcidlogo]{atlasdoc}
\usepackage{atlaspackage}
\usepackage{atlasbiblatex}
 
\usepackage{atlasphysics}

\graphicspath{{logos/}{figures/}}
 
\usepackage{ANA-SUSY-2023-01-PAPER-defs}
 
\usepackage{float}
\floatstyle{plaintop}
\restylefloat{table}
\usepackage{verbatim}
\usepackage[figuresright]{rotating}

\AtlasTitle{Search for direct production of electroweakinos in final states with one lepton, jets and missing transverse momentum in \(pp\) collisions at $\sqrt{s}=13$~\TeV\ with the ATLAS detector}
 
\AtlasJournal{JHEP}
\AtlasAbstract{ Searches for electroweak production of wino-like chargino pairs, $\tilde{\chi}^{+}_{1}\tilde{\chi}^{-}_{1}$, and of wino-like chargino and next-to-lightest neutralino, $\tilde{\chi}^{\pm}_{1}\tilde{\chi}^{0}_{2}$, are presented. The models explored assume that the charginos decay into a $W$ boson and the lightest neutralino, $\tilde{\chi}^{\pm}_1 \rightarrow W^{\pm} \tilde{\chi}^{0}_{1}$.  The next-to-lightest neutralinos are degenerate in mass with the chargino and decay to $\tilde{\chi}^{0}_{1}$ and either a $Z$ or a Higgs boson, $\tilde{\chi}^{0}_{2} \rightarrow Z \tilde{\chi}^{0}_{1}$ or $h \tilde{\chi}^{0}_{1}$.
The searches exploit the presence of a single isolated lepton and missing transverse momentum from the $W$ boson decay products and the lightest neutralinos,
and the presence of jets from hadronically decaying $Z$ or $W$ bosons or from the Higgs boson decaying into a pair of $b$-quarks.
The searches use 139~fb$^{-1}$ of $\sqrt{s}= 13$~\TeV\ proton--proton collisions data collected by the ATLAS detector at the Large Hadron Collider between 2015 and 2018. No deviations from the Standard Model expectations are found, and 95\% confidence level exclusion limits are set.  Chargino masses ranging from 260 to 520~\GeV\ are excluded for a massless $\tilde{\chi}^{0}_{1}$ in chargino pair production models. Degenerate chargino and next-to-lightest neutralino masses ranging from 260 to 420~\GeV\ are excluded for a massless $\tilde{\chi}^{0}_{1}$ for $\tilde{\chi}^{0}_{2} \rightarrow Z \tilde{\chi}^{0}_{1}$. For decays through an on-shell Higgs boson and for
mass-splitting between $\tilde{\chi}^{\pm}_{1} / \tilde{\chi}^{0}_{2}$ and $\tilde{\chi}^{0}_{1}$ as small as the Higgs boson mass,
mass limits are improved by up to 40~\GeV\ in the range of 200--260~\GeV\ and 280--470~\GeV\ compared to previous ATLAS constraints.
 
}

\author{The ATLAS Collaboration}
 
\AtlasRefCode{SUSY-2023-01}
 
\PreprintIdNumber{CERN-EP-2023-174}

\AtlasJournal{JHEP}
\AtlasJournalRef{JHEP 12 (2023) 167}
\AtlasDOI{10.1007/JHEP12(2023)167}

\hypersetup{pdftitle={ATLAS document},pdfauthor={The ATLAS Collaboration}}
 
\begin{document}
 
\maketitle

\section{Introduction }

The Standard Model (SM) is a strongly predictive effective theory, however it is not able to explain some observed phenomena, such as the abundance of dark matter and its nature, the matter--antimatter asymmetry, and the hierarchy problem~\cite{Sakai:1981gr,Dimopoulos:1981yj,Ibanez:1981yh,Dimopoulos:1981zb}. The ATLAS and CMS discovery of the SM Higgs boson~\cite{HIGG-2012-27,CMS-HIG-12-028,HIGG-2014-14,HIGG-2015-07} confirmed the mechanism of electroweak symmetry breaking and heightened attention on the hierarchy problem.
Supersymmetric (SUSY)~\cite{Golfand:1971iw,Volkov:1973ix,Wess:1974tw,Wess:1974jb,Ferrara:1974pu,Salam:1974ig} extensions to the SM can solve the hierarchy problem by introducing  a new symmetry that predicts bosonic (fermionic) partners for the fermions (bosons) of the SM.
In an \emph{R}-parity~\cite{Farrar:1978xj} conserving model, the SUSY particles are produced in pairs and the lightest SUSY particle (LSP) is a viable dark-matter candidate~\cite{Goldberg:1983nd,Ellis:1983ew}, as it is stable and weakly interacting.
 
The SUSY partners of the Higgs bosons and the SM electroweak gauge bosons, collectively called electroweakinos, are the higgsinos, winos (partners of the $\mathrm{SU(2)_L}$ gauge fields), and bino (partner of the $\mathrm{U(1)}$ gauge field). The electroweakino mass eigenstates are referred to as charginos $\tilde{\chi}_{i}^{\pm}$ $(i=1,2)$, linear combinations of higgsino and wino fields, and neutralinos $\tilde{\chi}_{j}^{0}$ $(j=1,2,3,4)$, linear combinations of higgsino, wino and bino fields. These are ordered in increasing
value of their masses.
 
Natural SUSY scenarios~\cite{Barbieri:1987fn,deCarlos:1993yy} predict that the lightest electroweakino mass be close to the electroweak scale. If squarks and sleptons (partners of the quarks and leptons) are heavier than a few \TeV\ and hence decoupled, they cannot be produced at the Large Hadron Collider (LHC).
The dominant SUSY production mechanism at the LHC may be the direct production of electroweakinos. SUSY models with light electroweakinos can also explain the observed discrepancy in the $g-2$ measurement compared to the SM predictions~\cite{PhysRevD.61.095004,PhysRevD.61.095004,HEINEMEYER2006265}. In the models considered in this paper, the compositions of the lightest chargino ($\tilde{\chi}_{1}^{\pm}$) and next-to-lightest neutralino ($\tilde{\chi}_{2}^{0}$) are wino-like and the two particles are nearly mass degenerate, while the lightest neutralino ($\tilde{\chi}_{1}^{0}$) is assumed to be a bino-like particle and the LSP\@. Two different SUSY processes are targeted in this paper: $\chinoonep\chinoonem$ and \chinoonepm\ninotwo\ pair production.

Three searches performed by the ATLAS Collaboration for the direct production of electroweakinos in proton--proton ($pp$) collisions produced at the LHC at $\sqrt{s} = 13$~\TeV\ are presented. The first analysis is designed to be sensitive to the direct pair-production of two charginos, referred to as the C1C1-WW model, where the charginos decay via $\chinoonepm \rightarrow W^{\pm} \ninoone$; the other two analyses are   designed to be sensitive to the associated production of nearly mass-degenerate charginos and next-to-lightest neutralinos,  latter decaying into the $\tilde{\chi}_{1}^{0}$ and either a $Z$ boson ($\ninotwo \rightarrow Z~ \ninoone$) or a SM-like Higgs boson ($\ninotwo \rightarrow h~ \ninoone$)~\cite{Fayet:1976et,Fayet:1977yc,Gunion:1987}, referred to as the C1N2-WZ and C1N2-Wh models, respectively.

\begin{figure}[h!]
\centering
\subfloat[]{
\centering\includegraphics[width=0.32\textwidth]{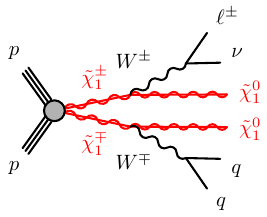}}
\subfloat[]{
\centering\includegraphics[width=0.32\textwidth]{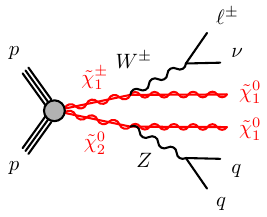}}
\subfloat[]{
\centering\includegraphics[width=0.32\textwidth]{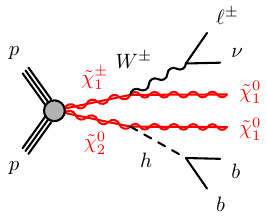}}
\caption{Feynman Diagrams for (a) electroweakino $\chinoonep\chinoonem$ and (b, c) $\chinoonep\ninotwo$ pair-production. One  $\chinoonepm$~decays into  a $\ninoone$~ and a $W$ boson that further decays leptonically. The other $\chinoonepm$ decays into a $\ninoone$~ and a $W$ boson that further decays hadronically (a). The $\ninotwo$~ decays into a $\ninoone$~ and either a $Z$ boson that further decays hadronically (b), or into a Higgs boson decaying into a pair of $b$-quarks (c).
}
\label{fig:diagrams}
\end{figure}

The target signature, in all scenarios, is a single isolated light lepton (electron or muon) produced by one of the $W$ decays, or by $\tau$-leptons produced in $W$ decays, and missing transverse momentum (\mpt) from LSPs and neutrinos. In the C1C1-WW and C1N2-WZ scenarios, due to the large momentum carried by the intermediate bosons, the jets are expected to be semi-boosted, or fully boosted. Thus up to three jets are required for these two models, which are produced by the hadronic decay of either a $W$ (in the $\chinoonep\chinoonem$ case) or a $Z$ (in the \chinoonepm\ninotwo\ case), and the hadronic radiation. In the C1N2-Wh model, the Higgs boson candidates are identified through their decay into a pair of $b$-quarks ($h \rightarrow b\bar{b}$) and two jets originating from the fragmentation of $b$-quarks, called $b$-jets, are required.
A set of simplified SUSY models~\cite{Alwall:2008ag,Alves:2011wf} is used to optimise the search and interpret the results. In these models the branching ratios of $\chinoonepm \rightarrow W \ninoone$ and $\ninotwo \rightarrow  Z \ninoone$ or $\ninotwo \rightarrow  h \ninoone$ are assumed to be 100\%. The branching ratios of $W$, $Z$ and Higgs bosons follow the SM predictions. Feynman diagrams of the processes under consideration are shown in Figure~\ref{fig:diagrams}.
 
Previous searches for electroweakino production at the LHC targeting $WW$, $WZ$ and intermediate states, and different lepton multiplicity in the final states, have been reported by the ATLAS~\cite{SUSY-2017-01,SUSY-2018-05, SUSY-2018-32, SUSY-2018-41, SUSY-2019-09-PAPER} and CMS~\cite{CMS-SUS-19-012,CMS-SUS-20-001,CMS-SUS-21-002} collaborations. This analysis is the first ATLAS search targeting final states with exactly one lepton, and profiting from the use of jet-substructure information for $W$ and $Z$ boson identification in large-$\mathrm{R}$ jets to target boosted regimes. The kinematic configurations where the decay products are boosted provide a handle to reduce the background.
In the case of decays via $Wh$, stringent constraints have been set by the ATLAS~\cite{ATLAS:2020pgy} and CMS~\cite{CMS:2021few} collaborations exploiting the $h \rightarrow b\bar{b}$ decay mode and multiple decay modes of the Higgs boson, respectively. This analysis targets final states with mass-splitting between the chargino and the LSP, $\Delta m (\chinoonepm, \tilde{\chi}^{0}_{1})$, between $m_h$ and around 250~\GeV, exploiting the $h \rightarrow b\bar{b}$ decay mode and the usage of dedicated boosted decision tree (BDT) discriminants. The BDT-based approach improves the sensitivity in the complex compressed phase-space where cut-and-count analyses suffer due to the similar kinematics of signal and SM backgrounds, especially from \ttbar\ and $Wt$ events.

\section{ATLAS detector}

The ATLAS detector~\cite{PERF-2007-01} at the LHC covers nearly the entire solid angle around the collision point.\footnote{ATLAS uses a right-handed coordinate system with its origin at the nominal interaction point in the centre of the detector. The positive $x$-axis is defined by the direction from the interaction point to the centre of the LHC ring, with the positive $y$-axis pointing upwards, while the beam direction defines the $z$-axis. Cylindrical coordinates $(r,\phi)$ are used in the transverse plane, $\phi$ being the azimuthal angle around the $z$-axis. The pseudorapidity $\eta$ is defined in terms of the polar angle $\theta$ by $\eta=-\ln\tan(\theta/2)$. Rapidity is defined as $y$ = $0.5 \ln[(E + p_z)/(E-p_z)]$ where $E$ denotes the energy and $p_z$ is the component of the momentum along the beam direction. The angular distance $\Delta R$ is defined as $\sqrt{(\Delta \eta)^2 + (\Delta \phi)^2}$.}
It consists of an inner tracking detector surrounded by a thin superconducting solenoid, electromagnetic and hadron calorimeters,
and a muon spectrometer incorporating three large superconducting air-core toroidal magnets.
 
The inner-detector system (ID) is immersed in a \qty{2}{\tesla} axial magnetic field
and provides charged-particle tracking in the range \(|\eta| < 2.5\).
The high-granularity silicon pixel detector covers the vertex region and typically provides four measurements per track,
the first hit normally being in the insertable B-layer (IBL) installed before Run~2~\cite{ATLAS-TDR-19,PIX-2018-001}.
It is followed by the silicon microstrip tracker (SCT), which usually provides eight measurements per track.
These silicon detectors are complemented by the transition radiation tracker (TRT),
which enables radially extended track reconstruction up to \(|\eta| = 2.0\).
The TRT also provides electron identification information
based on the fraction of hits (typically 30 in total) above a higher energy-deposit threshold corresponding to transition radiation.
 
The calorimeter system covers the pseudorapidity range \(|\eta| < 4.9\).
Within the region \(|\eta|< 3.2\), electromagnetic calorimetry is provided by barrel and
endcap high-granularity lead/liquid-argon (LAr) calorimeters,
with an additional thin LAr presampler covering \(|\eta| < 1.8\)
to correct for energy loss in material upstream of the calorimeters.
Hadron calorimetry is provided by the steel/scintillator-tile calorimeter,
segmented into three barrel structures within \(|\eta| < 1.7\), and two copper/LAr hadron endcap calorimeters.
The solid angle coverage is completed with forward copper/LAr and tungsten/LAr calorimeter modules
optimised for electromagnetic and hadronic energy measurements respectively.
 
The muon spectrometer (MS) comprises separate trigger and
high-precision tracking chambers measuring the deflection of muons in a magnetic field generated by the superconducting air-core toroidal magnets.
The field integral of the toroids ranges between \num{2.0} and \qty{6.0}{\tesla\metre}
across most of the detector.
Three layers of precision chambers, each consisting of layers of monitored drift tubes, cover the region \(|\eta| < 2.7\),
complemented by cathode-strip chambers in the forward region, where the background is highest.
The muon trigger system covers the range \(|\eta| < 2.4\) with resistive-plate chambers in the barrel, and thin-gap chambers in the endcap regions.
 
Interesting events are selected by the first-level trigger system implemented in custom hardware,
followed by selections made by algorithms implemented in software in the high-level trigger~\cite{TRIG-2016-01}.
The first-level trigger accepts events from the \qty{40}{\MHz} bunch crossings at a rate below \qty{100}{\kHz},
which the high-level trigger further reduces in order to record events to disk at about \qty{1}{\kHz}.
 
An extensive software suite~\cite{ATL-SOFT-PUB-2021-001} is used in data simulation, in the reconstruction
and analysis of real and simulated data, in detector operations, and in the trigger and data acquisition
systems of the experiment.


\section{Data and simulated events}

The analyses presented in this paper use 139~$\mathrm{fb}^{-1}$ of $pp$ collision data provided by the LHC and collected between 2015 and 2018 by the ATLAS detector, at a centre-of-mass energy of 13~\TeV\ and with an interval of 25~ns between consecutive crossings of proton bunches.  The average number of interactions per bunch crossing (pile-up) was $\langle\mu\rangle=20$ in 2015--2016,  $\langle\mu\rangle=38$ in 2017 and $\langle\mu\rangle=37$ in 2018.
The uncertainty in the combined 2015--2018 integrated luminosity is 1.7\%~\cite{ATLAS-CONF-2019-021}, obtained using the LUCID-2 detector~\cite{LUCID2} for the primary luminosity measurements.
 
Signal selection efficiencies and SM backgrounds are evaluated using Monte Carlo (MC) simulated event samples. The signal Monte Carlo samples were processed with a  complete simulation of the response of the detector provided by \textsc{Geant~4}~\cite{Agostinelli:2002hh}, or fast simulation~\cite{SOFT-2010-01} which relies on a parameterisation of the calorimeter response~\cite{ATL-PHYS-PUB-2010-013}.
A varying number of inelastic $pp$ interactions was overlaid on the hard-scattering event for all simulated events to model the multiple proton--proton interactions in the same and nearby bunch crossings. The pile-up events were generated with \pythia~8.186~\cite{Sjostrand:2007gs} using the NNPDF2.3LO parton distribution function (PDF) set~\cite{Ball:2012cx} and the A3 set of tuned parameters (tune)~\cite{ATL-PHYS-PUB-2016-017}. The simulated events were processed with the same trigger, reconstruction and identification algorithms used for data. Dedicated correction factors are applied to simulation to account for differences between data and MC simulated events.
 
The simulated backgrounds considered in the analyses are: $t\bar{t}$ pair production; single-top production ($s$-channel, $t$-channel, and associated $Wt$ production); $W/Z$+jets production; $t\bar{t}$ production with an electroweak boson ($t\bar{t}+V$); Higgs boson production (\ttbar+$h$, $Vh$); diboson ($WW$, $WZ$, $ZZ$) and multiboson ($VVV$ where $V=W,Z$) production.
The simulated ggF and VBF Higgs samples are not used as these processes are taken into account in the diboson samples. A further overlap removal is applied to avoid double counting between $Vh$ and diboson samples.
Different MC event generators were used depending on the simulated processes.
All simulated background processes were normalised to the best available theoretical calculation of their respective cross-sections.
The samples for $W$ and $Z$ boson production associated with jets ($W/Z$+jets) were simulated using \SHERPA. The modelling includes up to two partons at next-to-leading order (NLO), normalised to next-to-next-to-leading order (NNLO) for the inclusive cross-section, and five partons at leading order (LO) using Comix~\cite{Gleisberg:2008fv} and OpenLoops~\cite{Cascioli:2011va,Denner:2016kdg} and merged with the \SHERPA parton shower~\cite{Schumann:2007mg} according to the ME+PS@NLO prescription~\cite{Hoeche:2011fd,Hoeche:2012yf,Catani:2001cc,Hoeche:2009rj} using the set of tuned parameters developed by the \SHERPA authors. \SHERPA 2.2.1 is used in the C1N2-Wh analysis and \SHERPA 2.2.11 is used in the C1N2-WZ and C1C1-WW analyses. The event generators, the parton shower and hadronisation routines, and the underlying-event parameter tunes and PDF sets used in simulating the SM background processes, along with the accuracy of the theoretical cross-sections, are all summarised in Table~\ref{tab:MCgen}.
 
For all MC samples showered with \pythia, the EvtGen~v1.2.0~\cite{Lange:2001uf} program was used to simulate the properties of the bottom- and charm-hadron decays.
Systematic uncertainties associated with the different background-specific configurations of the MC generators are estimated by using MC samples produced without detector simulation. The uncertainties include variations of the renormalisation and factorisation scales, the CKKW-L~\cite{Lonnblad:2012ix} matching scale, and different PDF sets and fragmentation/hadronisation models. A detailed discussion of the uncertainties related to the MC modelling is presented in Section~\ref{sec:syst}.

The SUSY signal samples were simulated using \madgraph~v2.6.2~\cite{Alwall:2014hca}  and \pythia~8.230 with the A14~\cite{ATL-PHYS-PUB-2014-021} set of tuned parameters for the modelling of the parton showering (PS), hadronisation and underlying event. The matrix element (ME) calculation was performed at tree level and includes the emission of up to two additional partons. The ME--PS matching was performed using the CKKW-L prescription, with a matching scale set to one quarter of the chargino and next-to-lightest neutralino mass. The NNPDF2.3LO~\cite{Ball:2012cx} PDF set was used.

Signal cross-sections are calculated at NLO accuracy in the strong coupling constant, adding the resummation of soft gluon emission at next-to-leading-logarithmic accuracy (NLO+NLL)~\cite{Beenakker:1999xh,Debove:2010kf,Fuks:2012qx,Fuks:2013vua,Fiaschi:2018hgm}. The nominal cross-section and its uncertainty are taken as the midpoint and half-width of an envelope of cross-section predictions using different PDF sets and factorisation and renormalisation scales, as described in Ref.~\cite{Borschensky:2014cia}.
The simplified models considered for electroweakino production rely on two parameters: for \chinoonep\chinoonem, they are the masses of the \chinoonepm\ and the \ninoone; for \chinoonepm\ninotwo, they are the masses of \chinoonepm / \ninotwo \ (considered to be degenerate) and \ninoone. They also depend on the branching ratio $\mathcal{B}$ of the SUSY particles decays: \chinoonep\ decays into $W$\ninoone\ with $\mathcal{B}=100$\% whilst two separate sets of signal samples were produced for the  \chinoonepm\ninotwo\ process, with \ninotwo\ decaying into a \ninoone \
and either a $Z$ boson or a Higgs boson. In each case, a 100\% branching ratio was assumed for the \ninotwo \ decay.
The production cross-section of the process \chinoonep\chinoonem\ (\chinoonepm\ninotwo ) decreases from 903~fb (1807~fb) to 0.62~fb (1.34~fb) with increasing $m (\chinoonepm/\tilde{\chi}^{0}_{2})$ from 200 to 1000~\GeV.

\begin{table}
\begin{center}
\footnotesize
\resizebox{\textwidth}{!}{
\begin{tabular}{l | c  c  c  c  c }
\hline 
Process & Generator & Parton shower and        & Tune & PDF & Cross-section \\
&           & hadronisation &      &     &               \\
\hline 
$t\bar{t}$  & \POWHEGBOX v2~\cite{Alioli:2010xd,Frixione:2007vw,Frixione:2007nw,Nason:2004rx}    & \pythia 8.230~\cite{Sjostrand:2007gs}   & A14~\cite{ATL-PHYS-PUB-2014-021}     & NNPDF2.3LO~\cite{Ball:2012cx}   & NNLO+NNLL~\cite{Czakon:2011xx}\\
 
Single top  & \POWHEGBOX v2~\cite{Re:2010bp,Frederix:2012dh,Alioli:2009je}    & \pythia 8.230   & A14     & NNPDF2.3LO   & NLO+NNLL~\cite{Aliev:2010zk}\\
 
$W/Z$+jets  & \SHERPA 2.2.1 \&
2.2.11~\cite{Gleisberg:2008ta}    & \SHERPA 2.2.1 \& 2.2.11   & \SHERPA standard & NNPDF3.0NNLO  & NNLO~\cite{Anastasiou:2003ds} \\
 
Diboson     & \SHERPA 2.2.1~\cite{Gleisberg:2008ta} \& 2.2.2     & \SHERPA 2.2.1 \& 2.2.2   & \SHERPA standard & NNPDF3.0NNLO  & NLO~\cite{Anastasiou:2003ds}\\
 
Multiboson    & \SHERPA 2.2.1 \& 2.2.2   & \SHERPA 2.2.1 \& 2.2.2   & \SHERPA standard & NNPDF3.0NNLO  & NLO~\cite{Anastasiou:2003ds}\\
 
$t\bar{t}+V$&\madgraph~v2.3.3 & \pythia 8.210    & A14     & NNPDF2.3LO   &NLO~\cite{Baernreuther:2012ws} \\
 
\ttbar+ $h$       & \POWHEGBOX v2   & \pythia 8.230    & AZNLO~\cite{STDM-2012-23}     & CTEQ6L1~\cite{Nadolsky:2008zw} & NLO~\cite{deFlorian:2016spz}\\
 
$Vh$       & \POWHEGBOX v2   & \pythia 8.212    &  A14     & NNPDF2.3LO   &NLO~\cite{deFlorian:2016spz}  \\
 
\hline 
\end{tabular}
}
\caption{Simulated background MC samples used in this analysis with the corresponding matrix element and parton shower generators, underlying-event tune, PDF set, and cross-section order in $\alpha_s$.}
\label{tab:MCgen}
\end{center}
\end{table}


\section{Event reconstruction}

Events are selected if they have at least one reconstructed vertex with two or more associated tracks each with the transverse momentum $\pt >500$~\MeV. If multiple vertices are associated with an event the primary vertex (PV) is defined as the one with the highest scalar sum of the squared transverse momenta of the associated tracks~\cite{ATL-PHYS-PUB-2015-026}. A set of baseline quality criteria are applied to reject events with non-collision backgrounds or detector noise~\cite{ATLAS-CONF-2015-029}.
 
Candidate jets and leptons have two levels of classification: `baseline' and `signal'. Baseline objects have a lower purity but higher acceptance and are used for the computation of the missing transverse momentum and solving possible reconstruction ambiguities. Signal objects are a subset of baseline objects and are used in the definition of the regions of interest of the searches.
 
All electron candidates are reconstructed from energy deposits in the electromagnetic calorimeter that are matched to charged-particle tracks in the ID~\cite{Aad:2019tso}.
Baseline electron candidates are required to have $\pt>7$~\GeV\ and $|\eta|<2.47$, excluding the transition region between the barrel and endcap calorimeters ($1.3<|\eta|<1.52$), and satisfy the identification requirements of the `loose' operating point provided by a likelihood-based algorithm, described in Ref.~\cite{Aad:2019tso}.
The longitudinal impact parameter $z_0$ relative to the PV is required to satisfy $|z_0\sin\theta| <0.5$ mm.
Discrimination between electrons and converted photons is achieved by observing the number of hits in the innermost pixel layer.
Signal electrons are required to satisfy stricter identification criteria: they are required to satisfy a `tight' likelihood operating point selection and the significance of the transverse impact parameter $d_0$ must satisfy $|d_0/\sigma(d_0)|<5$.
Signal electron candidates are further refined using a multivariate likelihood discriminant, in order to discriminate against electron candidates coming from hadronic
jets, photon conversions and heavy-flavor hadron decays. Electron candidates with $\pt < 75$~\GeV\ use a looser selection on the likelihood output value (\emph{PLVLoose} working point), otherwise the candidates are required to satisfy more stringent selection(\emph{PLVTight} isolation working point), an analogue procedure has been used in Ref.~\cite{PhysRevD.97.072003}.

Muon candidates are reconstructed from matching tracks in the ID and muon spectrometer, refined through a global fit that uses the hits from both subdetectors~\cite{PERF-2015-10}.
Baseline muon candidates are required to have $\pt>6$~\GeV\ and $|\eta|<2.7$, $z_0$ is required to satisfy $|z_0\sin\theta| <0.5$~mm and the `medium' identification criteria.
Signal muon candidates are required to satify stricter requirements on pseudorapidity and impact parameter, $|\eta|<2.5$ and $|d_0/\sigma(d_0)|<3$.
Signal muon candidates are required to satisfy the \emph{PLVLoose} isolation working point if they have $\pt <75$ \GeV, and the \emph{PflowTightVarRad} isolation working point otherwise~\cite{MUON-2018-03}.
Finally a veto is applied on signal muons to reject events with a poorly measured charge-to-momentum ratio $\frac{\sigma(q/p)}{(q/p)}>0.4$.

Jets are reconstructed from three-dimensional topological energy clusters~\cite{PERF-2014-07}~ in the calorimeters using the anti-$k_t$ algorithm with a radius parameter $R=0.4$~\cite{Cacciari:2008gp}.
Baseline jet candidates are required to have $|\eta|<4.5$ and $\pt>20$~\GeV.
Signal jets are required to have $|\eta|<2.8$ and $\pt>30$~\GeV.
To suppress jets from pile-up interactions, signal jet candidates with $|\eta|<2.4$ and $\pt<60$~\GeV\ are required to be matched to the PV through the jet vertex tagger (JVT), a tagging algorithm that identifies jets originating from the PV using track information~\cite{PERF-2014-03, FastJet}, using the \emph{tight} working point. Additionally, jets are calibrated following the criteria in Ref.~\cite{ATL-PHYS-PUB-2015-015}, which, among other things, includes corrections to the jet energy and resolution.
 
To exploit the high \pt\ phase space, large-$\mathrm{R}$ jets are used for the C1C1-WW and C1N2-WZ models to reconstruct highly boosted $W$ and $Z$ bosons by utilising the substructure of collimated objects. Large-$\mathrm{R}$ jets are reconstructed with the same algorithm (anti-$k_t$) as standard jets, but with a large radius parameter of $R=1.0$. To reduce the pile-up contributions to the large-$\mathrm{R}$ jets, a jet trimming algorithm~\cite{Krohn:2009th} is employed, with the $R_{\mathrm{sub}}$ and $f_{\mathrm{cut}}$ parameters set to 0.2 and 0.05, respectively, to refine the jet reconstruction, removing low \pt\ radiation and allowing the parton sub-jets inside the large-$\mathrm{R}$ jets to be identified,  Large-$\mathrm{R}$ jets with \pt\ $>$ 200 \GeV\ and $\vert\eta\vert < 2.0$ are calibrated using ATLAS prescriptions~\cite{ATLAS-CONF-2016-035}, and are identified as possible $W$ or $Z$ candidates using dedicated taggers designed to identify  $W$ and $Z$ bosons at 50\% tagging efficiency~\cite{JETM-2018-03,ATL-PHYS-PUB-2020-017}.
 
Jets originating from the hadronisation of a $b$-quark are identified ($b$-tagged) via a multivariate algorithm that combines information from the impact parameters of displaced tracks and topological properties of secondary and tertiary decay vertices reconstructed within the jet. The $b$-tagging relies on the \emph{DL1r} tagger~\cite{FTAG-2019-07}. The full distribution of the tagger score is used in a procedure referred to as pseudo-continuous $b$-tagging, allowing a more fine-grained calibration of the $b$-tagged jets. The score is divided into five bins defined by fixed $b$-tagging efficiency working points and the distribution edge points (interpreted as the working points at 100\% and 0\% efficiency). The $b$-tagged jets are defined using a working point providing a 77\% efficiency for $b$-hadron identification in \ttbar\ simulated events. The variable quantifying the likelihood of a jet to be $b$-tagged (\emph{$b$-quantile}) according to the pseudo-continuous $b$-tagging procedure is used in the analysis targeting the C1N2-Wh model.
 
To resolve the reconstruction ambiguities between electrons, muons, and jets, an overlap removal procedure is applied to baseline objects.
First, any electron sharing the same ID track with a muon is rejected. If it shares the same ID track with another electron, the one with lower \pt is discarded.
Next, jets are rejected if they lie within  $\Delta R = 0.2$ of a muon or if the muon is matched to the jet through ghost association~\cite{ghostassociation:2008gn}.
Subsequently, electrons within a cone of size $\Delta R = \min(0.4,0.04+10$ \GeV$/\pt)$ around a jet are removed.
Lastly, muons within a cone, defined in the same way as for electrons, around any remaining jet are removed.
 
The missing transverse momentum \mpt, and its magnitude \met,  are reconstructed by using the set of reconstructed and fully calibrated baseline objects, i.e., electrons, muons, photons and jets described above. Baseline photons~\cite{PERF-2017-02} are defined as those that satisfy \pt~$> 25$~\GeV, $|\eta| <2.37$, and the \emph{tight} identification criteria. The determination of the missing transverse momentum also includes a soft term consisting of tracks that are not associated with any reconstructed object. In the searches described here, the \emph{tight} working point is used for the missing transverse momentum~\cite{PERF-2016-07,ATLAS-CONF-2018-023}.


\section{Analysis strategy and event selection}


Three sets of signal regions (SRs) are defined in the analyses, with each set targeting one of the three models considered for electroweakino production and decay. All event selections defined for these regions require that the events were recorded with single lepton (electron and muon) triggers~\cite{TRIG-2018-01, TRIG-2018-05}. The offline lepton $\pt$ thresholds are set to ensure that the selected events are in the plateau region of the corresponding trigger efficiency distribution.  The offline trigger $\pt$  threshold values increased over the years due to the increase in luminosity, going from 25 (21)~\GeV\ to 27 (27.3)~\GeV\ for electron (muon) events.
 
Signal signatures have one leptonically decaying $W$ boson, one hadronically decaying $W$ or $Z$ boson or a Higgs boson decaying into $b$-quarks, and missing transverse momentum due to the $\ninoone$ and neutrinos escaping detection.
Hence events are required to have exactly one signal electron or muon, one to three (two to three) signal jets, allowing for an additional jet from initial- or final-state radiation and large (moderate) \met, for the C1C1-WW and C1N2-WZ (C1N2-Wh) scenarios.
 
A main feature of this analysis is that in the C1C1-WW and C1N2-WZ cases all events are additionally required to contain at least one large-$\mathrm{R}$ jet. This complements previous results ~\cite{SUSY-2017-01} and allows boosted $W$ or $Z$ boson decays to be probed. 
Different boson tagging schemes are employed for different signal scenarios:  W tagging is applied for the $\chinoonep\chinoonem$ scenario, while Z tagging is applied for the \chinoonepm\ninotwo\  scenario. On the other hand, final states with small mass-splitting between the chargino and the LSP (still above the Higgs boson mass) are targeted in the case of the C1N2-Wh models. This complements the analysis presented in Ref. ~\cite{ATLAS:2020pgy}, optimised for  large mass-splittings and small LSP masses. No significant boost is expected for the Higgs boson, and two $b$-tagged jets are required to identify the Higgs boson candidate. A BDT, described in the following, is used as the final discriminant.
 
The SRs for the three analyses are defined through selections that suppress background contributions and maximize the sensitivity to signal. The numbers of residual SM events are then estimated with the aid of MC simulated samples and using a profile likelihood fit~\cite{Baak:2014wma} as detailed in Section~\ref{sec:bkg}. Normalisation factors of the MC samples corresponding to the SM processes expected to contribute the most to the event yields in the SRs are left free to float. A set of control regions (CR), specific for each analysis, are designed to aid in the SM backgrounds evaluation. The likelihood (one for each analysis) is finally built as the product of Poissonian terms for each CR and, when assessing the discovery (model-independent) or exclusion (model-dependent) sensitivity to new physics model, SR. 
 
A set of kinematic variables is built from the physics objects introduced in the previous section and used to define the event selection for the SRs and CRs. They are described in the following.

\begin{itemize}
 
\item The transverse mass, \mt, is defined from the lepton transverse momentum ${\boldsymbol p}_\mathrm{T}^{\ell}$ and the missing transverse momentum \mpt\ as
\begin{linenomath}
\begin{equation}
m_{\mathrm{T}} = \sqrt{2 p_{\mathrm{T}}^{\ell} \met
(1-\cos[\Delta\phi({\boldsymbol p}_{\mathrm{T}}^{\ell},\mpt)])},
\nonumber
\end{equation}
\end{linenomath}
where $\Delta\phi({\boldsymbol p}_{\mathrm{T}}^{\ell},\mpt)$ is the azimuthal angle between ${\boldsymbol p}_{\mathrm{T}}^{\ell}$ and \mpt.
For $W$+jets and semileptonic \ttbar\ events in which one on-shell $W$ boson decays leptonically, this observable has an upper endpoint at the $W$ boson mass, while for signal events the \mt\ distribution extends significantly above $m_{W}$.
A requirement is placed on the upper value of $\Delta\phi({\boldsymbol p}_{\mathrm{T}}^{\ell},\mpt)$ for the analysis targeting the C1C1-WW and C1N2-WZ models to reject background with a high momentum lepton and soft jets, where the angle between lepton and \mpt\ can be large in $\phi$.

\item The missing transverse energy significance, $\sigma_{\met}$~\cite{ATLAS-CONF-2018-038}, is defined as the log-likelihood ($\mathcal{L}$) ratio of measuring the total observed transverse momentum to the likelihood of the null hypothesis,
\begin{eqnarray}
\sigma_{\met} = \sqrt{ 2\ln\left[\frac{\mathrm{max}_{\boldsymbol{p}_{\mathrm{T}}^{\mathrm{inv}}\neq 0}\mathcal{L}\left(\met | \boldsymbol{p}_{\mathrm{T}}^{\mathrm{inv}}\right)}{\mathrm{max}_{\boldsymbol{p}_{\mathrm{T}}^{\mathrm{inv}}= 0}\mathcal{L}\left(\met | \boldsymbol{p}_{\mathrm{T}}^{\mathrm{inv}}\right)}\right]}.
\end{eqnarray}
A high value indicates that the measured \met\ value is not compatible with resolution effects alone and suggests that the event is more likely to contain objects escaping detection, which happens more in the signal events than the background events.

\item Throughout the analyses, variables denoted by $m_{\alpha\beta}$ are invariant masses of particles $\alpha$ and $\beta$. In particular, the invariant mass of the two leading (highest $\pt$) jets, \mjj, is required to be in a range around the $W$ or $Z$ mass 
as signal events are expected to emit an on-shell $W$ or $Z$ boson and have a mass peak in the \mjj\ distribution. The invariant mass of the two jets with highest $b$-tag weight and consistent with being a $b$-jet, \mbb, is required to be close to the Higgs boson mass for C1N2-Wh models. The invariant mass of the lepton and the jets with highest $b$-tag weight is denoted by $m_{\ell b^{}_{i}}$, with $i=1,2$ and referring to the first, second leading $b$-jet, respectively. This observable provides good discrimination against \ttbar\ and single-top background events.

\item The effective mass, $\meff$, is defined as the scalar sum of the lepton transverse momentum, the signal jets' transverse momenta, and the missing transverse momentum,
\begin{eqnarray}
\meff = \ptl + \sum\limits_{\mathrm{jets}}p_{\mathrm{T}} + \met.
\end{eqnarray}
In the design of SRs targeting exclusion sensitivity for the C1C1-WW and C1N2-WZ models, two \meff\ regions are constructed to target low and high signal 
mass differences between the $\chinoonepm / \ninotwo$ and the $\ninoone$.
 
\item The $am^{}_{\mathrm{T2}}$~\cite{Lester:2014yga} is referred to as the asymmetric stransverse mass. The stransverse mass, $m^{}_{\mathrm{T2}}$ is a generalization of the transverse mass applied to signatures where pair-produced parent particles decay semi-invisibly. The $am^{}_{\mathrm{T2}}$ is used if the parent particles decay with different (asymmetric) masses of the invisible particles. For the C1N2-Wh models, where the visible parts of the signal decay are the two $b$-jets and the lepton and the invisible parts are the neutralinos and the neutrino, $am^{}_{\mathrm{T2}}$ is used as a discriminant to reject \ttbar\ contributions. The jets are ordered by their $b$-tag weight as $j_1$ and $j_2$, so that the $am^{}_{\mathrm{T2}}$ is defined as:
\begin{equation}
am^{}_{\mathrm{T2}} = {\mathrm{min}}(m_{\mathrm{T2}}(\ell+j_1,j_2),m_{\mathrm{T2}}(\ell+j_2,j_1)),
\end{equation}
where $m_{\mathrm{T2}}$ is defined as $\mathrm{min}[\mathrm{max}(m_\mathrm{T}^2(p_{\alpha},p),m_\mathrm{T}^2(p_{\beta},q))]$, with $p_{\alpha}$ either $p_{\ell}+p_{j_1}$ or $p_{\ell}+p_{j_2}$ and $p_{\beta}$ either $p_{j_2}+p_{\ell}$ or $p_{j_1}+p_{\ell}$, as the momenta of the visible parts of the decay branches, and $p$ and $q$ are the possible transverse momenta of the invisible particles in the branches. The minimisation is conducted by selecting values for $p$ and $q$ such that their vector sum is equal to the missing transverse momentum.
 
\item The variable $m_{\mathrm{CT}}$\cite{Tovey:2008ui}, referred as contransverse mass, has similar properties to the stransverse mass and is defined as:
\begin{equation}\label{eq:mct}
m_{\mathrm{CT}}^2 = (E_T(\alpha) + E_T(\beta))^2 - (p_T(\alpha) - p_T(\beta))^2,
\end{equation}
where $\alpha$ and $\beta $ are defined as above and $p_T(\alpha)$ and $p_T(\beta)$ are their transverse components. Similarly to $am^{}_{\mathrm{T2}}$, the \mct\ variable is also used for the C1N2-Wh models to reject top-quark SM background contributions, since for signal, the two $b$-jets arise from the same particle (the Higgs boson), while for \ttbar\ and single top-quark production they arise from two different particles.

\end{itemize}
 
SRs are then constructed through selections on these quantities or, for the C1N2-Wh models, using a BDT\@. Their definition follows two approaches: the exclusion SRs are designed for setting model-dependent exclusion limits (`excl.'); the discovery SRs are constructed for model-independent limits and null-hypothesis tests (`disc.' for discovery).
Once SRs are defined, the signal and background yield estimates are computed. The strategy is detailed in Section~\ref{sec:bkg}. The systematic uncertainties, fit and results are then discussed in the following sections.
 
\subsection{C1C1-WW and C1N2-WZ SRs definition}
An overview of the SR definitions targeting the C1C1-WW and C1N2-WZ models is provided in Table~\ref{tab:SR}. The main difference between chargino--chargino and chargino--neutralino signal scenarios is the large-$\mathrm{R}$ jet boson-tagging type. Three separate classes of SRs are defined for each scenario, using \mt\ to target regions sensitive to the increasing mass differences
between the $\chinoonepm$ (and its mass-degenerate $\ninotwo$ wino partner) and the $\ninoone$. These regions are labelled as SRLM, SRMM and SRHM to indicate low (LM), medium (MM) and high (HM) mass differences, respectively.
The requirements on \mt\ make the three regions mutually exclusive.

For both the C1C1-WW and C1N2-WZ models, each LM, MM and HM exclusion SR is further split into two \meff\ bins,
thus providing six exclusion SR bins in total per model for a simultaneous two-dimensional fit in \mt\ and \meff.
The multi-bin approach enhances the sensitivity to a range of SUSY scenarios with different properties. The missing transverse energy significance is optimized separately for low and high \meff\ bins.
In the low \meff\ bin, the \mjj\ reconstructed from two resolved jets is required to be close to the mass of the $W$ or $Z$ boson. This is to improve the sensitivity in a semi-boosted regime where the large-$\mathrm{R}$ jet would catch most of the boson decay products but often two jets are resolved.
The high \meff\ bin is to target a fully boosted topology hence there is  no additional mass constraint on the resolved jets.
For the \chinoonepm\ninotwo\ model, the acceptance times efficiency is ~0.37\% in SRHM for a 600~\GeV\ \chinoonepm/\ninotwo\ mass and massless \ninoone.
For the \chinoonep\chinoonem\ model, the acceptance times efficiency is ~0.31\% in SRMM for a 600~\GeV\ \chinoonepm\ mass and massless \ninoone.
 
The discovery SRs are defined such that the various \meff\ bins are merged for each of the three SRs per model, and selections on \mjj\ and $\sigma_{\met}$ are optimized for the best signal sensitivity at a benchmark point for each  \meff\ bin.


\begin{table}[t!]
\begin{center}
\begin{tabular}{c|c|c|c|c|c|c|c|c|c|c|c|c}
\hline
\hline
Variable &\multicolumn{6}{c|}{\textbf{C1C1-WW model}} & \multicolumn{6}{c}{\textbf{C1N2-WZ model}}  \\
&\multicolumn{2}{c}{\textbf{SRLM}} & \multicolumn{2}{c}{\textbf{SRMM}} & \multicolumn{2}{c|}{\textbf{SRHM}}  & \multicolumn{2}{c}{\textbf{SRLM}} & \multicolumn{2}{c}{\textbf{SRMM}} & \multicolumn{2}{c}{\textbf{SRHM}}  \\
\hline
\hline
$N_{\mathrm{lep}}$ (\pt $> 25$~GeV) & \multicolumn{12}{c}{1} \\
$N_{\mathrm{jet}}$ (\pt $> 30 $~GeV) &  \multicolumn{12}{c}{1--3} \\
$N_{\mathrm{large-R jet}}$ (\pt $>250$~GeV)  &  \multicolumn{12}{c}{$\geq$ 1} \\
\met[GeV] &  \multicolumn{12}{c}{$>$ 200} \\
\dPhimetlep &  \multicolumn{12}{c}{$<$ 2.6} \\
 
Large-$\mathrm{R}$ jet type & \multicolumn{6}{c|}{$W$ tagged} & \multicolumn{6}{c}{$Z$ tagged}\\

\mt[GeV]  &\multicolumn{2}{c}{120--200}& \multicolumn{2}{c}{200--300}& \multicolumn{2}{c|}{$>$ 300}&\multicolumn{2}{c}{120--200}& \multicolumn{2}{c}{200--300}& \multicolumn{2}{c}{$>$ 300}  \\
\hline
& \multicolumn{12}{c}{Exclusion SR}\\
\hline
 
\meff[GeV] (excl.)  & \multicolumn{6}{c|}{[600--850, $>$ 850]}   & \multicolumn{6}{c}{[600--850, $>$ 850]}\\
\mjj[GeV]  (excl.) & \multicolumn{6}{c|}{[70--90,  -- ]}   & \multicolumn{6}{c}{[80--100,  -- ]} \\
$\sigma_{\met}$ (excl.)&  \multicolumn{6}{c|}{[$>$ 12, $>$ 15]} &   \multicolumn{6}{c}{[$>$ 12, $>$ 12]}\\
 
\hline
& \multicolumn{12}{c}{Discovery SR}\\
\hline
\meff[GeV] (disc.) & \multicolumn{2}{c}{$>$ 600} & \multicolumn{2}{c}{$>$ 600} & \multicolumn{2}{c|}{$>$ 850} & \multicolumn{2}{c}{$>$ 600} & \multicolumn{2}{c}{$>$ 850} & \multicolumn{2}{c}{$>$ 850}\\
\mjj[GeV] (disc.)  & \multicolumn{2}{c}{-} & \multicolumn{2}{c}{-} & \multicolumn{2}{c|}{-}   & \multicolumn{2}{c}{80--100} & \multicolumn{2}{c}{-} &\multicolumn{2}{c}{-} \\
$\sigma_{\met}$ (disc.) &  \multicolumn{2}{c}{$>$ 15}& \multicolumn{2}{c}{$>$ 15}&\multicolumn{2}{c|}{$>$ 15}&\multicolumn{2}{c}{$>$ 12}&\multicolumn{2}{c}{$>$ 12}&\multicolumn{2}{c}{$>$ 12}   \\

\hline

\hline
\hline
\end{tabular}
\end{center}
\caption{Overview of the selection criteria for the exclusion SRs and the discovery SRs used in C1C1-WW and C1N2-WZ models.
For exclusion SRs, they are further divided into two \meff\ bins. The selection on \mjj\ and $\sigma_{\met}$ varies for low and high \meff\ bins.  For discovery SRs, one SR is defined per \mt\ region. The symbol `--' indicates no additional requirement.}
\label{tab:SR}
\end{table}


\subsection{C1N2-Wh SR definition}
 
The analysis targets scenarios characterised by $\Delta m (\chinoonepm, \tilde{\chi}^{0}_{1})$ of at least $m_h$, where selections on individual variables are expected to be sub-optimal to separate the signal from SM production processes due to the similar kinematic properties of SUSY and SM background events. Consequently, a  multivariate approach, where the discriminating power of multiple observables is exploited at once, is expected to increase the sensitivity. A BDT is implemented in the analysis by making use of the \textsc{XGBoost} (XGB)~\cite{Chen_2016} framework.  The training procedure uses events that satisfy an initial selection that requires exactly one signal lepton with $p_T > 27$~\GeV, two to three jets with $p_T > 30$~\GeV, exactly two $b$-tagged jets, \met $> 50$~\GeV, $\sigma_{\met} > 5$ and \mbb\ in the range of 50--200~\GeV.
A set of object-based and event-based variables (30 in total) are used in the training. Object-based variables include the $p_T$, $\eta$ and $\phi$ of the lepton and the jets, and the $b$-quantile of the jets. Event-based observables include \mbb, $am^{}_{\mathrm{T2}}$, \mct, $m_{\mathrm{T}}$, $\sigma_{\met}$, $m_{\ell b^{}_{i},i=1,2}$ and radial distances between pairs of visible objects.
 
Events are classified in five different categories: three corresponding to the main backgrounds processes ($t\bar{t}$, single-top and $W$+jets), one including all remaining minor background processes ($Z$+jets, diboson, rare processes), and one grouping together the signal samples in the region $m_h < \Delta m (\chinoonepm, \tilde{\chi}^{0}_{1}) < 200$~\GeV. The grouping of multiple signal samples increases the statistical power and is enabled by the similarity of the kinematic properties of the SUSY models of interest. A one-versus-rest multi-classification procedure was used, wherein each class is fitted against all the other classes producing output scores containing the predicted probability of an event being in that class.
This method is more effective in discriminating the signal from the dominant backgrounds than using a binary
signal versus background classifier. This also has the additional benefit of having background-processes classification scores that can be used to increase the purity of different backgrounds whilst building control and validation regions.
 
The output score $w^{}_{\mathrm{sig}}$ denotes the signal class output score and is used in the definition of the SRs. The scores of the background classes are used in the definition of CRs and validation regions (VR). Tools to interpret the BDT learning process are used to identify the most relevant observables.  The \mbb\ variable has the most predictive power for signal, as expected since it is used to identify the Higgs boson candidate. The $m_{\mathrm{T}}$ and the $am^{}_{\mathrm{T2}}$ are, on the other hand, the most predictive variables for $W$+jets and $t\bar{t}$ events, respectively.
 
The final selection for the analysis targeting the C1N2-Wh model requires $w^{}_{\mathrm{sig}}>0.91$ (where the BDT score is defined between 0 and 1) and more stringent requirements on the invariant mass of the two $b$-tagged jets, $95<$\mbb$<140$~\GeV, and on the missing transverse energy significance, $\sigma_{\mathrm{\met}}>8$. As an example, the acceptance times efficiency is around ~0.1\%  for a 350~\GeV\ \chinoonepm/\ninotwo\ mass and a 150~\GeV\ \ninoone\ mass. When assessing the exclusion sensitivity for the signal-plus-background hypothesis, four $w^{}_{\mathrm{sig}}$ bins are used in the likelihood fit: [0.91, 0.928), [0.928, 0.948), [0.948, 0.964), [0.964, 1], referred to as SRXGB Bin 1--4. Three discovery signal regions are defined, integrated over the signal score with increasing thresholds on it, namely [0.928,1], [0.948,1] and [0.964, 1].


\section{Background estimation}
\label{sec:bkg}

 
Dominant SM background sources in the SRs depend on the analysis. For analyses targeting the C1C1-WW and C1N2-WZ models, $W$+jets and diboson production events constitute the main background (46\%--73\% and 16\%--39\%, respectively, depending on the SR). Subdominant SM background contributions originate from $Z$+jets, single-top, multiboson, \ttbar~+~$V$, \ttbar+$h$ and $Vh$. For the analysis targeting C1N2-Wh models, \ttbar and single-top (almost exclusively $Wt$ production) processes are dominant (about 30\%--35\% each across all bins), followed by $W$+jets (15\%--20\%). Remaining contributions originate from $Z$+jets, diboson, \ttbar~+~$V$ and Higgs boson production processes.
 
The main background contributions are estimated by using partially data-driven techniques through the set of CRs designed to be mutually exclusive and non-overlapping with the SRs (across and within the three analyses), and characterised by negligible expected signal contributions for the models of interest.
The expected background yield in each SR is determined in the profile likelihood fit using the `background-only fit' approach~\cite{Baak:2014wma}. With this fit, the normalisation of the major backgrounds is adjusted to match the data in CRs with negligible signal contamination. A probability density function is defined for each CR\@. The inputs are the observed event yield and the predicted background yield from simulation, with Poisson statistical uncertainties and systematic uncertainties (detailed in Section~\ref{sec:syst}) as nuisance parameters. The nuisance parameters are constrained by Gaussian distributions with widths corresponding to the sizes of the uncertainties. Systematic uncertainties account for bin-to-bin correlations, with normalisation and nuisance parameters correlated in all regions. The product of all the probability density functions forms the likelihood, which is maximised by adjusting the normalisation and nuisance parameters. The resulting normalisation factors are then used to correct the expected yields of the corresponding backgrounds in the various SRs. The extrapolation of the adjusted normalisation and nuisance parameters to the SRs is checked in VRs, which kinematically resemble the SRs but are expected to have low signal contamination, and do not overlap with either CRs or SRs.
 
\subsection{C1C1-WW and C1N2-WZ Control and Validation regions}
 
For the diboson background, single-lepton processes ($\ell vvv$) and dilepton processes ($\ell \ell vv$) contribute equally to the backgrounds in the signal regions. The $\ell vvv$ ($\ell \ell vv$) process is marked as diboson1l (diboson2l) in the following yield tables and kinematic figures. The diboson $\ell \ell vv$ entering in the SRs are events with two real leptons present in the decay chain, where one lepton fails the signal lepton requirement, or escapes detection.
The $\ell \ell vv$ background is estimated and validated in the two-lepton control and validation regions.
The crucial variable to enrich the diboson background contribution in the corresponding CR is the dilepton invariant mass, which is required to be consistent with the SM $Z$ boson mass.
Further selection criteria for \met, $\sigma_{\met}$, and  $\Delta\phi({\boldsymbol p}_{\mathrm{T}}^{\ell},\mpt)$ are defined similarly to the SRs, but with less stringent bounds, to enhance the number of events. An additional veto on the \mjj\ variable minimizes the potential overlap with a complementary chargino and neutralino search with two leptons and two jets in the final states performed by ATLAS~\cite{SUSY-2018-05}, to allow future statistical combinations of different channels targeting the same SUSY production processes.
In addition to the these selections, the control region DB2LCR  requires  \mt in the range of 50--200~\GeV\ and the validation region DB2LVR requires \mt\ in the range of 200--350~\GeV\ .

The single-lepton diboson process $\ell vvv$ has one lepton and missing energy in the final state, the kinematic behaviour of which is identical to $W$+jets background. A set of shared control and validation regions, the WDB1L regions, are designed for these two processes. The CR is defined with a selection similar to the SRs, but with \mt\ in the range of 50--80~\GeV\ and with inverted $\sigma_{\met}$ requirements. A $b$-jet veto is applied to reduce heavy flavour contamination. Two sets of VRs are defined: the VR1 validates the extrapolation from the CR to the SRs in \mt, and the VR2 validates the extrapolation from the CR to the SRs in $\sigma_{\met}$ and \mt. The control and validation regions share the same \meff\ binning as the signal regions.
The \ttbar control and validation regions, namely TCR, TVR1, and TVR2, have the same selections as the WDB1L regions, except for the requirement of at least one $b$-tagged jet.

 
\begin{table}[b!]
\begin{center}
\small
\begin{tabular}{c|c|c}
\hline
\hline
Variable            & \multicolumn{2}{c}{\textbf{DB2L}} \\
& \textbf{CR} & \textbf{VR} \\
\hline
\hline
$N_{\mathrm{lep}}$ (\pt $> 25$~GeV) & \multicolumn{2}{c}{2} \\ 
$N_{\mathrm{jet}}$ (\pt $> 30 $~GeV) & \multicolumn{2}{c}{1--3} \\ 
$N_{\mathrm{b-jet}}$ (\pt $> 30 $~GeV) & \multicolumn{2}{c}{0} \\ 
\met[GeV] & \multicolumn{2}{c}{$> 200 $} \\ 
\dPhimetlep & \multicolumn{2}{c}{$< 2.9$} \\ 
\mll[GeV] & \multicolumn{2}{c}{70--100} \\ 
\mjj veto [GeV] & \multicolumn{2}{c}{75--95} \\ 
$\sigma_{\met}$ & $> 12$ & $> 10$ \\
\mt[GeV] & 50--200 & 200--350 \\
\hline
\hline
\end{tabular}
\end{center}
\caption{Overview of the CR and VR definitions for diboson $\ell \ell vv$ backgrounds. The $N_{\mathrm{lep}}$ variable provides the orthogonality to the SR.} 
\label{tab:CRVR2L}
\end{table}

\begin{table}[t!]
\begin{center}
\small
\begin{tabular}{c|c|c|c}
\hline
\hline
Variable            & \multicolumn{3}{c}{\textbf{WDB1L} and \textbf{T}} \\
&  \textbf{CR}  & \textbf{VR1} & \textbf{VR2} \\
\hline
\hline
$N_{\mathrm{lep}}$ (\pt $> 25$~GeV) & \multicolumn{3}{c}{1} \\
$N_{\mathrm{jet}}$ (\pt $> 30$~GeV)&  \multicolumn{3}{c}{1--3} \\
$N_{\mathrm{b-jet}}$ (\pt $> 30$~GeV) & \multicolumn{3}{c}{0 for WDB1L; $> 0$ for Top} \\
$N_{\mathrm{large-R jet}}$ (\pt $> 250$~GeV)  &  \multicolumn{3}{c}{$\geq$ 1} \\
\met[GeV] & \multicolumn{3}{c}{$>$ 200 } \\
\dPhimetlep & \multicolumn{3}{c}{$<$ 2.9} \\
Large-$\mathrm{R}$ jet type & \multicolumn{3}{c}{W-tagged}\\
\meff[GeV]  & \multicolumn{3}{c}{[600--850, $>$ 850]}\\
$\sigma_{\met}$ & \multicolumn{1}{c|}{$< 12$} & \multicolumn{1}{c|}{$< 12$} & \multicolumn{1}{c}{$> 12$} \\
\mt[GeV] & \multicolumn{1}{c|}{50--80} & \multicolumn{1}{c|}{$ > 80$} & \multicolumn{1}{c}{50--120} \\
 
\hline
\hline
\end{tabular}
\end{center}
\caption{Overview of the CR and VR definitions for $W$+jets, diboson $\ell vvv$ and $\ttbar$ backgrounds. They share the same CR and VR definitions except for number of $b$-tagged jets requirement. The \mt\ variable provides the orthogonality to the SR.}
\label{tab:CRVR1L}
\end{table}


A summary of all CR and VR selection criteria is reported in Tables~\ref{tab:CRVR2L} and \ref{tab:CRVR1L}. The $W$+jets ($\ell vvv$) purity is 42\%-56\% (13\%-21\%) in WDB1LCR. The \ttbar\ purity is 58\%--77\%  in TCR and $\ell \ell vv$ purity is ~58\% in DB2LCR.

Sub-dominant background processes, such as $Z$+jets, single-top, multiboson, \ttbar+$V$, \ttbar+$h$ and $Vh$, which have no dedicated control regions, are normalised to the cross-sections indicated in Table~\ref{tab:MCgen}. Similarly to the dominant backgrounds, their expected yields in the SRs are subject to statistical and systematic uncertainties. Backgrounds with misidentified (fake) leptons such as jets misreconstructed as a lepton, and events with leptons originating from a jet produced by heavy-flavour quarks or from photon conversions, are estimated by using a matrix method as described in Ref.~\cite{EGAM-2019-01}, and found to be negligible in all regions.
 
\subsection{C1N2-Wh Control and Validation regions}
 
The multi-class BDT approach results in a classifier-output score for each of the background categories. Only the three categories representing the dominant backgrounds (\ttbar, single-top, and $W$+jets associated production) are considered, and selections on their output scores ($w^{}_{\ttbar}$, $w^{}_{\mathrm{st}}$ and $w^{}_{W+\mathrm{jets}}$, respectively) are applied to define the CRs after the initial common selection. Table~\ref{tab:CR_Wh_def} shows the definition of the CRs.  The selections on the output scores are defined to maximize the purity of the CR for the targeted background. To reduce the contamination from signals of interest to a negligible level, the $w^{}_{\mathrm{sig}}$ score is also required to be low. The purity of the CRs obtained with these selections are 95\%, 56\% and 72\%  for $t\bar{t}$, single-top and $W$+jets, respectively. In the case of the single-top CR, most of the remaining events arise from $t\bar{t}$ production.
 
\begin{table}[t!]
\begin{center}
\begin{tabular}{l|c|c|c}
\hline \hline
Variable & \multicolumn{3}{|c}{Regions} \\ 		\hline
\met[GeV] & \multicolumn{3}{|c}{$> 50$} \\
$N_{\mathrm{lep}}$ ($\pt > 27$~GeV) & \multicolumn{3}{|c}{1} \\
$N_{\mathrm{jet}}$ ($\pt > 30$~GeV) & \multicolumn{3}{|c}{2--3} \\
$N_{\mathrm{b-jet}}$ ($\pt > 30$~GeV) & \multicolumn{3}{|c}{2} \\
$m^{}_{\mathrm{bb}}$~[GeV]               & \multicolumn{3}{|c}{$\in[50,200]$} \\
$\sigma_{\met}$                   & \multicolumn{3}{|c}{$>5$}\\
\hline\hline
& CR\ttbar (\textbf{CRttXGB}) & CR single-top (\textbf{CRstXGB}) & CR W+jets (\textbf{CRWXGB})\\ \hline
$w^{}_{\mathrm{sig}}$                & $\in[0.2,0.3]$ & $\in[0,0.2]$ & $\in[0.0,0.2]$ \\
$w^{}_{\mathrm{t\bar{t}}}$             & $>0.73$ & -- & -- \\
$w^{}_{\mathrm{st}}$                 & $<0.2$ & $>0.45$ & $<0.2$ \\
$w^{}_{\mathrm{W+jets}}$             & $<0.4$ & -- & $>0.65$ \\
\hline\hline
& VR \ttbar (\textbf{VRttXGB}) & VR single-top (\textbf{VRstXGB}) & VR W+jets (\textbf{VRWXGB})\\
\hline
 
$w_{\mathrm{sig}}$                & $\in[0.4,0.9]$ & $\in[0.2,0.9]$ &  $\in[0.2,0.9]$ \\
$w_{\mathrm{\ttbar}}$             & $>0.4$ & -- & -- \\
$w_{\mathrm{st}}$                 & $<0.2$ & $>0.2$ & $<0.2$ \\
$w_{\mathrm{W+jets}}$                  & $<0.4$ & -- & $>0.4$ \\ \hline\hline
 
\end{tabular}
\end{center}
\caption{Definition of the CRs and VRs used to estimate \ttbar, single-top and $W$+jets background processes.}
\label{tab:CR_Wh_def}
\end{table}
 
VRs are defined to validate the extrapolation from CRs. Each have events selected with tightened scores towards their respective background class and a higher signal classification score that approaches the SRs range. 
The requirements on the signal score are such that the validation regions are orthogonal to both the CRs (by the lower bound) and the SR (by the upper bound). A selection on the background classification scores is maintained in VRs to isolate the extrapolation from each of the control regions and to reduce potential signal contamination to reasonable levels ($<$10\% for all models). The definition of the validation regions are also given in Table~\ref{tab:CR_Wh_def}.
 
Similarly to the C1C1-WW and C1N2-WZ models analyses, sub-dominant background processes, such as $Z$+jets, diboson, multiboson, \ttbar+$V$, \ttbar+$h$ and $Vh$, are normalised to their respective cross-sections and their expected yields in the SRs are subject to statistical and systematic uncertainties. Background contributions from misidentified leptons are also evaluated with the matrix method and found negligible.


\section{Systematic uncertainties}
\label{sec:syst}

The background yield in the SRs is subject to theoretical and experimental systematic uncertainties. The source of these systematic uncertainties for all simulated signal and background processes are evaluated and presented in this section.
 
The experimental uncertainties are related to the jet energy scale (JES), jet energy resolution (JER), $b$-tagging, \met modelling, lepton reconstruction and identification,  pile-up, and JVT\@.
The dominant uncertainties across all analyses and SRs arise from JES and JER, which are parameterized as a function of the \pt\ and $\eta$ of the jet, the pile-up conditions, and the jet flavour composition~\cite{PERF-2016-04}. The uncertainties arising from the large-$\mathrm{R}$ jet-boson tagging are grouped into JES and JER systematic uncertainties, for the C1C1-WW and C1N2-WZ SRs. The impact of uncertainties on the efficiencies and mis-tag rates of the $b$-tagging algorithm is relevant for the C1N2-Wh SRs and is estimated by varying, as a function of \pt, $\eta$ and jet flavour, the scale factors used to correct the MC simulation, in a range that reflects the uncertainty in their measurement~\cite{FTAG-2018-01,ATLAS-CONF-2018-055}.
The \met modelling systematic uncertainties are estimated by propagating the uncertainties in the energy and momentum scale of each of the objects entering the calculation, and the uncertainties in the soft term's resolution and scale~\cite{ATLAS-CONF-2018-023}.
The evaluation of the lepton reconstruction and identification uncertainties is performed using $Z\rightarrow\ell^+\ell^-$, $J/\psi\rightarrow\ell^+\ell^-$ samples ~\cite{EGAM-2018-01,PERF-2015-10}.
The procedure of pile-up reweighting is applied to the simulation to match the number of reconstructed vertices to the data. The pile-up uncertainty is estimated by performing a reweighting in which $\langle \mu\rangle$ is varied by $\pm4\%$.
 
Uncertainties in the modelling of the SM background processes from MC simulation are profiled for dominant backgrounds in dedicated control regions, where the systematic uncertainties only have an impact on the extrapolation factors, while for sub-dominant backgrounds they are entirely estimated from simulation and affect the inclusive cross-section for each process and the acceptance of the analysis selection in all regions. They are assumed to be fully correlated across signal regions within the same analysis, but uncorrelated between different processes.
 
Theoretical uncertainties in the \ttbar and single-top backgrounds are dominant for the analysis targeting C1N2-Wh models, but also relevant for the others. They take into account uncertainties due to modelling of the hard-scattering, evaluated through a comparison between the nominal \POWHEGBOX +\PYTHIA 8 sample and the alternative a\textsc{MC@NLO} +\PYTHIA 8 sample, and uncertainties arising from the parton shower and hadronisation models, derived from comparisons between samples generated with \POWHEGBOX +\PYTHIA 8 and \POWHEGBOX +\HERWIG 7~\cite{Bellm:2015jjp}. Variations of the renormalisation and factorisation scales (scaled up and down by a factor of two),
the initial- and final-state radiation parameters and PDF sets are also considered. The uncertainty assigned to the interference between single-top $Wt$ and $t\bar{t}$ production~\cite{ATL-PHYS-PUB-2016-004} is obtained by comparing  diagram removal (DR) and diagram subtraction (DS) samples, modelled by \POWHEGBOX +\PYTHIA 8 for the C1C1-WW and C1N2-WZ channels. In the case of the C1N2-Wh analysis, the $Wt$ predictions of the DS sample in some of the CRs and VRs are significantly lower than those of the nominal sample and the data, such that a systematic uncertainty estimation by comparison is not possible. A conservative 35\% uncertainty is assumed for the uncertainty in the interference between the $Wt$ and $t\bar{t}$ processes, following studies reported in Ref.\cite{ATLAS:2020pgy} and Ref.\cite{ATLAS:2018qmw}.
 
The diboson modelling uncertanties are among the dominant uncertainties in the C1C1-WW and C1N2-WZ channels, and are studied separately for the single-lepton and the dilepton processes. They are evaluated by studying the envelope of the variations of the renormalisation and factorisation scales. Variations of the renormalisation and factorisation scales are also applied to $W/Z$+jets, multiboson, \ttbar+$V$, \ttbar+$h$, and $Vh$. The PDF uncertainties are considered following the PDF4LHC15 recommendations~\cite{Butterworth:2015oua}.
 
For $W/Z$+jets, the resummation (QSF) and matching scale (CKKW-L)~\cite{Lonnblad:2011xx} for the $W/Z$+jets are estimated by varying the scale parameters up and down for the \SHERPA generator. Further, for \SHERPA 2.2.11 $W/Z$+jets samples, the electroweak NLO correction uncertainties are assigned to account for the impact of applying different correction methods.
An overall 5\%  systematic uncertainty in the inclusive cross-section is assigned for the $Z$+jets samples~\cite{ATL-PHYS-PUB-2016-002} and similar cross-section uncertainties, 5\%--10\%, are also assigned for other sub-dominant  background contributions.

The variations of the parameters corresponding to the factorisation, renormalisation and CKKW-L matching scales in a\textsc{MC@NLO} +\PYTHIA 8 samples provide the uncertainties for the two simplified signal models considered.
 
The dominant systematic uncertainties in the background estimates for the signal regions are presented in Tables~\ref{systematicssummaryWW}, \ref{systematicssummaryWZ} and \ref{systematicssummaryWh}. Theoretical and experimental uncertainties are shown for each of the dominant background contribution. The uncertainties in the scale factor fits to the control regions are listed as `Normalisation of dominant backgrounds'. For the SRs targeting the C1C1-WW and C1N2-WZ models, they contribute around 6\%--7\% across regions. For the analysis targeting the C1N2-Wh model, they are  around 9\%--16\% and are dominated by \ttbar\ and single-top backgrounds.
The largest individual experimental uncertainty amounts to 4\%--17\%  depending on the SR for C1C1-WW and C1N2-WZ\@. For the C1N2-Wh analysis, the dominant experimental uncertainty arises from the JER (12\%--20\%) followed by the JES and $b$-tagging. The MC statistical uncertainties contribute up to 19\% depending on the SR and analysis.

\begin{table}
\begin{center}
\setlength{\tabcolsep}{0.0pc}
\begin{tabular*}{\textwidth}{@{\extracolsep{\fill}}lccc}
\noalign{\smallskip}\hline\noalign{\smallskip}{\textbf{C1C1-WW model}}                                    & SRLM           & SRMM            & SRHM            \\
\noalign{\smallskip}\hline\noalign{\smallskip}
Total background expectation             &  $22.0$        &  $9.2$        &  $15.7$       \\
\noalign{\smallskip}\hline\noalign{\smallskip}
Total background systematic uncertainty              & $\pm 3.2\ $        & $\pm 2.5\ $        & $\pm 3.0\ $             \\
\noalign{\smallskip}\hline\noalign{\smallskip}
\noalign{\smallskip}\hline\noalign{\smallskip}
\multicolumn{4}{c}{Theoretical systematic uncertainties}\\
\noalign{\smallskip}\hline\noalign{\smallskip}
\ttbar\          & $\pm 1.1\ $           & $\pm 0.25\  $           & $\pm 0.15\ $       \\
Single top       & $\pm 0.31\ $            & $\pm 0.08\ $           & $\pm 0.35\  $       \\
$W$+jets         & $\pm 0.4\ $            & $\pm 0.15\ $           & $\pm 0.32\ $       \\
Diboson          & $\pm 0.29\ $            & $\pm 0.24\ $           & $\pm 0.26\ $       \\
Other backgrounds         & $\pm 0.10\ $   & $\pm 0.07\  $           & $\pm 0.08\  $       \\
\noalign{\smallskip}\hline\noalign{\smallskip}
\multicolumn{4}{c}{MC statistical uncertainties}\\
\noalign{\smallskip}\hline\noalign{\smallskip}
MC statistical uncertainty         & $\pm 2.1\ $          & $\pm 1.6\  $          & $\pm 2.1\  $       \\
\noalign{\smallskip}\hline\noalign{\smallskip}
\multicolumn{4}{c}{Uncertainties in the background normalisation}\\
\noalign{\smallskip}\hline\noalign{\smallskip}
Normalisation of dominant backgrounds         & $\pm 1.5\  $          & $\pm 0.6\ $          & $\pm 1.1\  $       \\
\noalign{\smallskip}\hline\noalign{\smallskip}
\multicolumn{4}{c}{Experimental systematic uncertainties}\\
\noalign{\smallskip}\hline\noalign{\smallskip}
Jet energy resolution       & $\pm 1.1\  $             & $\pm 1.0\ $          & $\pm 1.4\ $       \\
Jet energy scale            & $\pm  1.7\ $             & $\pm 1.5\  $           & $\pm 1.1\ $       \\
\met                        & $\pm 0.5\  $             & $\pm 0.26\  $           & $\pm 0.6\ $       \\
Lepton uncertainties        & $\pm 0.4\ $             & $\pm 0.10\ $           & $\pm 0.5\ $       \\
Pile-up/JVT                      & $\pm 0.10\ $             & $\pm 0.21\ $           & $\pm 0.23\ $       \\
\noalign{\smallskip}\hline\noalign{\smallskip}
\end{tabular*}
\end{center}
\caption[Breakdown of uncertainty in background estimates]{
Breakdown of the dominant systematic uncertainties in background estimates in the various exclusion signal regions for the C1C1-WW model.
The individual uncertainties can be correlated, and do not necessarily add up in quadrature to
the total background uncertainty. 
\label{systematicssummaryWW}}
\end{table}

\begin{table}
\begin{center}
\setlength{\tabcolsep}{0.0pc}
\begin{tabular*}{\textwidth}{@{\extracolsep{\fill}}lccc}
\noalign{\smallskip}\hline\noalign{\smallskip}{\textbf{C1N2-WZ model}}                                    & SRLM            & SRMM            & SRHM            \\
\noalign{\smallskip}\hline\noalign{\smallskip}
Total background expectation             &  $29$        &  $12.7$        &  $17.0$       \\
\noalign{\smallskip}\hline\noalign{\smallskip}
Total background systematic uncertainty           & $\pm 4\ $        & $\pm 2.5\ $        & $\pm 2.9\ $             \\
\noalign{\smallskip}\hline\noalign{\smallskip}
\noalign{\smallskip}\hline\noalign{\smallskip}
\multicolumn{4}{c}{Theoretical systematic uncertainties}\\
\noalign{\smallskip}\hline\noalign{\smallskip}
\ttbar\          & $\pm 0.9\ $            & $\pm 0.29\ $           & $\pm 0.20\ $       \\
Single top       & $\pm 1.1\ $            & $\pm 0.24\ $           & $\pm 0.5\ $       \\
$W$+jets         & $\pm 0.6\ $            & $\pm 0.22\ $           & $\pm 0.4\ $       \\
Diboson          & $\pm 0.5\ $            & $\pm 0.24\ $           & $\pm 0.6\ $       \\
Other backgrounds         & $\pm 0.15\ $   & $\pm 0.18\ $           & $\pm 0.09\ $       \\
\noalign{\smallskip}\hline\noalign{\smallskip}
\multicolumn{4}{c}{MC statistical uncertainties}\\
\noalign{\smallskip}\hline\noalign{\smallskip}
MC statistical uncertainty         & $\pm 2.5\ $          & $\pm 1.4\ $          & $\pm 2.1\ $       \\
\noalign{\smallskip}\hline\noalign{\smallskip}
\multicolumn{4}{c}{Uncertainties in the background normalisation}\\
\noalign{\smallskip}\hline\noalign{\smallskip}
Normalisation of dominant backgrounds         & $\pm 2.0\ $          & $\pm 0.8\ $          & $\pm 1.2\ $       \\
\noalign{\smallskip}\hline\noalign{\smallskip}
\multicolumn{4}{c}{Experimental systematic uncertainties}\\
\noalign{\smallskip}\hline\noalign{\smallskip}
Jet energy resolution       & $\pm 1.1\  $             & $\pm 1.3\ $           & $\pm 0.8\ $       \\
Jet energy scale            & $\pm 1.3\ $             & $\pm 1.0\ $           & $\pm 1.3\ $       \\
\met                        & $\pm 0.5\ $             & $\pm 0.6\ $           & $\pm 0.07\ $       \\
Lepton uncertainties        & $\pm 0.34\ $             & $\pm 0.23\ $           & $\pm 0.20\ $       \\
Pile-up/JVT                     & $\pm 0.06\ $            & $\pm 0.8\ $            & $\pm 0.11\ $       \\
\noalign{\smallskip}\hline\noalign{\smallskip}
\end{tabular*}
\end{center}
\caption[Breakdown of uncertainty in background estimates]{
Breakdown of the dominant systematic uncertainties in background estimates in the various exclusion signal regions for the C1N2-WZ model.
The individual uncertainties can be correlated, and do not necessarily add up in quadrature to the total background uncertainty. 
\label{systematicssummaryWZ}}
\end{table}

\begin{table}
\small{
\begin{center}
\setlength{\tabcolsep}{0.0pc}
\begin{tabular*}{\textwidth}{@{\extracolsep{\fill}}lcccc}
\noalign{\smallskip}\hline\noalign{\smallskip}{\small{\textbf{C1C1-Wh model}}} &  SRXGB Bin 1  & SRXGB Bin 2  & SRXGB Bin 3    & SRXGB Bin 4      \\
& [0.91, 0.928) & [0.928, 0.948) & [0.948, 0.964) & [0.964, 1] \\
 
\noalign{\smallskip}\hline\noalign{\smallskip}
Total background expectation        &  $9.4$        &  $5.7$    &  $4.2$   &  $2.2$       \\
 
\noalign{\smallskip}\hline\noalign{\smallskip}
 
Total background systematic uncertainty    & $\pm 2.1\ $  & $\pm 2.0\ $ & $\pm 1.4\ $ & $\pm 0.7\ $             \\
 
\noalign{\smallskip}\hline\noalign{\smallskip}
\noalign{\smallskip}\hline\noalign{\smallskip}
\multicolumn{5}{c}{Theoretical systematic uncertainties}\\
\noalign{\smallskip}\hline\noalign{\smallskip}
\ttbar\          & $\pm 1.1\  $   & $\pm 0.7\ $   & $\pm 0.5\ $  & $\pm 0.10\ $  \\ 
Single top       & $\pm 1.2\ $   & $\pm 0.9\  $  & $\pm 0.9\ $  &  $\pm 0.4\  $     \\
$W$+jets         & $\pm 0.17\ $  & $\pm 0.14\ $   & $\pm 0.12\ $  &   $\pm 0.04\ $  \\
Other backgrounds  & $\pm 0.14\ $   & $\pm 0.13\ $  & $\pm 0.13\  $ & $\pm 0.10\ $       \\
\noalign{\smallskip}\hline\noalign{\smallskip}
\multicolumn{5}{c}{MC statistical uncertainties}\\
\noalign{\smallskip}\hline\noalign{\smallskip}
MC statistical uncertainty  & $\pm 1.0\ $ & $\pm 0.8\ $ & $\pm 0.7\ $ & $\pm 0.4\ $\\
\noalign{\smallskip}\hline\noalign{\smallskip}
\multicolumn{5}{c}{Uncertainties in the background normalisation}\\
\noalign{\smallskip}\hline\noalign{\smallskip}
Normalisation of dominant backgrounds & $\pm 1.3\ $  & $\pm 0.9\ $  & $\pm 0.5\ $ & $\pm 0.19\ $      \\
\noalign{\smallskip}\hline\noalign{\smallskip}
\multicolumn{5}{c}{Experimental systematic uncertainties}\\
\noalign{\smallskip}\hline\noalign{\smallskip}
Jet energy resolution  & $\pm  1.1\ $ & $\pm 1.2\ $   & $\pm 0.6\ $ & $\pm 0.4\ $     \\
Jet energy scale       & $\pm 0.5\ $  & $\pm 0.31\ $   & $\pm 0.33\ $  &  $\pm 0.07\ $   \\
$b$-tagging    & $\pm 0.12\ $  & $\pm 0.8\ $   & $\pm 0.05\ $  & $\pm 0.06\ $       \\
Pile-up/JVT    & $\pm 0.4\ $  & $\pm 0.5\ $  & $\pm 0.29\ $  & $\pm 0.09\ $      \\
Lepton and \met uncertainties     & $\pm 0.05\ $  & $\pm 0.4\ $  & $\pm 0.14\ $  & $\pm 0.12\ $     \\
\noalign{\smallskip}\hline\noalign{\smallskip}
\end{tabular*}
\end{center}
}
\caption[Breakdown of uncertainty in background estimates]{
Breakdown of the dominant systematic uncertainties in background estimates in the various exclusion signal regions for the C1N2-Wh model.
The individual uncertainties can be correlated, and do not necessarily add up in quadrature to
the total background uncertainty. 
The category `Others' refers to diboson, \ttbar\+$W/Z$, Higgs boson and $Z$+jets events.
\label{systematicssummaryWh}}
\end{table}
 
\pagebreak
\clearpage


\section{Results}
\label{sec:result}

The observed event yield in each of the exclusion signal regions for the three analyses is summarized in Tables~\ref{table_results_C1C1_WW},  \ref{table_results_C1N2_WZ} and~\ref{table_results_C1N2_Wh} for the C1C1-WW, C1N2-WZ and C1N2-Wh models, respectively. Yields in data are reported with the corresponding SM predictions obtained from the background-only fit, where the predicted post-fit level of background is compared with the observed yields in the corresponding VRs and SRs. Two distinct fits are run, one for the C1C1-WW and C1N2-WZ analyses, and one for the analysis targeting C1N2-Wh models. In the former case, the normalisation factors to be applied to the MC predictions of the main SM backgrounds are ${ 0.81 }^{ +0.10 }_{ -0.09 }$  for \ttbar,  ${ 1.05 }^{ +0.09 }_{ -0.09 }$ for $W$+jets and diboson 1L, and ${ 1.22 }^{ +0.18 }_{ -0.18 }$ for diboson 2L. In the latter case, the normalisation factors are found to be $1.00\pm 0.29$  for \ttbar, $0.95 \pm 0.19$ for single top and $1.30\pm 0.05$  for $W$+jets production.
The large uncertainties in the \ttbar\ and single top normalisation factors are related to their large theoretical systematic uncertainties.
Overall the normalisation factors are found to be consistent across analyses for each background process within uncertainties.
For the $W$+jets normalisation factors, differences are expected because of the different requirement on the number of $b$-tagged  jets. MC predictions for the production of a $W$ or $Z$ boson and $b$-jets are consistently below data in SM cross-section measurements, and in agreement if no $b$-jets are required in the event~\cite{ATLAS:2020juj}.

The agreement between the observed and expected event yields in control, validation and exclusion signal regions is illustrated in Figures~\ref{fig:pullplot_SR_VR_CR} and~\ref{fig:pullWh}. No significant excesses are observed in data above the SM prediction.
 
\begin{figure}
\begin{center}
\subfloat[]{
\includegraphics[width=0.4\textwidth]{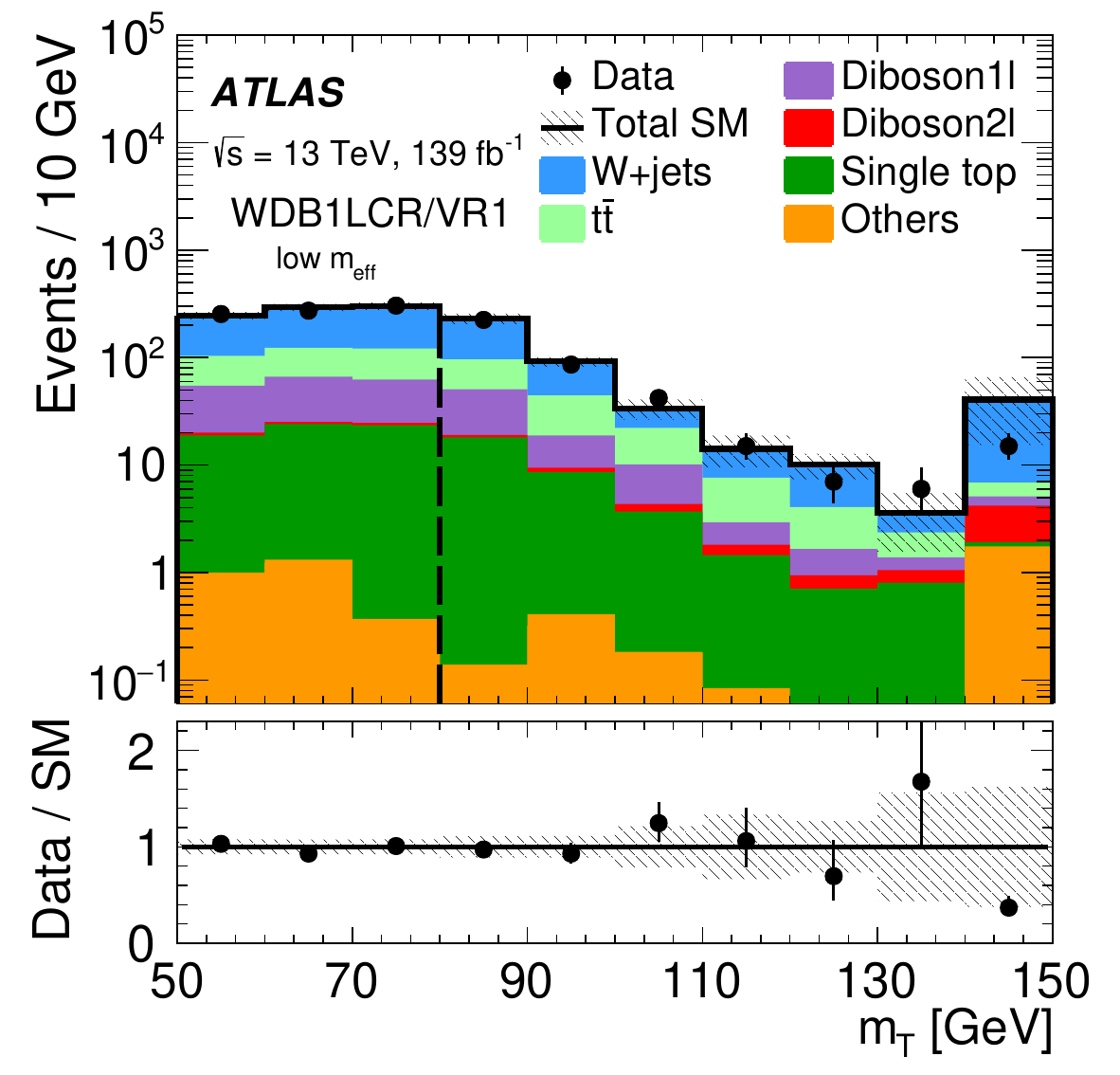}}
\subfloat[]{
\includegraphics[width=0.4\textwidth]{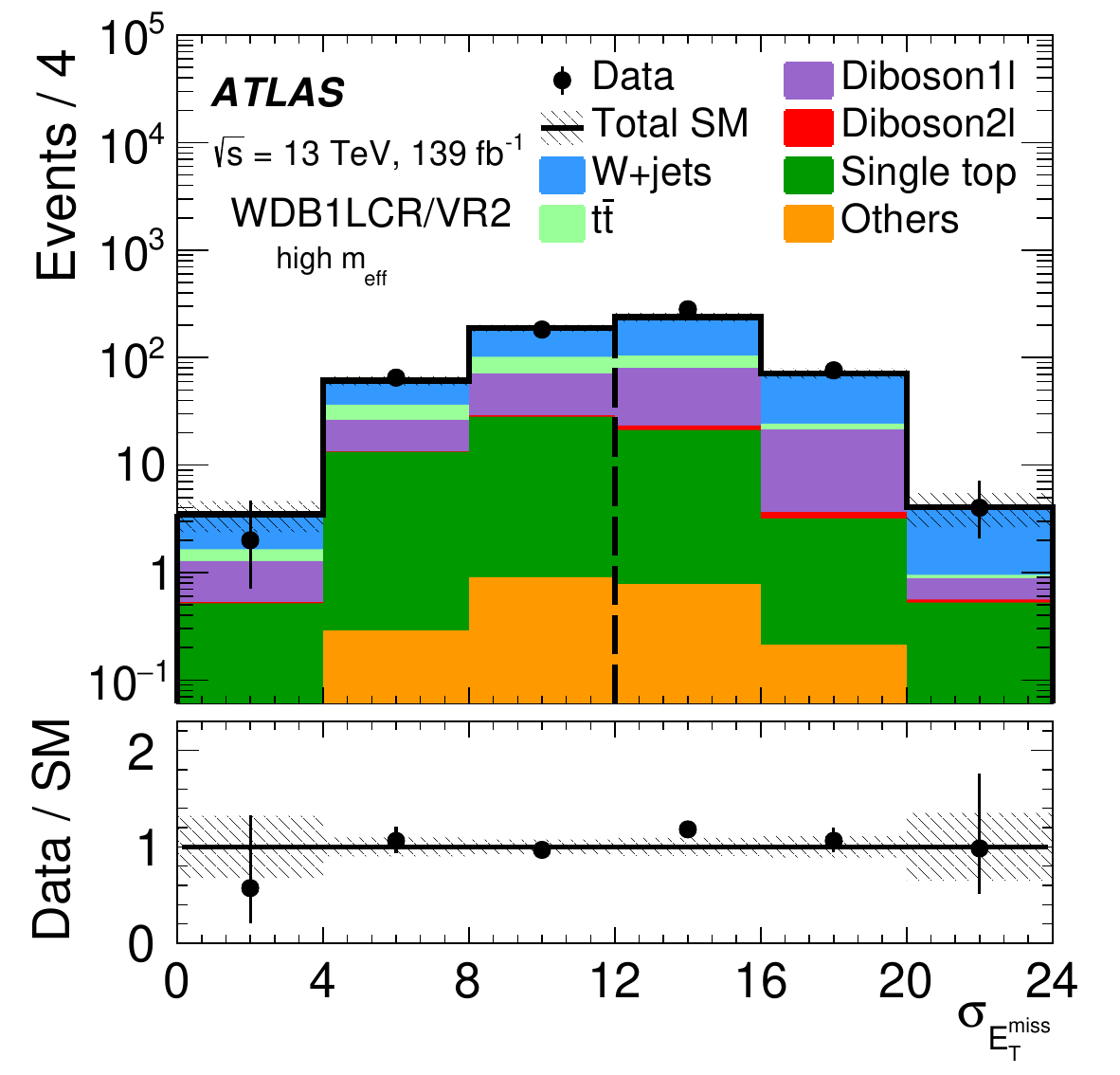}}
\quad
\subfloat[]{
\includegraphics[width=0.4\textwidth]{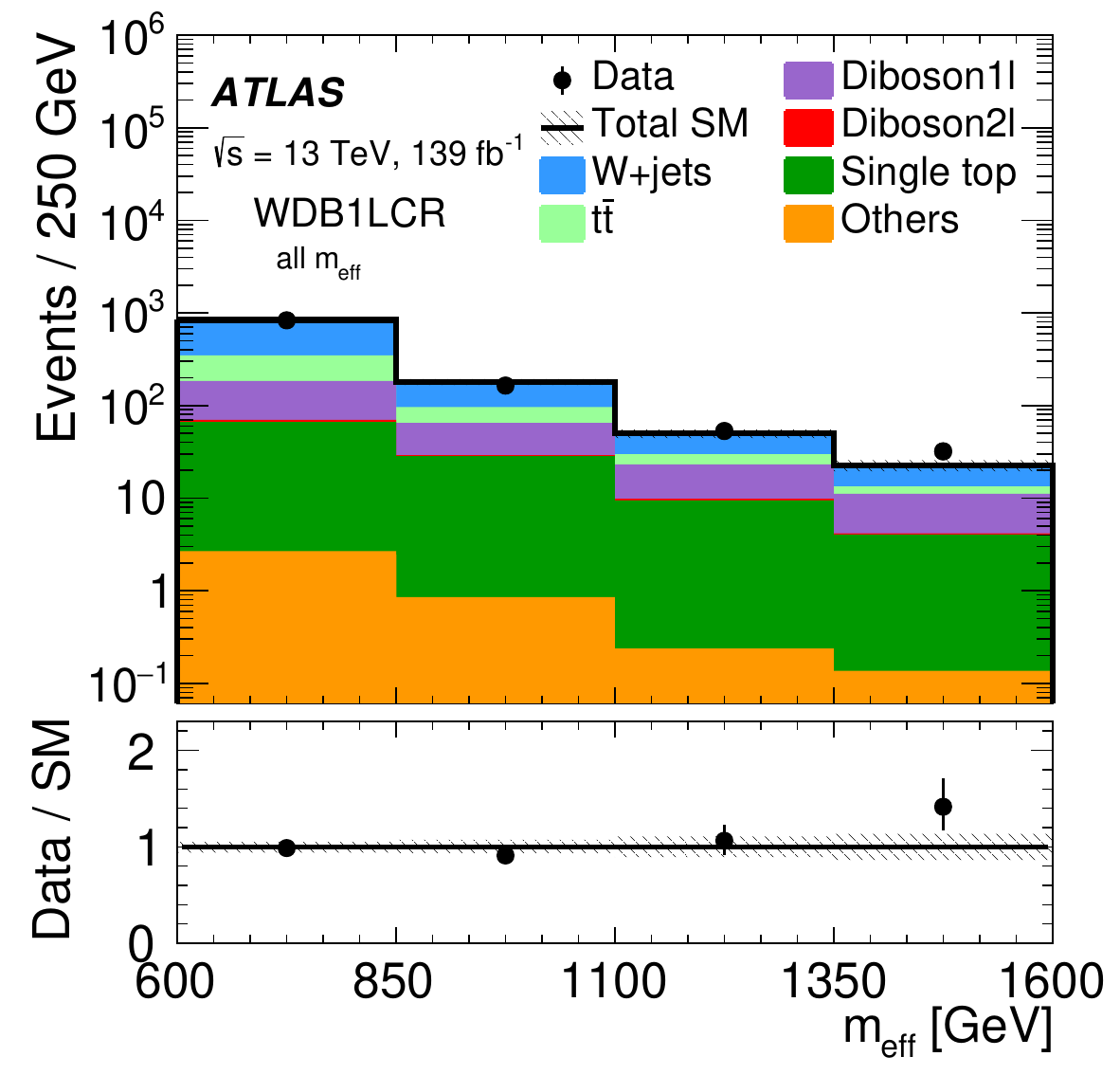}}
\end{center}
 
\caption{The post-fit (a) \mt\ distribution in WDB1L CR and VR1 for low \meff bin,
(b)  $\sigma_{\met}$ distribution in WDB1L CR and VR2 for high \meff bin
and (c) \meff\ distribution in WDB1L CR\@. The vertical dashed line separates the control and validation regions.  The uncertainty bands plotted include all statistical and systematic uncertainties. The overflow events, where present, are included in the last bin.
The lower panels of the plots show the ratio of the observed data to the total background prediction.}
\label{fig:CR1}
\end{figure}

\begin{figure}
\begin{center}
\subfloat[]{
\includegraphics[width=0.4\textwidth]{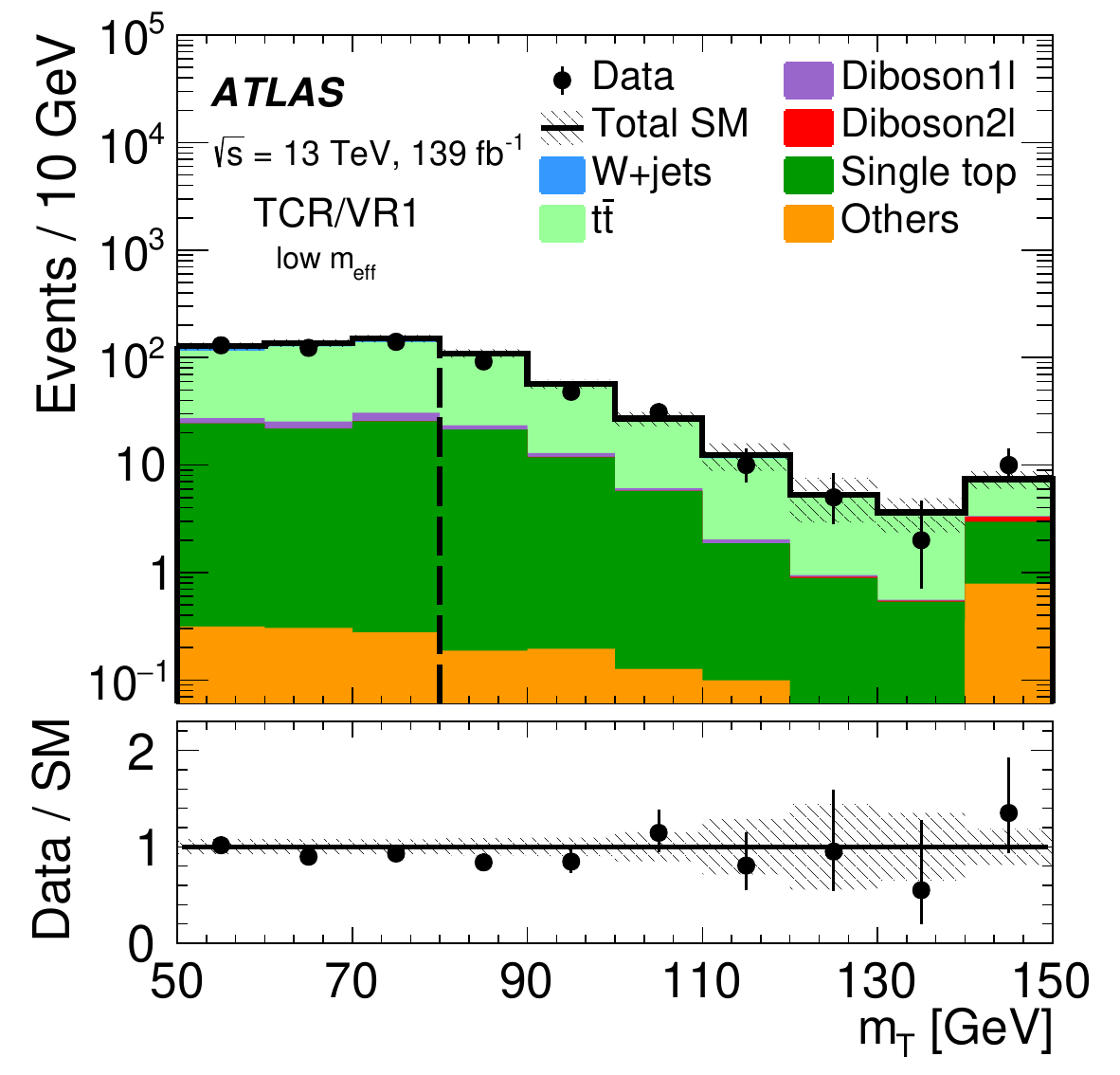}}
\subfloat[]{
\includegraphics[width=0.4\textwidth]{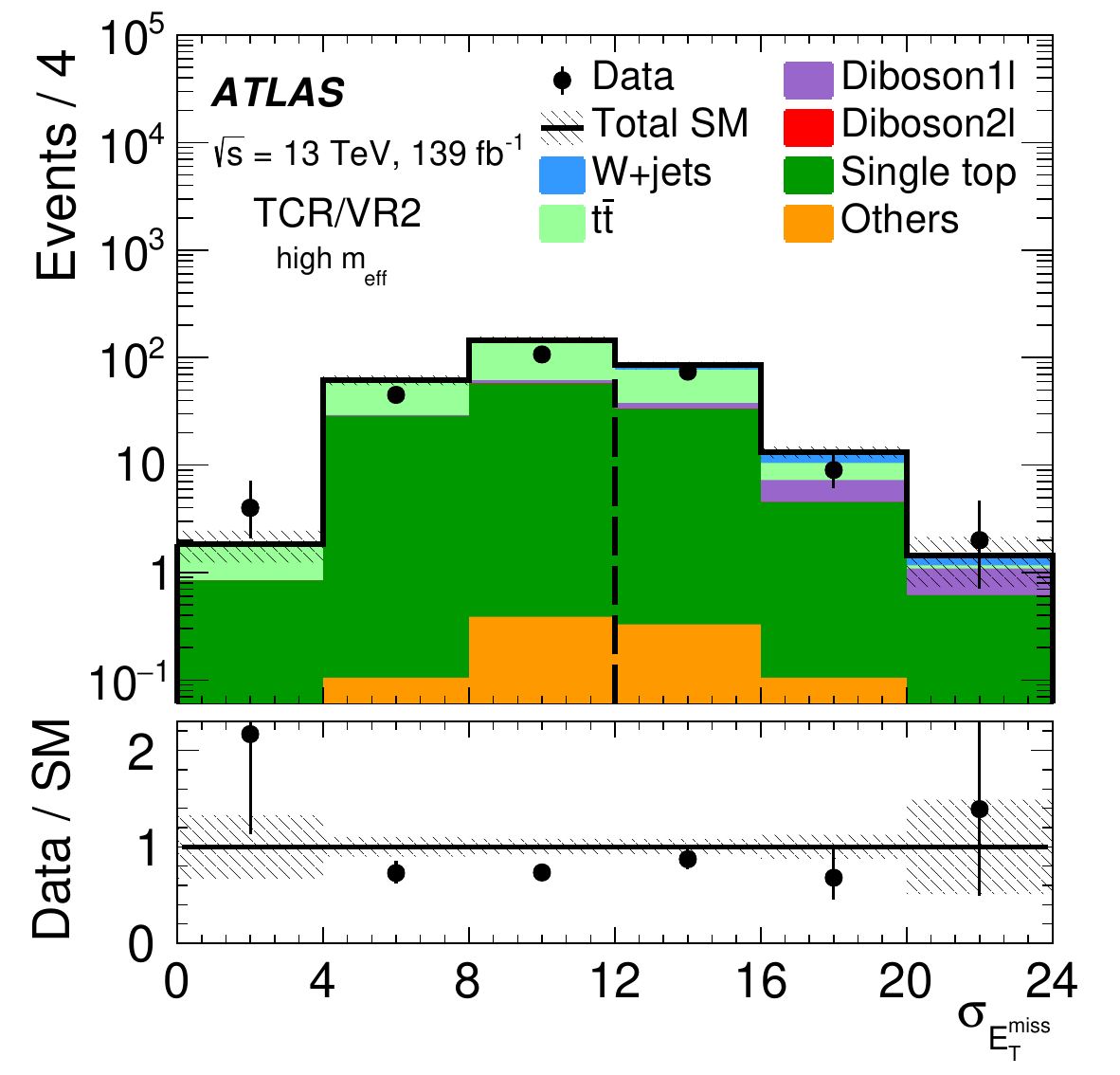}}
\quad
\subfloat[]{
\includegraphics[width=0.4\textwidth]{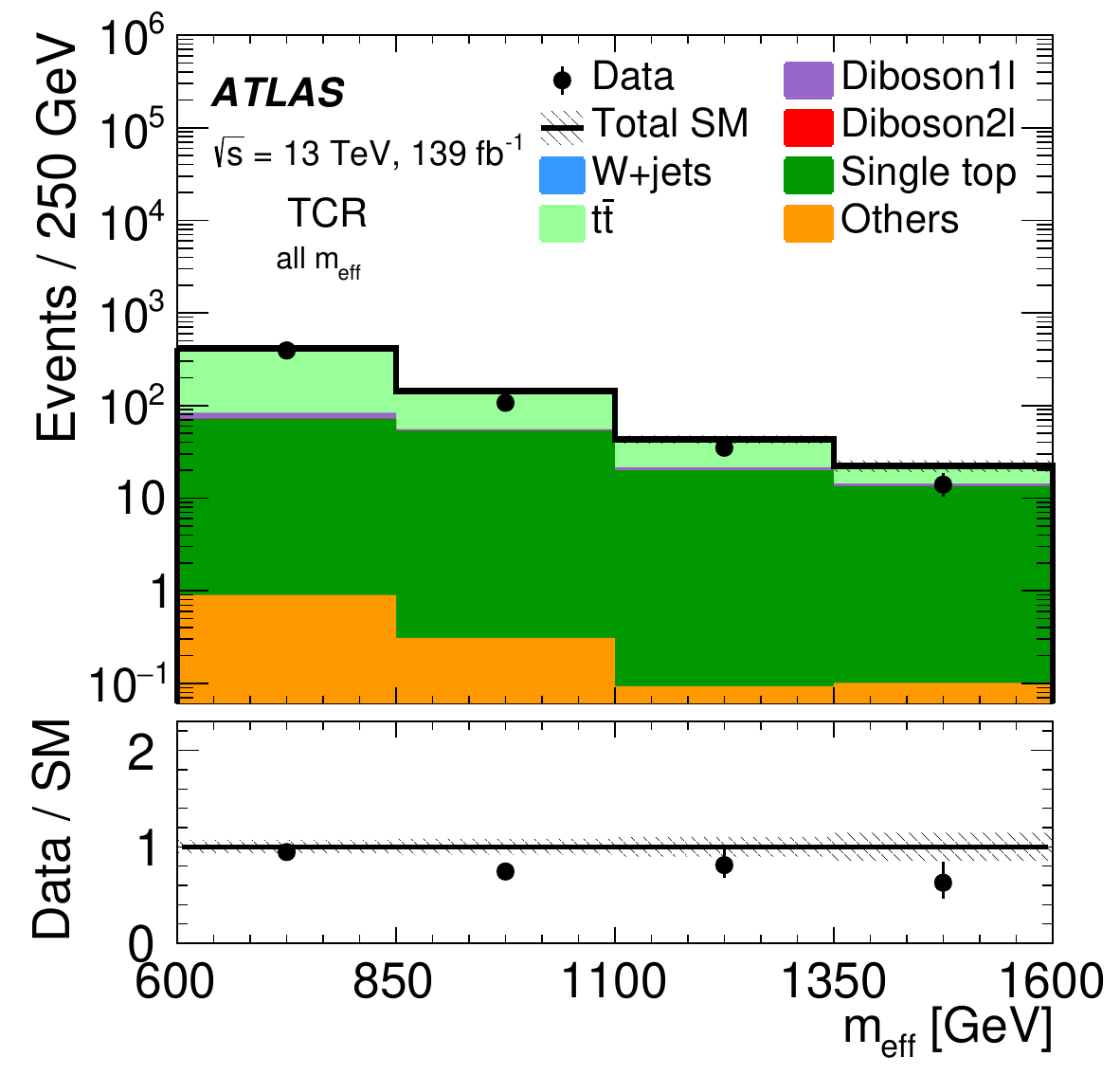}}
\end{center}
\caption{The post-fit (a) \mt\ distribution in \ttbar\ CR and VR1 for low \meff bin,
(b) $\sigma_{\met}$ distribution in \ttbar\ CR and VR2 for high \meff bin
and (c) \meff\ distribution in \ttbar\ CR\@.  The vertical dashed line separates the control and validation regions. The uncertainty bands plotted include all statistical and systematic uncertainties. The overflow events, where present, are included in the last bin.
The lower panels of the plots show the ratio of the observed data to the total background prediction.}
\label{fig:CR2}
\end{figure}

Figures~\ref{fig:CR1}--\ref{fig:CR2} present the post-fit \mt, $\sigma_{\met}$, and \meff\ distributions for the C1C1-WW and C1N2-WZ analyses compared with the data in the selected control and validation regions. The data and the background expectation in all validation regions agree well within around two standard deviations. Therefore no further systematic uncertainty is applied to the background estimate in the signal regions. Figure~\ref{fig:SRs} shows the post-fit \meff\ distributions in SRLM, SRMM, and SRHM for both the C1C1-WW and C1N2-WZ models. The uncertainty bands include all statistical and systematic uncertainties. The dashed lines represent the benchmark signal points.
 
\begin{figure}
\begin{center}
\subfloat[]{
\includegraphics[width=0.49\textwidth]{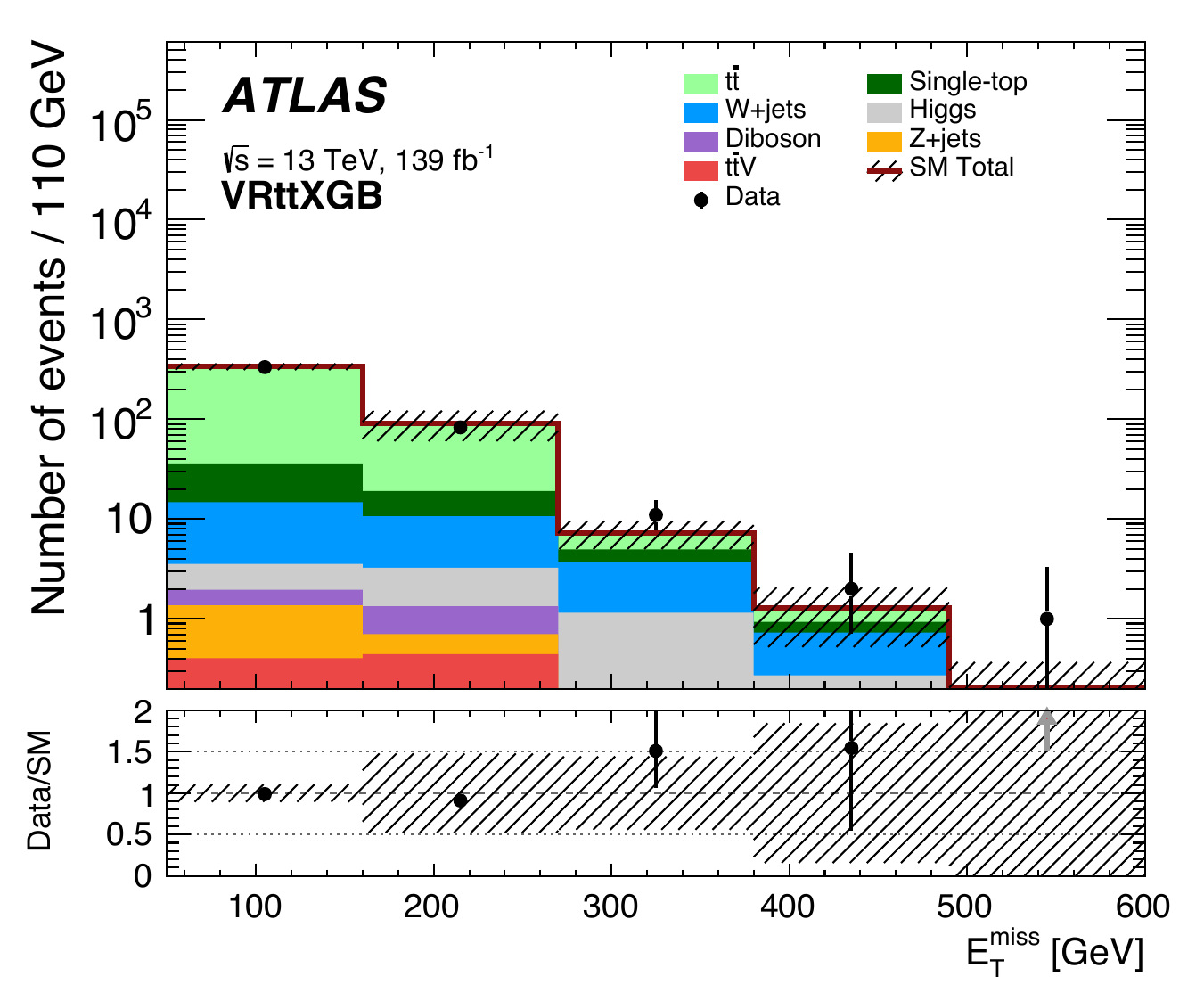}}
\subfloat[]{
\includegraphics[width=0.49\textwidth]{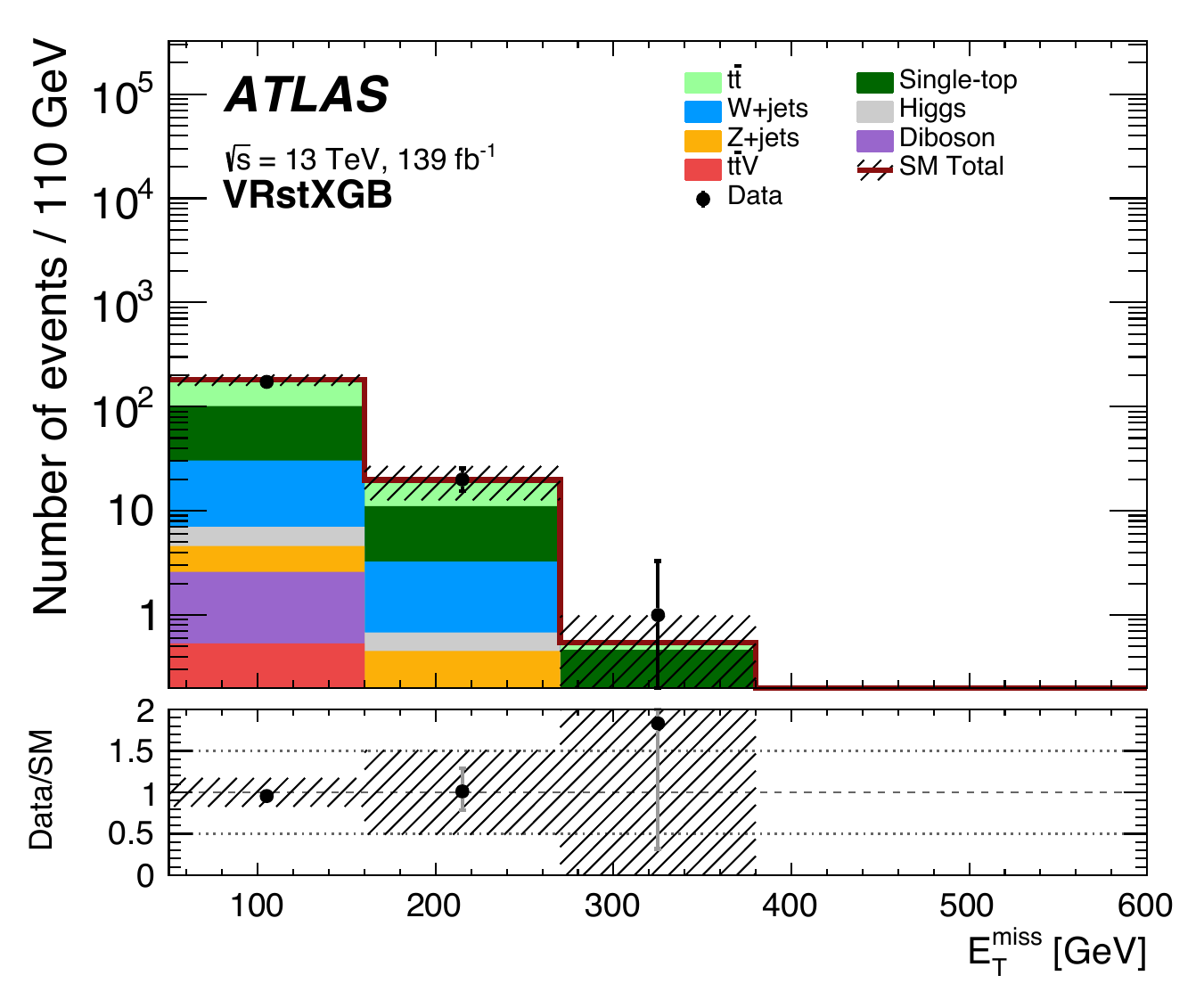}}
\quad
\subfloat[]{
\includegraphics[width=0.49\textwidth]{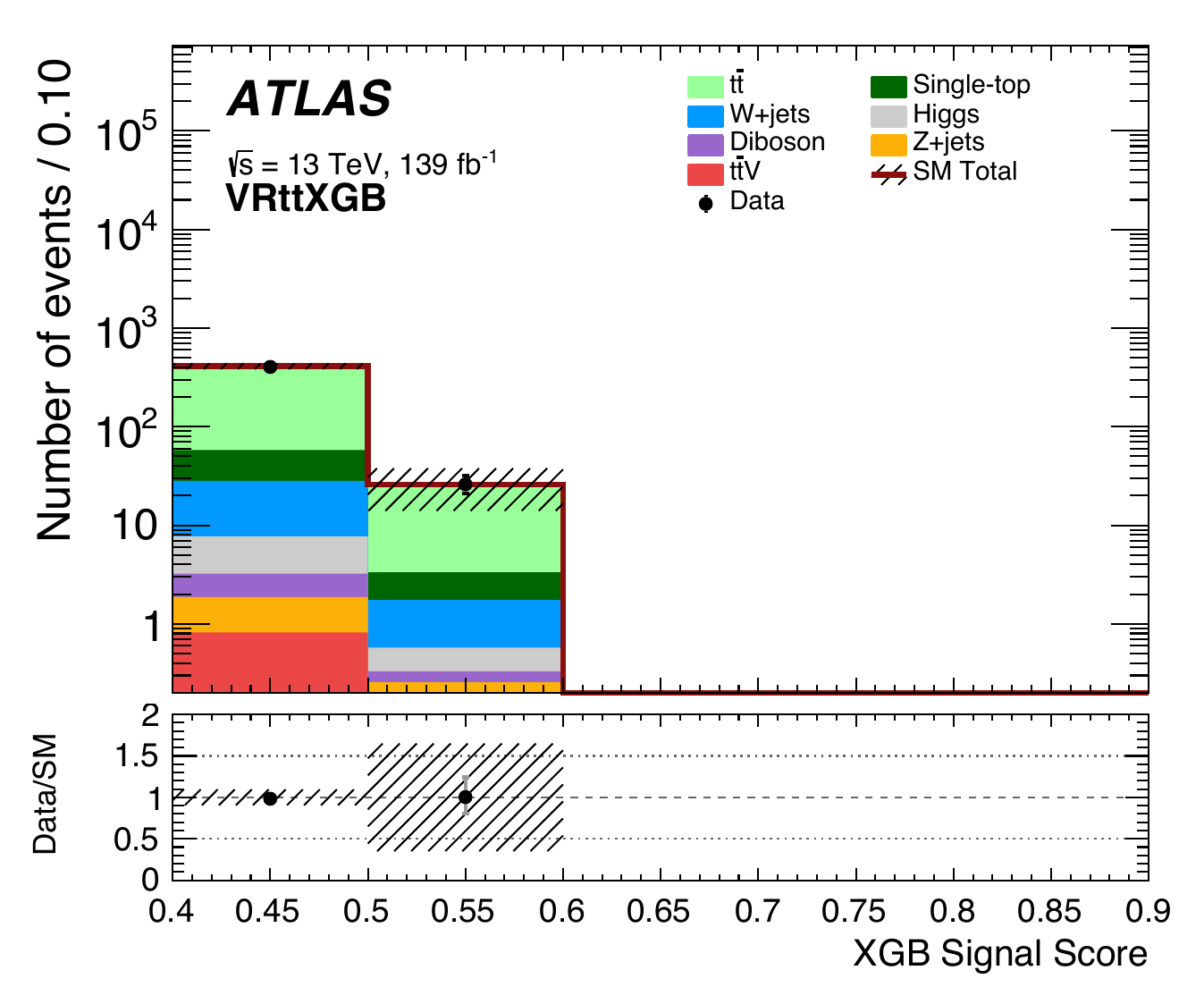}}
\subfloat[]{
\includegraphics[width=0.49\textwidth]{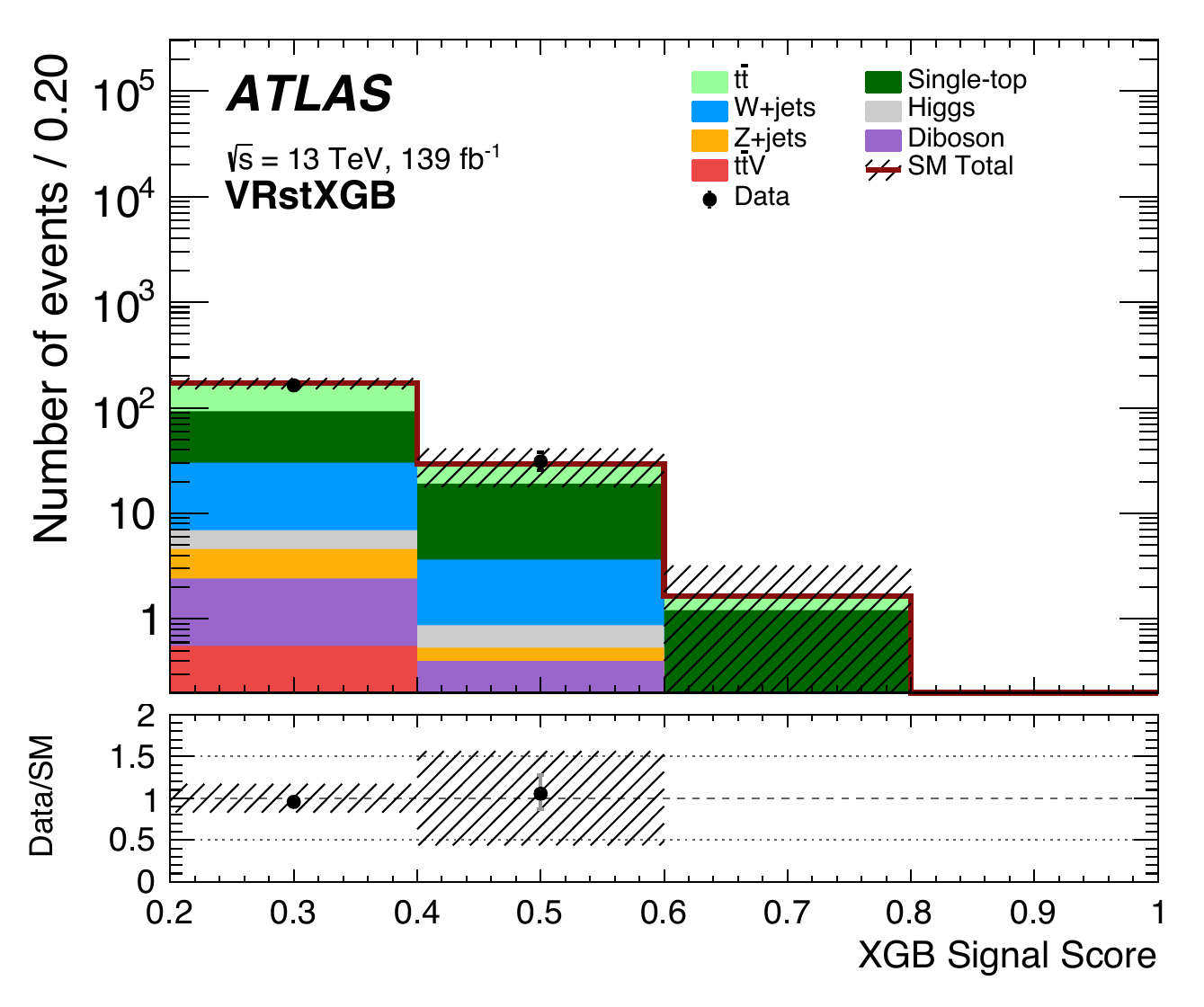}}
\end{center}
 
\caption{Post-fit \met\ and signal output score distributions: (a) and (c) in \ttbar, (b) and (d) in single-top VRs. The uncertainty bands plotted include all statistical and systematic uncertainties. The overflow events, where present, are included in the last bin.
The lower panels of the plots show the ratio of the observed data to the total background prediction.}
\label{fig:VR_Wh}
\end{figure}

\begin{figure}
\begin{center}
\subfloat[]{
\includegraphics[width=0.49\textwidth]{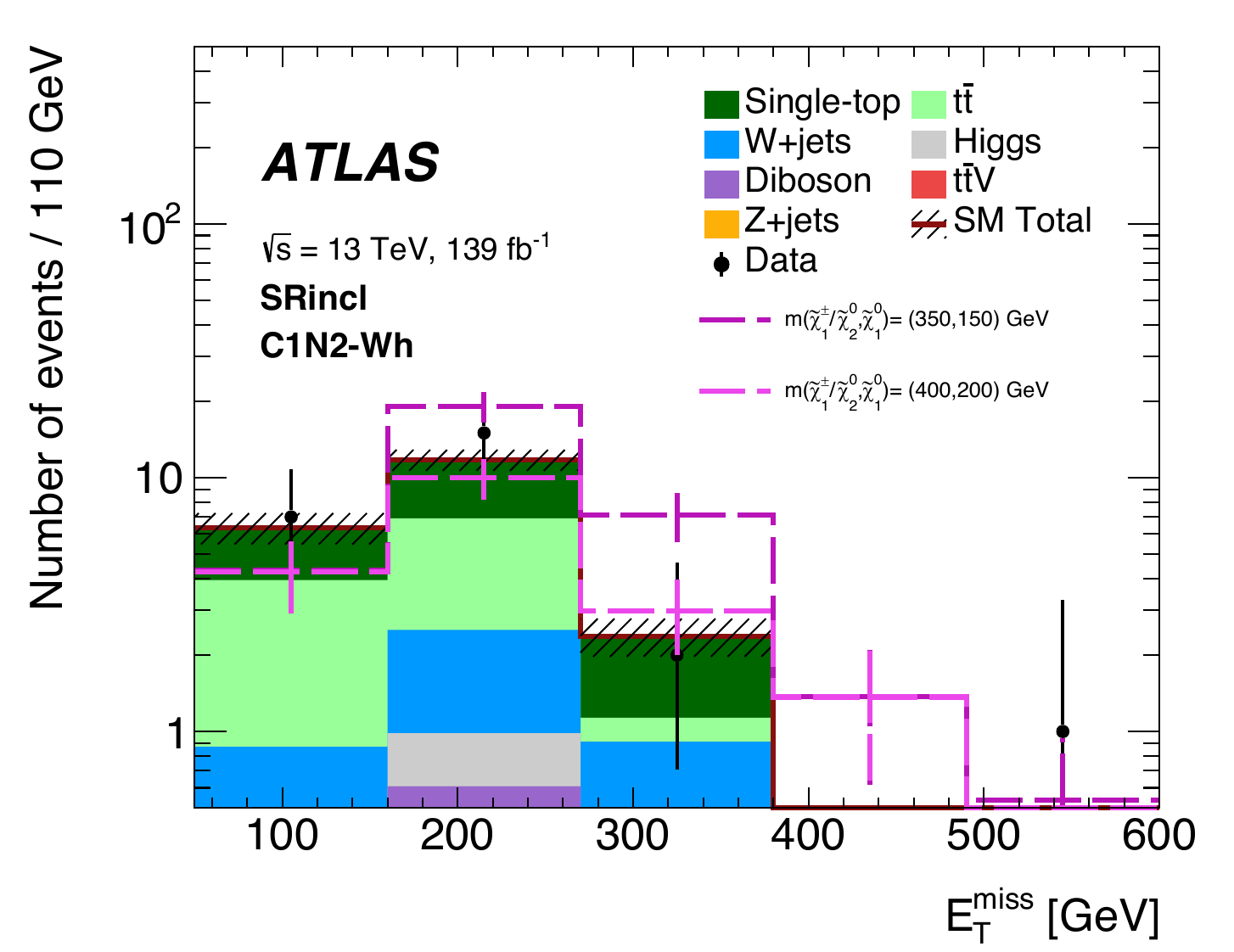}}
\subfloat[]{
\includegraphics[width=0.49\textwidth]{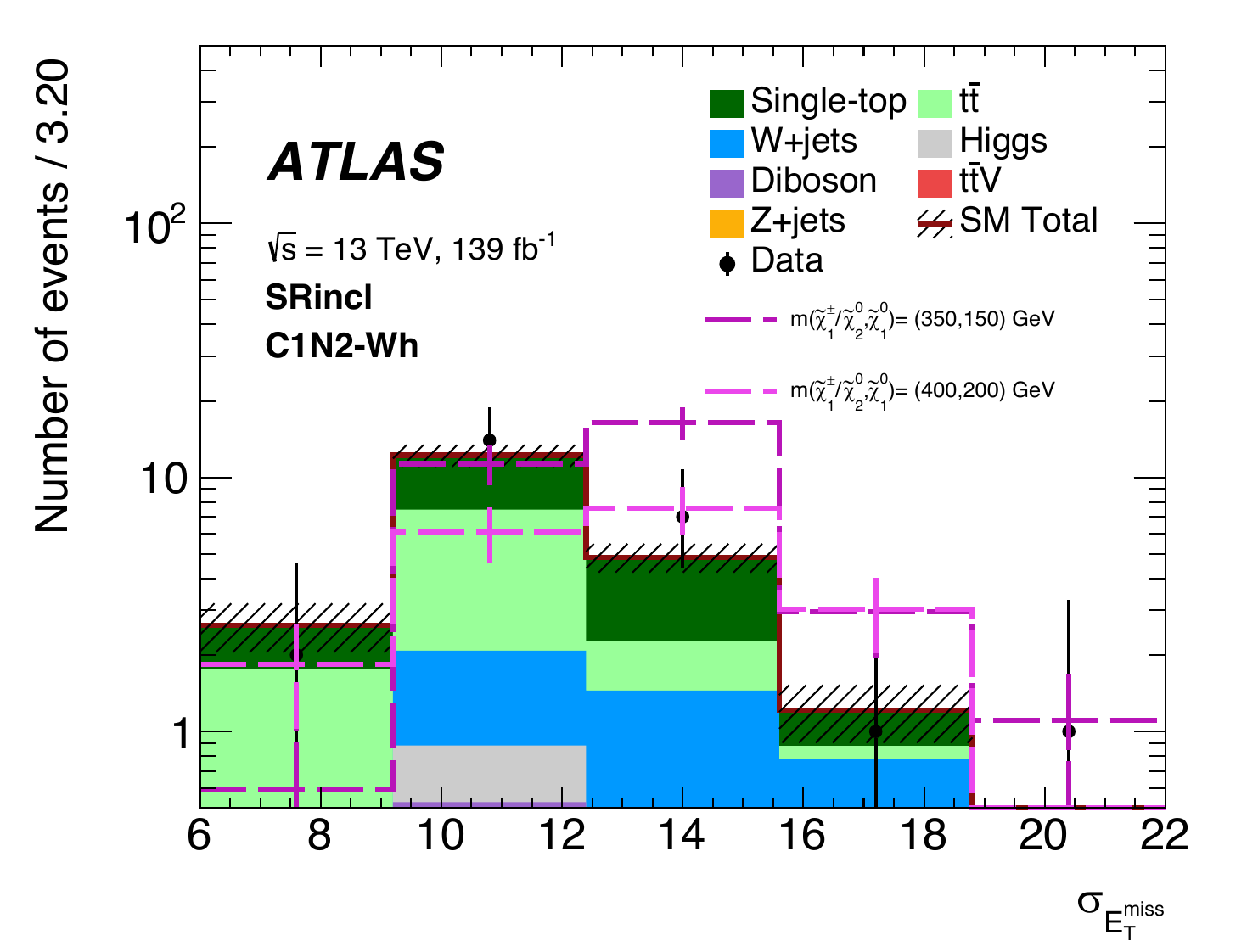}}
\quad
\subfloat[]{
\includegraphics[width=0.49\textwidth]{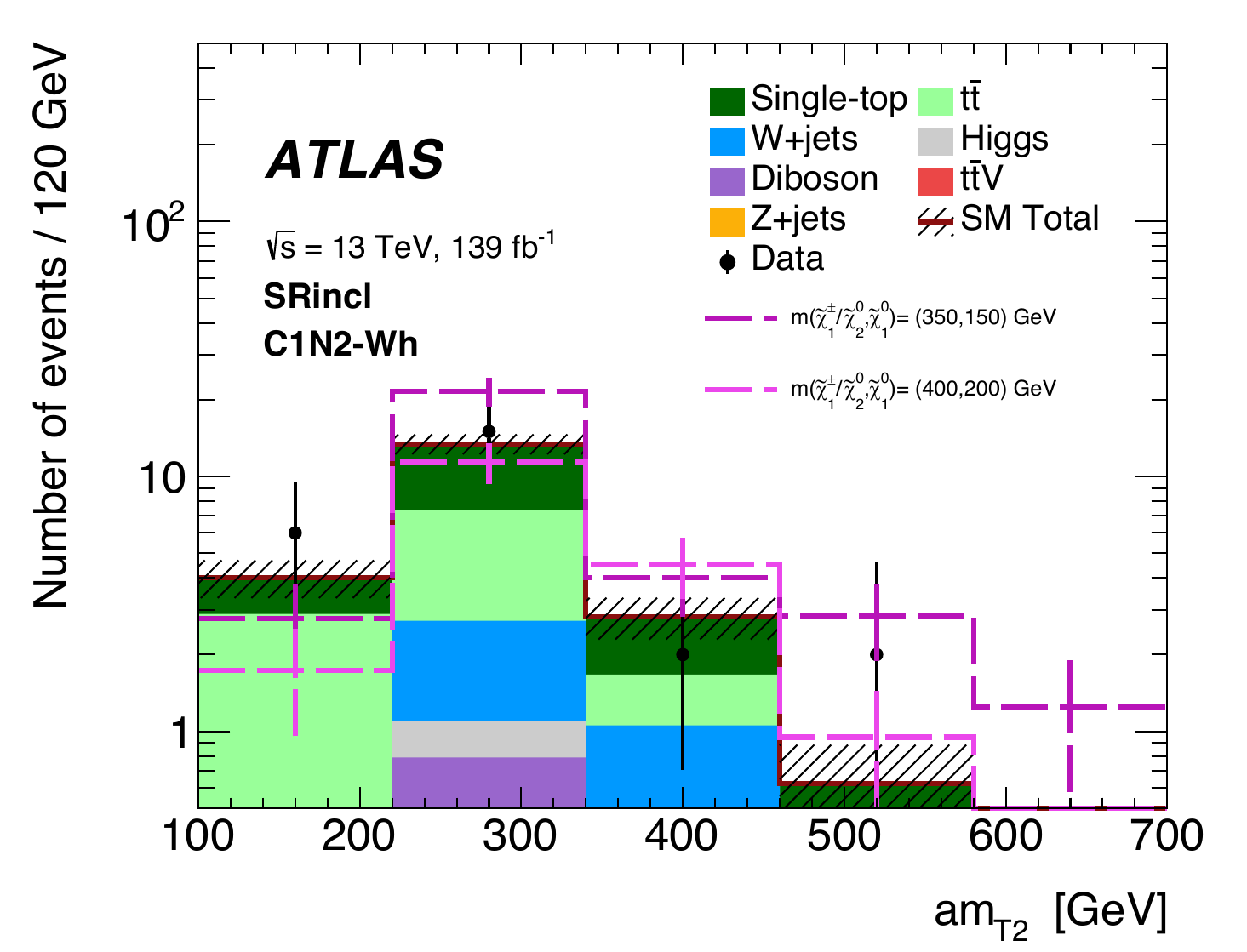}}
\subfloat[]{
\includegraphics[width=0.49\textwidth]{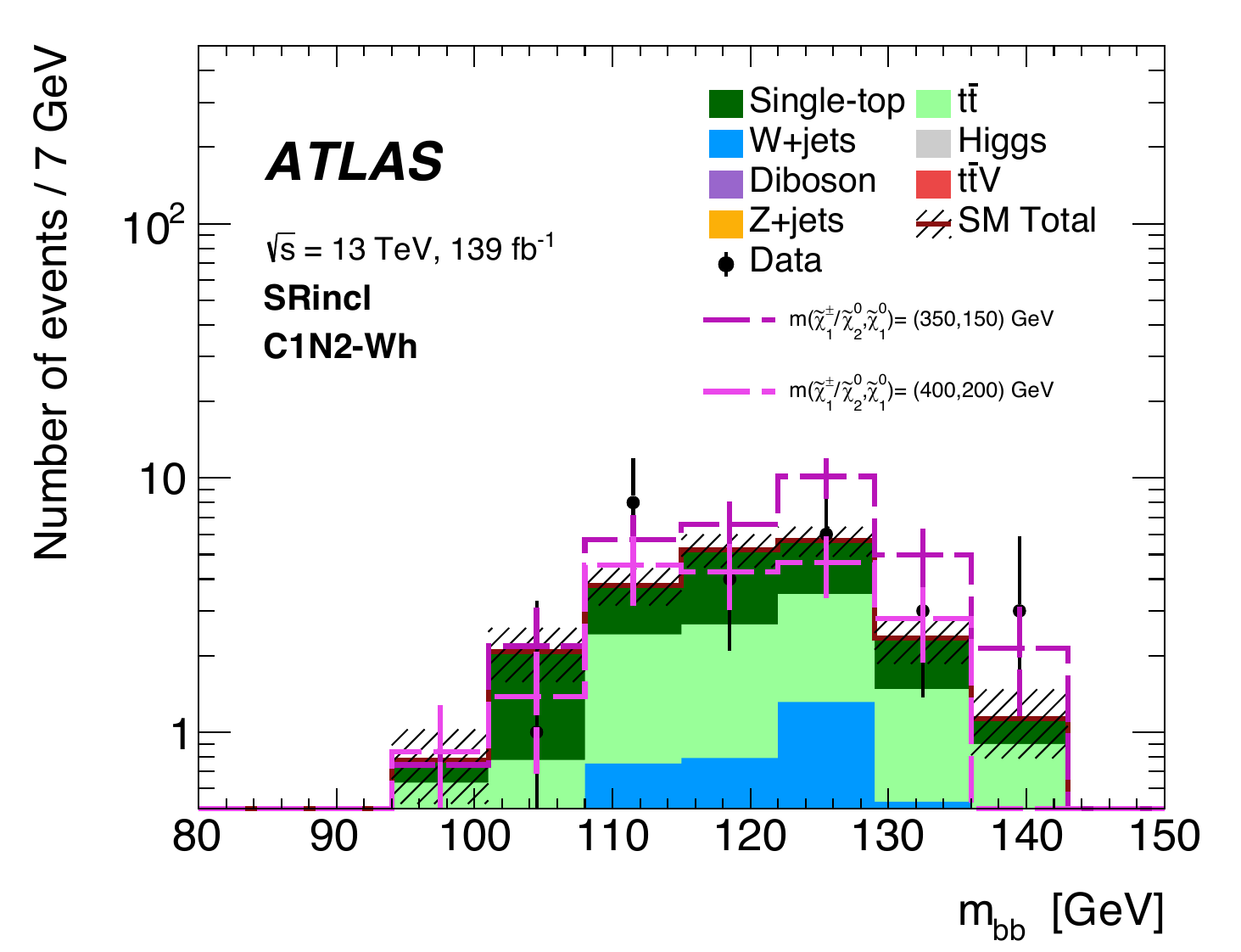}}\\
\subfloat[]{
\includegraphics[width=0.49\textwidth]{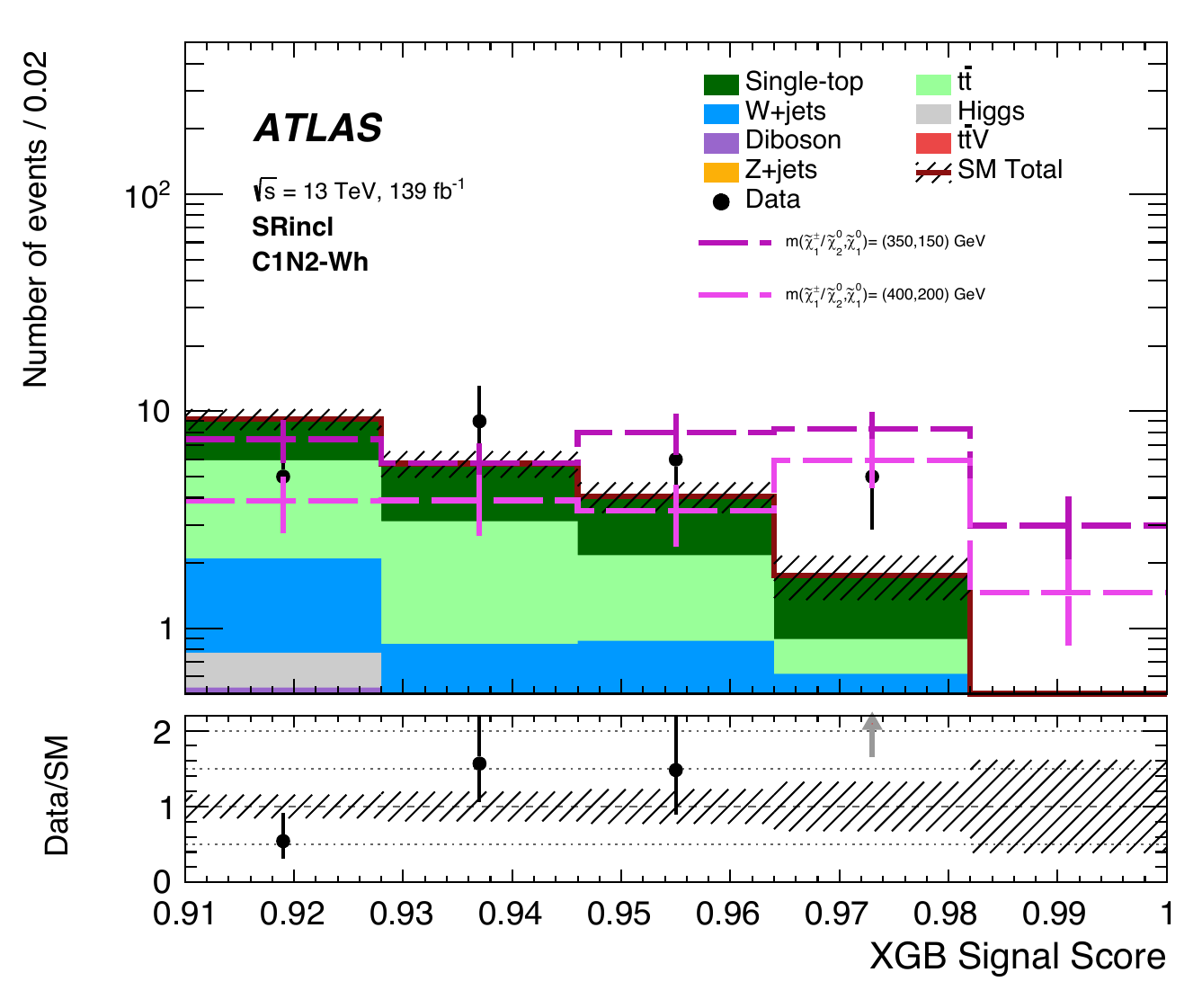}}
\end{center}
 
\caption{Post-fit distributions in the SRs (not binned in signal output score) for representative kinematic observables (a)  \met, (b) $\sigma_{\met}$, (c)  $am^{}_{\mathrm{T2}}$, (d)  \mbb and (e) the signal output score. Two representative SUSY signal models are overlaid for illustration. The uncertainty bands plotted include all statistical and systematic uncertainties. The overflow events, where present, are included in the last bin.}
\label{fig:SR_Wh}
\end{figure}

For the C1N2-Wh analysis, \met\ and output signal score distributions in \ttbar\ and single top validation regions are shown in Figure~\ref{fig:VR_Wh}. The data and the background expectation in all validation regions agree well within around one standard deviations. Figure ~\ref{fig:SR_Wh} shows the post-fit distributions for \met, $am^{}_{\mathrm{T2}}$, \mbb\, and $\sigma_{\met}$ in the inclusive SR, i.e., considering all SR bins. Events selected by the BDT as compatible with the signal of interest (high $w^{}_{\mathrm{sig}}$ score) have a moderate to large \met\ and $am^{}_{\mathrm{T2}}$, \mbb\ close to the Higgs boson mass and large $\sigma_{\met}$.

 
\begin{table}
\begin{center}
\setlength{\tabcolsep}{0.0pc}{\small
\newcolumntype{C}{@{}>{{}}c<{{}}@{}}
\begin{tabular*}{\textwidth}{@{\extracolsep{\fill}}l *3{C}}
\noalign{\smallskip}\hline\noalign{\smallskip}{\textbf{ C1C1-WW model}}                       & SRLM          &    Bin 0        & Bin 1          \\[-0.05cm]
&               &     $[600,850]$~\GeV\        & $>850$~\GeV\             \\[-0.05cm]
\noalign{\smallskip}\hline\noalign{\smallskip}
Observed events                       & $23$              & $16$              & $7$                     \\
\noalign{\smallskip}\hline\noalign{\smallskip}
Total SM background events         &  $22.0 \pm 3.2$          & $15.8 \pm 2.8$          & $6.2 \pm 1.0$              \\
\noalign{\smallskip}\hline\noalign{\smallskip}
\ttbar\           & $2.6 \pm 1.3$          & $2.2 \pm 1.2$          & $0.36 \pm 0.21$               \\
$W$+jets          & $13.6 \pm 2.7$          & $10.1 \pm 2.3$          & $3.5 \pm 0.8$              \\
$Z$+jets         & $0.10_{-0.10}^{+0.16}$          & $0.04_{-0.04}^{+0.13}$          & $0.054 \pm 0.034$              \\
Single-top       & $0.5 \pm 0.4$          & $0.19_{-0.19}^{+0.25}$          & $0.26 \pm 0.18$              \\
Diboson2l        & $1.7 \pm 0.5$          & $1.3 \pm 0.4$          & $0.43 \pm 0.12$               \\
Diboson1l        & $3.0 \pm 0.7$          & $1.5 \pm 0.5$          & $1.41 \pm 0.30$              \\
\ttbar+$V$       & $0.50 \pm 0.14$          & $0.30 \pm 0.10$          & $0.20 \pm 0.06$              \\
\ttbar+$h$       & $0.008 \pm 0.005$          & $0.005 \pm 0.005$          & $0.003 \pm 0.001$              \\
Multiboson       & $0.005 \pm 0.002$          & $-$          & $0.005 \pm 0.002$              \\

\noalign{\smallskip}\hline\noalign{\smallskip}
\noalign{\smallskip}\hline\noalign{\smallskip}{\textbf{ C1C1-WW model}}                       & SRMM          &    Bin 0        & Bin 1          \\[-0.05cm]
&               &     $[600,850]$~\GeV\        & $>850$~\GeV\              \\[-0.05cm]
\noalign{\smallskip}\hline\noalign{\smallskip}
Observed events                       & $11$              & $7$              & $4$                    \\
\noalign{\smallskip}\hline\noalign{\smallskip}
Total SM background events         & $9.2 \pm 2.5$          & $6.4 \pm 2.2$          & $2.7 \pm 0.9$              \\
\noalign{\smallskip}\hline\noalign{\smallskip}
\ttbar\          & $0.60 \pm 0.32$          & $0.44 \pm 0.25$          & $0.16 \pm 0.10$              \\
$W$+jets         & $5.3 \pm 2.1$          & $4.0 \pm 1.9$          & $1.3 \pm 0.5$              \\
$Z$+jets         & $0.11 \pm 0.04$          & $0.09 \pm 0.04$          & $0.020 \pm 0.010$              \\
Single-top       & $0.12 \pm 0.09$          & $0.09 \pm 0.07$          & $0.03_{-0.07}^{+0.03}$              \\
Diboson2l        & $1.4 \pm 0.4$          & $0.89 \pm 0.32$          & $0.49 \pm 0.21$              \\
Diboson1l        & $1.4 \pm 0.4$          & $0.80 \pm 0.28$          & $0.64 \pm 0.25$              \\
\ttbar+$V$       & $0.29 \pm 0.12$          & $0.16 \pm 0.09$          & $0.13 \pm 0.07$              \\
\ttbar+$h$       & $0.004 \pm 0.002$          & $0.004 \pm 0.002$          & $-$              \\
Multiboson       & $-$          & $-$          & $-$              \\

\noalign{\smallskip}\hline\noalign{\smallskip}
\noalign{\smallskip}\hline\noalign{\smallskip}{\textbf{ C1C1-WW model}}                       & SRHM          &    Bin 0        & Bin 1          \\[-0.05cm]
&               &     $[600,850]$~\GeV\        & $>850$~\GeV\              \\[-0.05cm]
 
\noalign{\smallskip}\hline\noalign{\smallskip}
Observed events                       & $16$              & $4$              & $12$                    \\
\noalign{\smallskip}\hline\noalign{\smallskip}
Total SM background events         & $15.7 \pm 3.0$          & $4.8 \pm 1.3$          & $10.8 \pm 2.5$              \\
\noalign{\smallskip}\hline\noalign{\smallskip}
\ttbar\          & $0.36 \pm 0.18$          & $0.19 \pm 0.11$          & $0.17 \pm 0.09$              \\
$W$+jets         & $11.3 \pm 2.5$          & $3.3 \pm 1.1$          & $8.0 \pm 2.2$              \\
$Z$+jets         & $0.17 \pm 0.07$          & $0.14 \pm 0.06$          & $0.029 \pm 0.018$              \\
Single-top       & $0.49 \pm 0.34$          & $0.23 \pm 0.17$          & $0.26 \pm 0.20$              \\
Diboson2l        & $1.5 \pm 0.4$          & $0.54 \pm 0.17$          & $0.99 \pm 0.28$              \\
Diboson1l        & $1.5 \pm 0.4$          & $0.39 \pm 0.14$          & $1.09 \pm 0.31$              \\
\ttbar+$V$       & $0.33 \pm 0.12$          & $0.037 \pm 0.035$          & $0.29 \pm 0.11$              \\
\ttbar+$h$       & $0.003 \pm 0.003$          & $-$          & $0.003 \pm 0.003$              \\
Multiboson       & $0.004 \pm 0.001$          & $-$          & $0.004 \pm 0.001$              \\

\noalign{\smallskip}\hline\noalign{\smallskip}
\end{tabular*}
}
\end{center}
\caption{
Observed event yields and the background expectation obtained from a background fit in the C1C1-WW model SRs with an integrated luminosity of 139~\ifb.
The first column with numbers stands for the yields in all bins. The second and third columns correspond to the low and high bins in \meff.
Uncertainties reported for the fitted background estimates combine statistical and systematic uncertainties.}
\label{table_results_C1C1_WW}
\end{table}
 
\begin{table}
\begin{center}
\setlength{\tabcolsep}{0.0pc}{\small
\newcolumntype{C}{@{}>{{}}c<{{}}@{}}
\begin{tabular*}{\textwidth}{@{\extracolsep{\fill}}l *3{C}}
\noalign{\smallskip}\hline\noalign{\smallskip}{\textbf{ C1N2-WZ model}}                       & SRLM          &    Bin 0        & Bin 1          \\[-0.05cm]
&               &     $[600,850]$~\GeV\        & $>850$~\GeV\             \\[-0.05cm]
 
\noalign{\smallskip}\hline\noalign{\smallskip}
Observed events                       & $26$              & $16$              & $10$                     \\
\noalign{\smallskip}\hline\noalign{\smallskip}
Total SM background events         & $29 \pm 4$          & $15.6 \pm 2.8$          & $13.0 \pm 1.8$             \\
\noalign{\smallskip}\hline\noalign{\smallskip}
\ttbar\          & $2.7 \pm 1.0$          & $1.6 \pm 0.7$          & $1.1 \pm 0.4$              \\
$W$+jets         & $18.1 \pm 2.8$          & $10.6 \pm 2.1$          & $7.5 \pm 1.2$              \\
$Z$+jets         & $0.06_{-0.06}^{+0.60}$          & $0.03_{-0.03}^{+0.66}$          & $0.03_{-0.03}^{+0.06}$               \\
Single-top       & $0.6_{-0.6}^{+1.0}$          & $0.16_{-0.16}^{+0.34}$          & $0.5_{-0.5}^{+0.8}$    \\
Diboson2l        & $2.4 \pm 0.6$          & $1.4 \pm 0.4$          & $0.96 \pm 0.28$              \\
Diboson1l        & $3.9 \pm 0.9$          & $1.3 \pm 0.4$          & $2.5 \pm 0.6$              \\
\ttbar+$V$       & $0.74 \pm 0.22$          & $0.37 \pm 0.13$          & $0.37 \pm 0.11$              \\
\ttbar+$h$       & $0.021 \pm 0.006$          & $0.011 \pm 0.004$          & $0.010 \pm 0.004$              \\
Multiboson       & $0.020 \pm 0.005$          & $-$          & $0.020 \pm 0.005$              \\

\noalign{\smallskip}\hline\noalign{\smallskip}
\noalign{\smallskip}\hline\noalign{\smallskip}{\textbf{ C1N2-WZ model}}                       & SRMM          &    Bin 0        & Bin 1          \\[-0.05cm]
&               &     $[600,850]$~\GeV\       & $>850$~\GeV\              \\[-0.05cm]
 
\noalign{\smallskip}\hline\noalign{\smallskip}
Observed events                       &$22$              & $13$              & $9$                     \\
\noalign{\smallskip}\hline\noalign{\smallskip}
Total SM background events         & $12.7 \pm 2.5$          & $7.0 \pm 2.2$          & $5.7 \pm 1.3$              \\
\noalign{\smallskip}\hline\noalign{\smallskip}
\ttbar\          & $0.9 \pm 0.4$          & $0.52 \pm 0.23$          & $0.41 \pm 0.20$              \\
$W$+jets         & $7.3 \pm 2.0$          & $4.4 \pm 1.9$          & $2.9 \pm 0.8$              \\
$Z$+jets         & $0.30 \pm 0.23$          & $0.12 \pm 0.12$          & $0.18 \pm 0.15$              \\
Single-top       & $0.14_{-0.14}^{+0.24}$          & $0.09_{-0.09}^{+0.16}$          & $0.05_{-0.05}^{+0.11}$      \\
Diboson2l        & $1.5 \pm 0.4$          & $0.66 \pm 0.28$          & $0.85 \pm 0.25$              \\
Diboson1l        & $1.9 \pm 0.5$          & $0.94 \pm 0.27$          & $0.95 \pm 0.30$              \\
\ttbar+$V$       & $0.54 \pm 0.17$          & $0.25 \pm 0.10$          & $0.28 \pm 0.10$              \\
\ttbar+$h$       & $0.026_{-0.026}^{+0.029}$          & $0.005 \pm 0.004$          & $0.022_{-0.022}^{+0.025}$              \\
Multiboson       & $0.008 \pm 0.002$          & $0.004 \pm 0.001$          & $0.004 \pm 0.001$              \\

\noalign{\smallskip}\hline\noalign{\smallskip}
\noalign{\smallskip}\hline\noalign{\smallskip}{\textbf{ C1N2-WZ model}}                       & SRHM          &    Bin 0        & Bin 1          \\[-0.05cm]
&               &     $[600,850]$~\GeV\        & $>850$~\GeV\              \\[-0.05cm]
 
\noalign{\smallskip}\hline\noalign{\smallskip}
Observed events                       & $26$              & $5$              & $21$                    \\
\noalign{\smallskip}\hline\noalign{\smallskip}
Total SM background events         & $17.0 \pm 2.8$          & $3.3 \pm 1.4$          & $13.7 \pm 2.3$              \\
\noalign{\smallskip}\hline\noalign{\smallskip}
\ttbar\          & $0.63 \pm 0.28$          & $0.23 \pm 0.15$          & $0.40 \pm 0.19$              \\
$W$+jets         & $11.6 \pm 2.1$          & $2.3 \pm 1.1$          & $9.3 \pm 1.6$              \\
$Z$+jets         & $0.045 \pm 0.022$          & $0.076 \pm 0.021$          & $-$             \\
Single-top       & $1.3 \pm 0.7$          & $0.014_{-0.014}^{+0.066}$          & $1.2 \pm 0.7$              \\
Diboson2l        & $1.5 \pm 0.7$          & $0.30 \pm 0.18$          & $1.2 \pm 0.5$              \\
Diboson1l        & $1.6 \pm 0.4$          & $0.35 \pm 0.15$          & $1.22 \pm 0.30$              \\
\ttbar+$V$       & $0.42 \pm 0.14$          & $0.019_{-0.019}^{+0.023}$          & $0.40 \pm 0.14$              \\
\ttbar+$h$       & $0.003 \pm 0.003$          & $-$          & $0.003 \pm 0.003$              \\
Multiboson       & $0.006_{-0.006}^{+0.011}$          & $-$          & $0.006_{-0.006}^{+0.011}$               \\

\noalign{\smallskip}\hline\noalign{\smallskip}
\end{tabular*}
}
\end{center}
\caption{
Observed event yields and the background expectation obtained from a background fit in the C1N2-WZ model SRs with an integrated luminosity of 139~\ifb.
The first column  with numbers stands for the yields in all bins. The second and third columns correspond to the low and high bins in \meff.
Uncertainties reported for the fitted background estimates combine statistical and systematic uncertainties.} 
\label{table_results_C1N2_WZ}
\end{table}
 
\begin{table}{\small
\begin{center}
 
\newcolumntype{C}{@{}>{{}}c<{{}}@{}}
\begin{tabular*}{\textwidth}{@{\extracolsep{\fill}}l *4{C}}
 
\toprule
\textbf{Yields}           & SRXGB Bin 1           & SRXGB Bin 2            & SRXGB Bin 3          & SRXGB Bin 4              \\
& [0.91, 0.928) & [0.928, 0.948) & [0.948, 0.964) & [0.964, 1] \\
 
\midrule
Observed events    & $5$              & $9$              & $6$              & $5$                    \\
\midrule
Total SM background events   & $9.4 \pm 2.1$   & $5.7 \pm 2.0$  & $4.2 \pm 1.4$  & $2.2 \pm 0.8$              \\
 
\midrule
\ttbar\  & $3.8 \pm 1.1$          & $2.3 \pm 1.4$          & $1.3 \pm 0.7$          & $0.30_{-0.30}^{+0.32}$       \\
 
Single-top     & $3.1 \pm 1.6$          & $2.5 \pm 1.2$          & $1.8 \pm 1.0$          & $0.8 \pm 0.5$            \\
$W$+jets      & $1.7 \pm 0.6$          & $0.5 \pm 0.4$          & $0.8 \pm 0.2$          & $0.8 \pm 0.2$              \\
 
Diboson   & $0.4 \pm 0.2$          & $0.14 \pm 0.09$          & $0.09 \pm 0.07$          & $0.05 \pm 0.02$              \\
\ttbar\+h, $Wh$         & $0.24 \pm 0.22$          & $0.18 \pm 0.05$          & $0.14 \pm 0.02$          & $0.15 \pm 0.03$              \\
\ttbar+$W/Z$         & $0.13 \pm 0.06$          & $0.04_{-0.04}^{+0.05}$          & $0.08 \pm 0.06$          & $0.05 \pm 0.02$              \\
Others & $0.04_{-0.04}^{+0.05}$ &  $0.05 \pm 0.04$ & $0.01\pm0.01$ & $0.02\pm0.01$ \\
\midrule
\end{tabular*}
\end{center}
}
 
\caption{
Observed event yields and the background expectation obtained from a background fit in the C1N2-Wh model SR with an integrated luminosity of 139~\ifb. Each column corresponds to a bin in $w^{}_{sig}$ score. Uncertainties reported for the fitted background estimates combine statistical and systematic uncertainties. The category `Others' refers to $Z$+jets and multiboson events. } 
\label{table_results_C1N2_Wh}
\end{table}


\FloatBarrier
\clearpage
\begin{figure}
\centering
\subfloat[]{
\includegraphics[width= 0.95\textwidth]{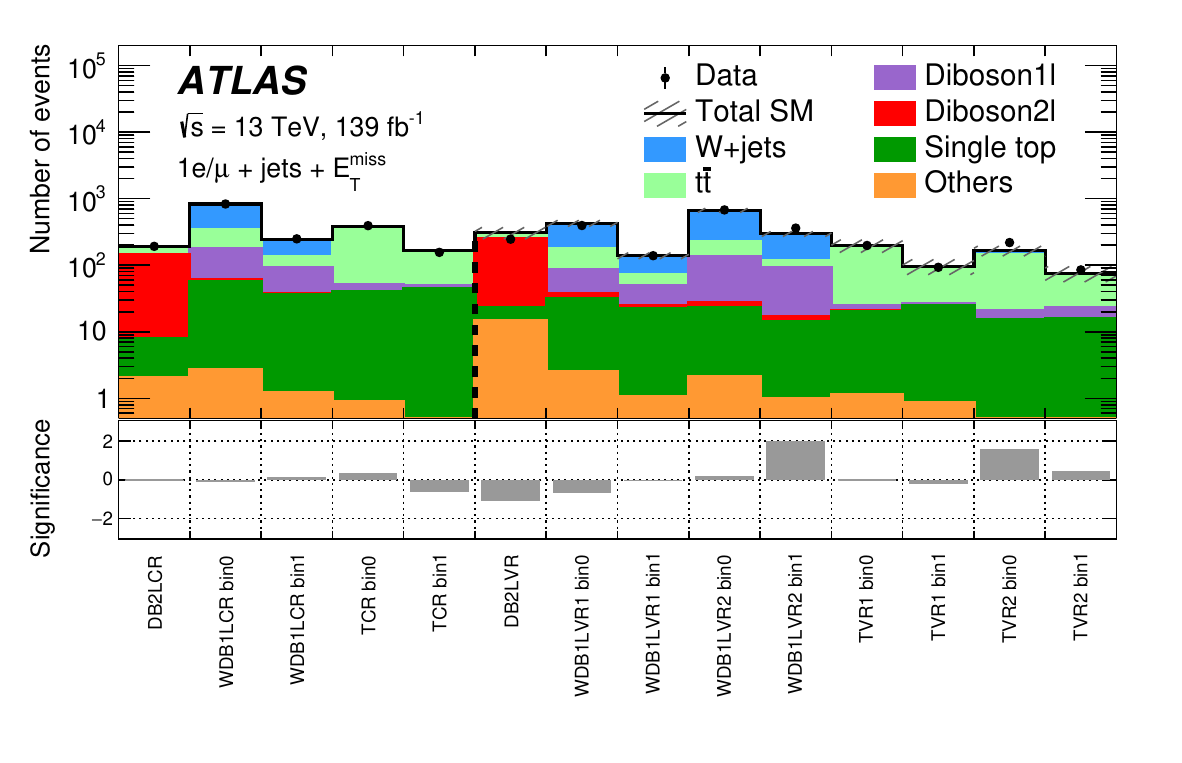}}
\quad
\subfloat[]{
\includegraphics[width= 0.95\textwidth]{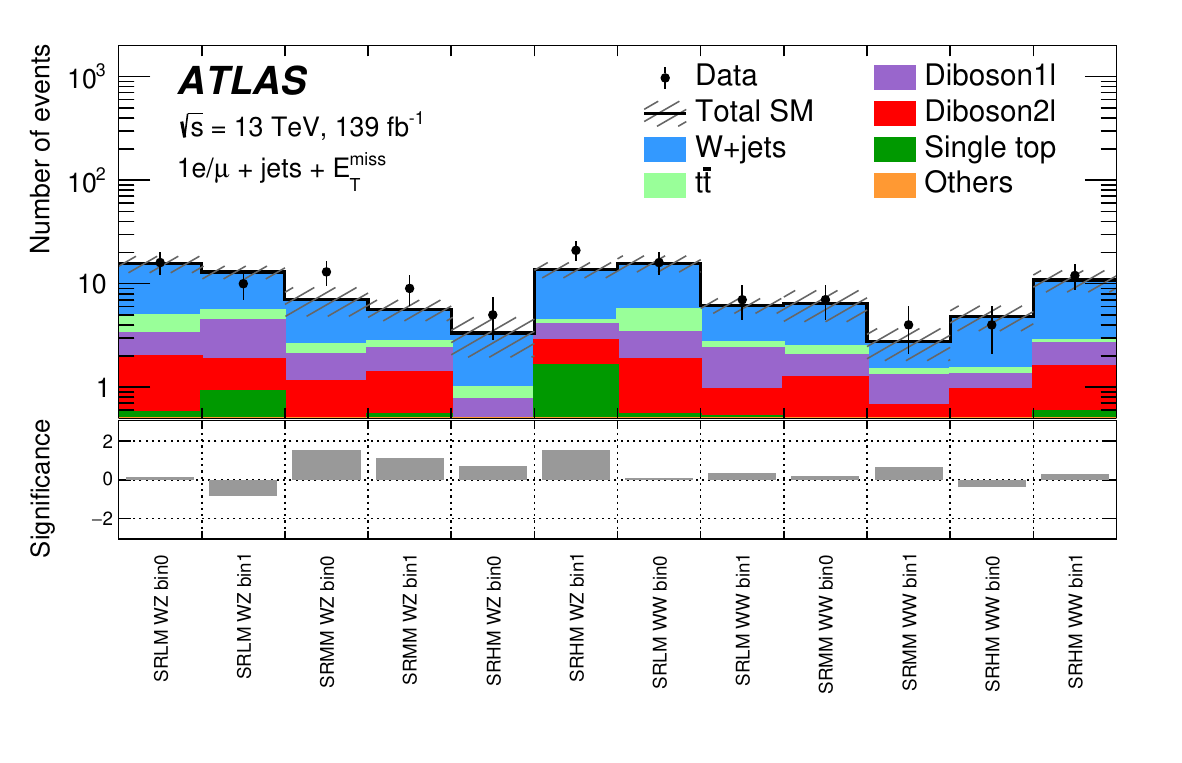}}
\caption{Comparison of the observed and expected event yields in  (a) the control and validation regions, and (b) the exclusion signal regions for the C1C1-WW and C1N2-WZ analyses. Uncertainties in the background estimates include both the statistical  and systematic uncertainties. The bottom panel shows the significance~\cite{2008NIMPA.595..480C} of the differences between the observed and expected yields.} 
\label{fig:pullplot_SR_VR_CR}
\end{figure}
 
\begin{figure}
\centering
\includegraphics[width=1.05\textwidth]{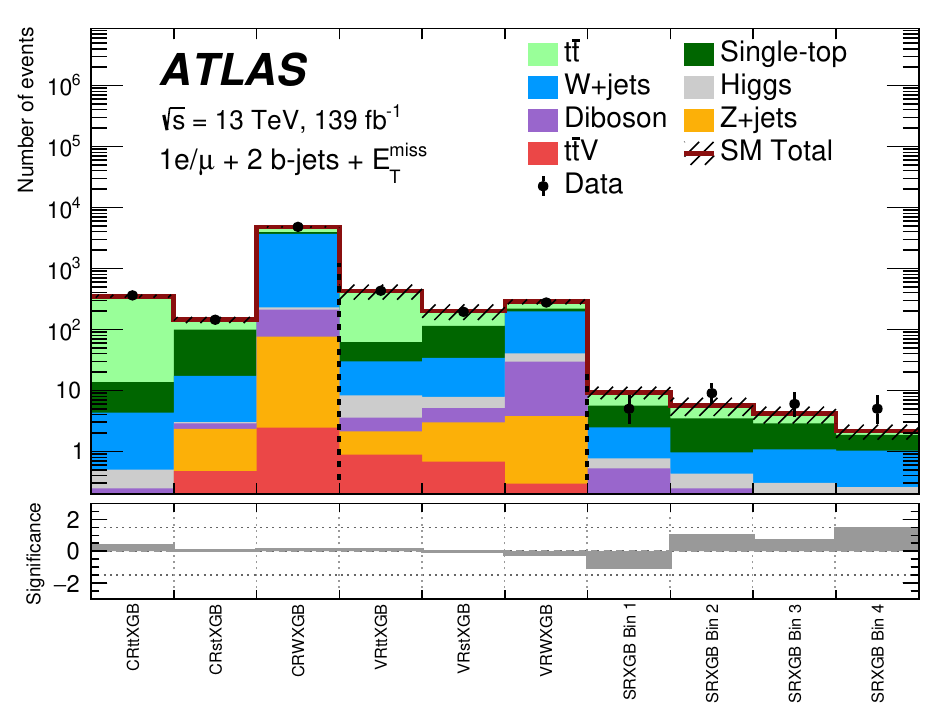}
\caption{Comparison of the observed and expected event yields in  control,  validation regions, and exclusion SR bins for the C1N2-Wh analysis. Uncertainties in the background estimates include both the statistical and systematic uncertainties. The bottom panel shows the significance~\cite{2008NIMPA.595..480C} of the differences between the observed and expected yields for control regions, validation regions and in the signal region in bins of $w_{\mathrm{sig}}$.}
\label{fig:pullWh}
\end{figure}

\begin{figure}
\begin{center}
\subfloat[]{
\includegraphics[width=0.4\textwidth]{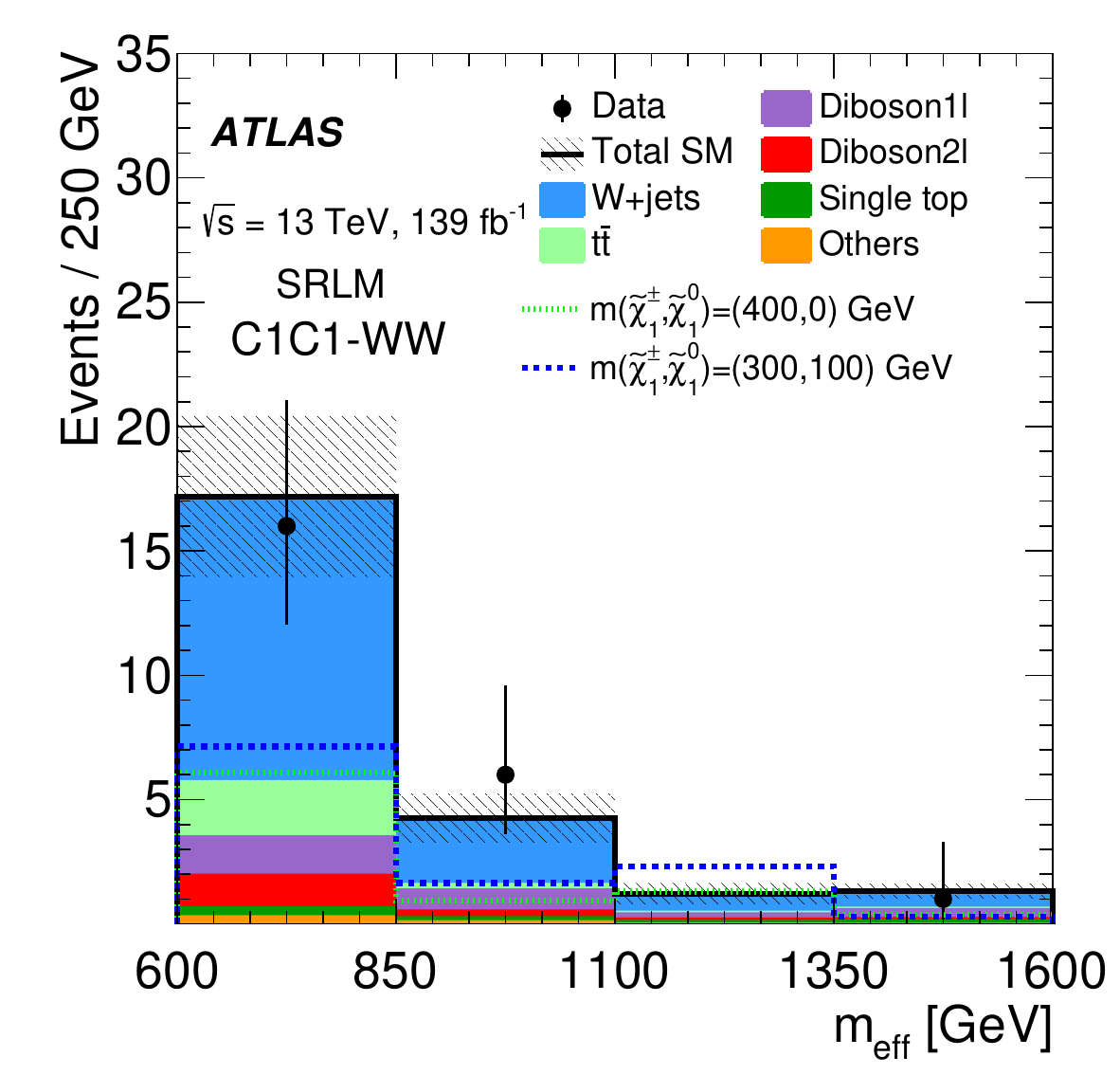}}
\subfloat[]{
\includegraphics[width=0.4\textwidth]{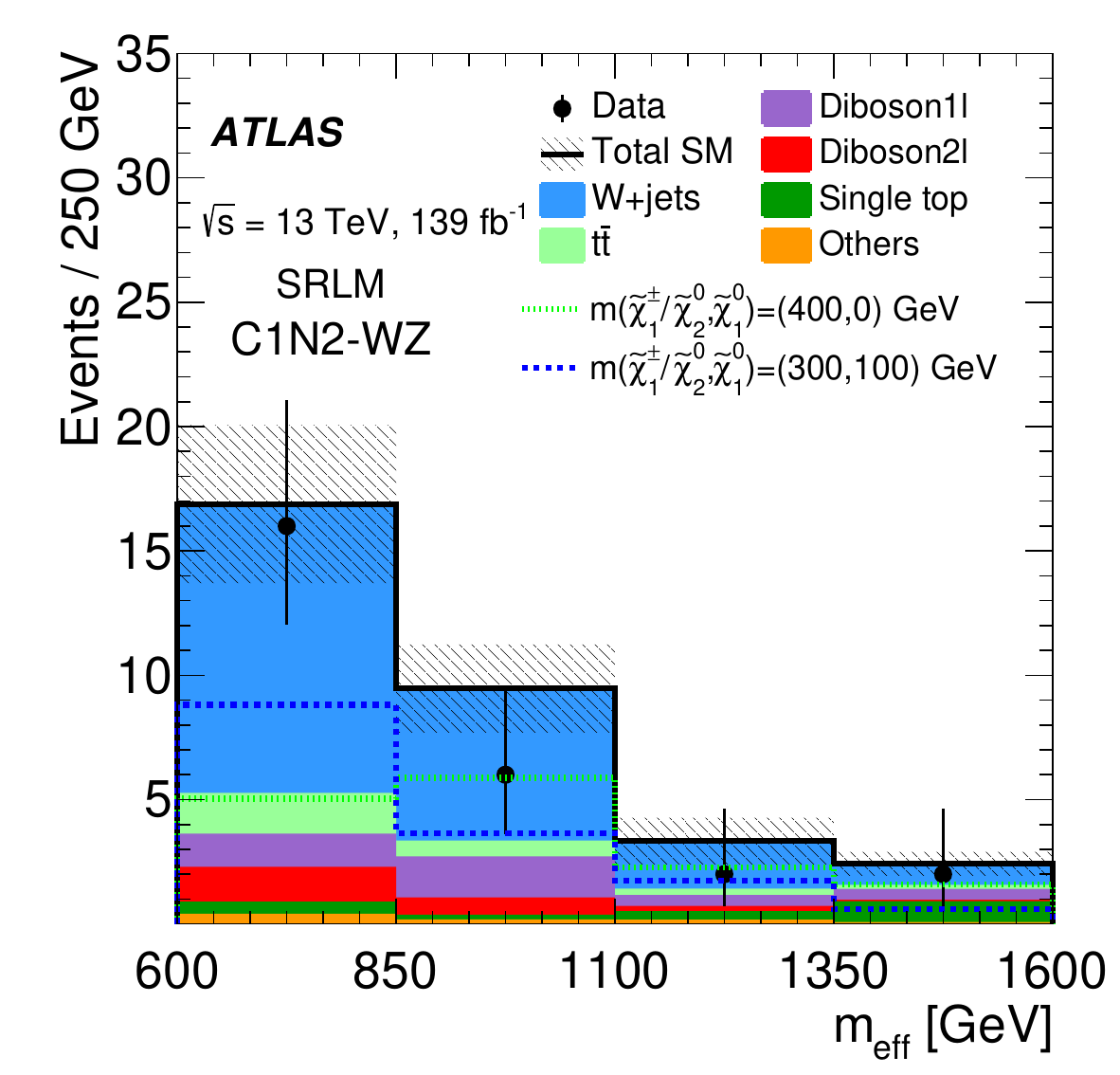}}
\quad
\subfloat[]{
\includegraphics[width=0.4\textwidth]{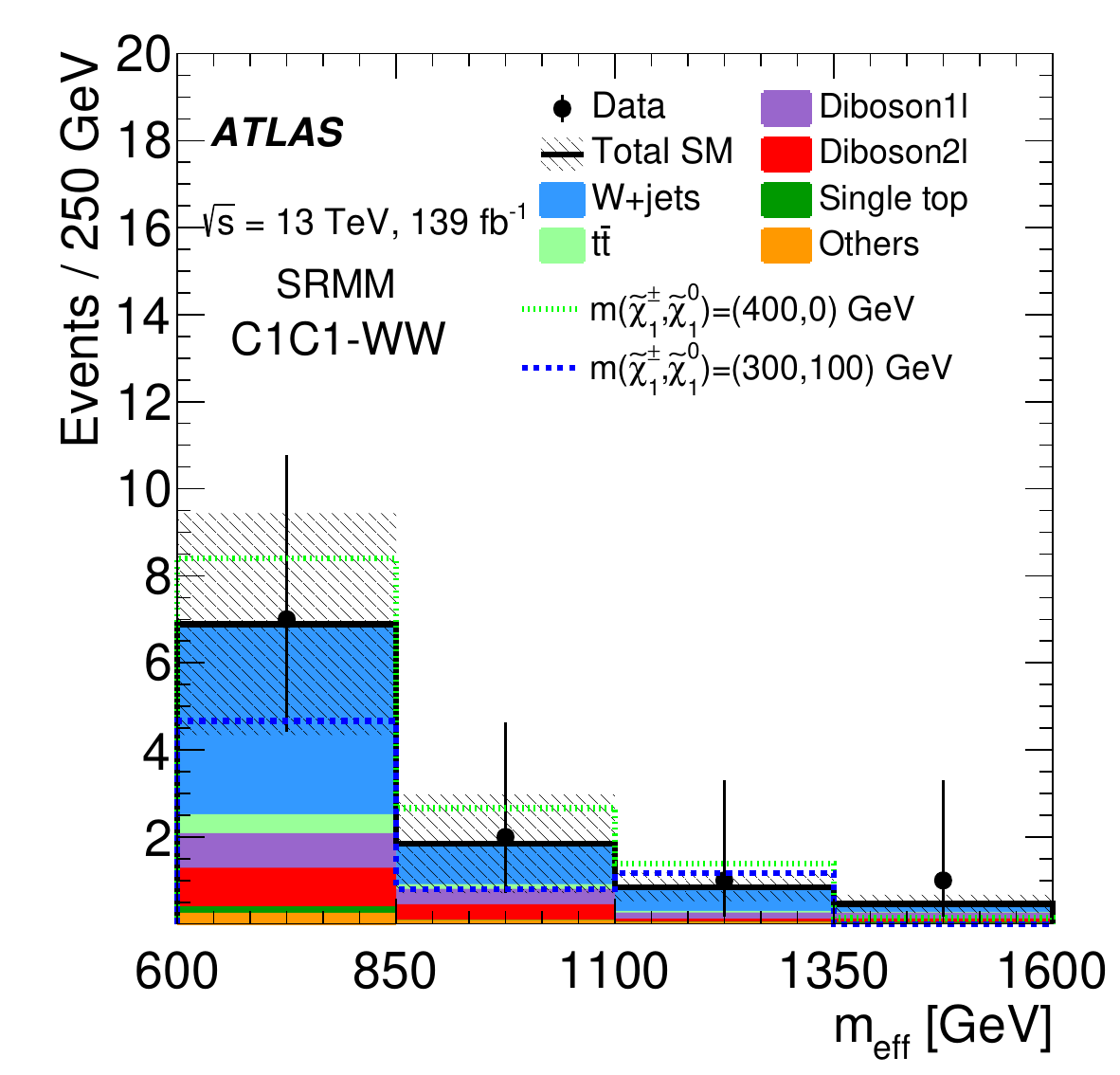}}
\subfloat[]{
\includegraphics[width=0.4\textwidth]{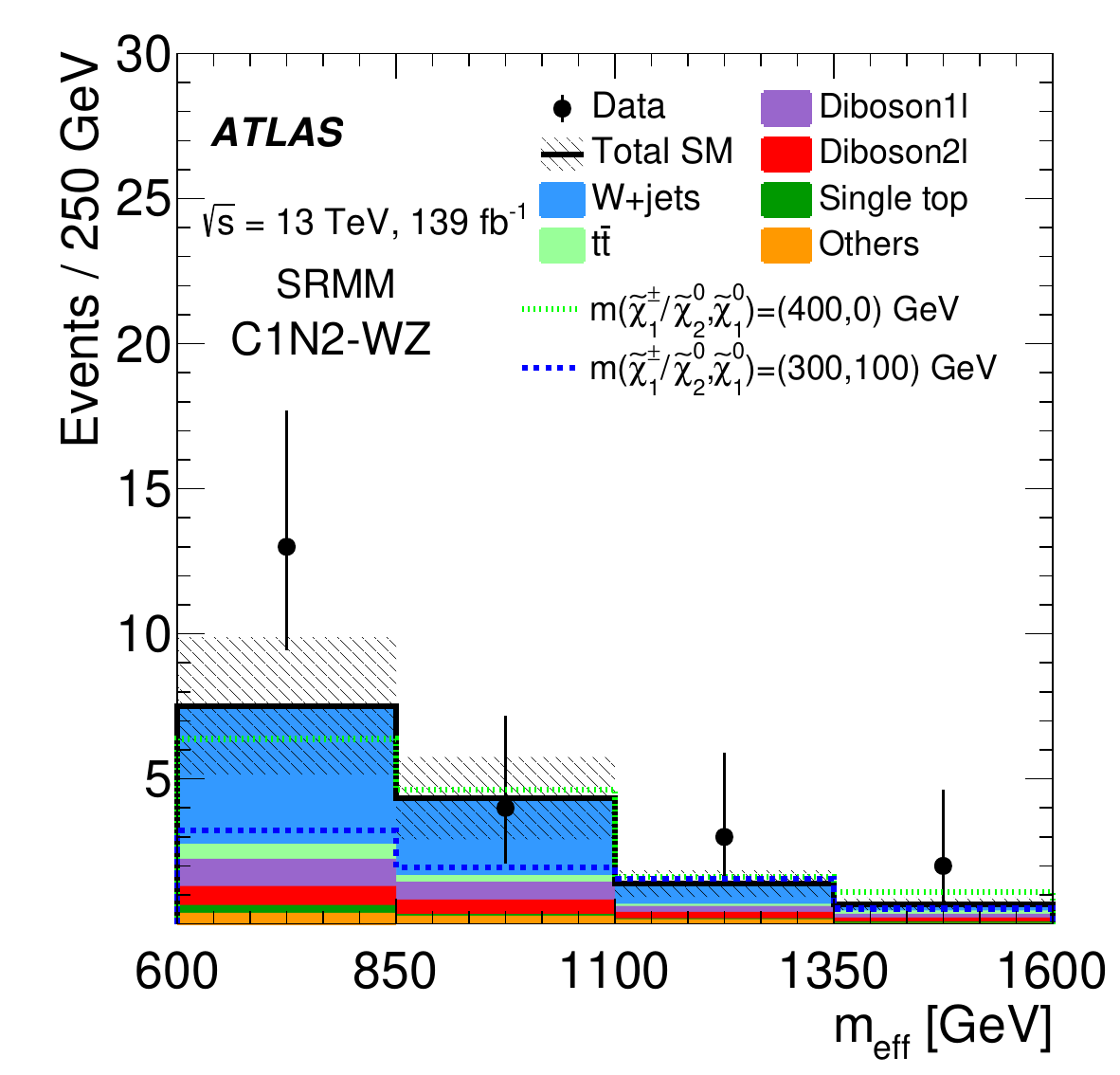}}
\quad
\subfloat[]{
\includegraphics[width=0.4\textwidth]{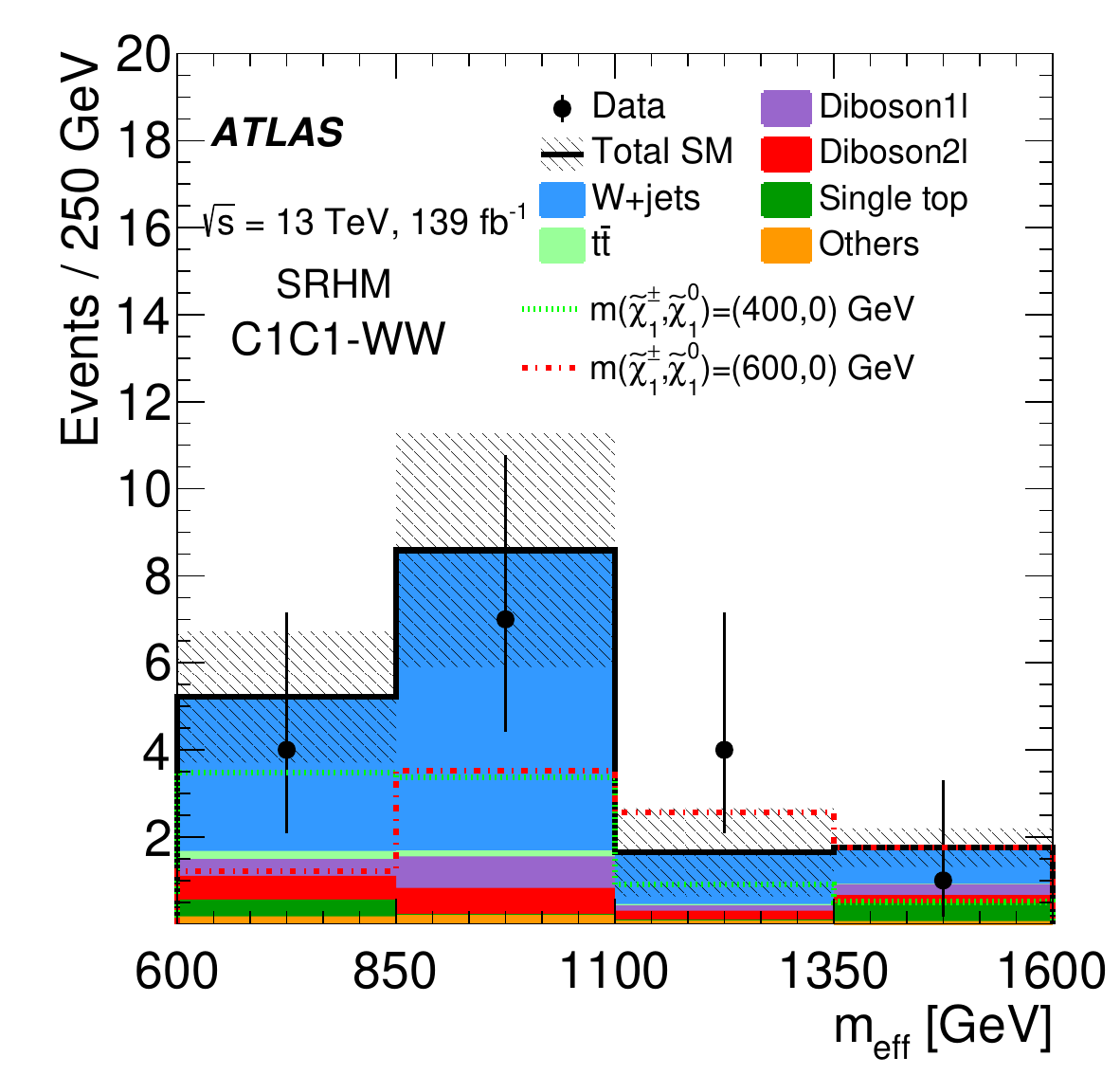}}
\subfloat[]{
\includegraphics[width=0.4\textwidth]{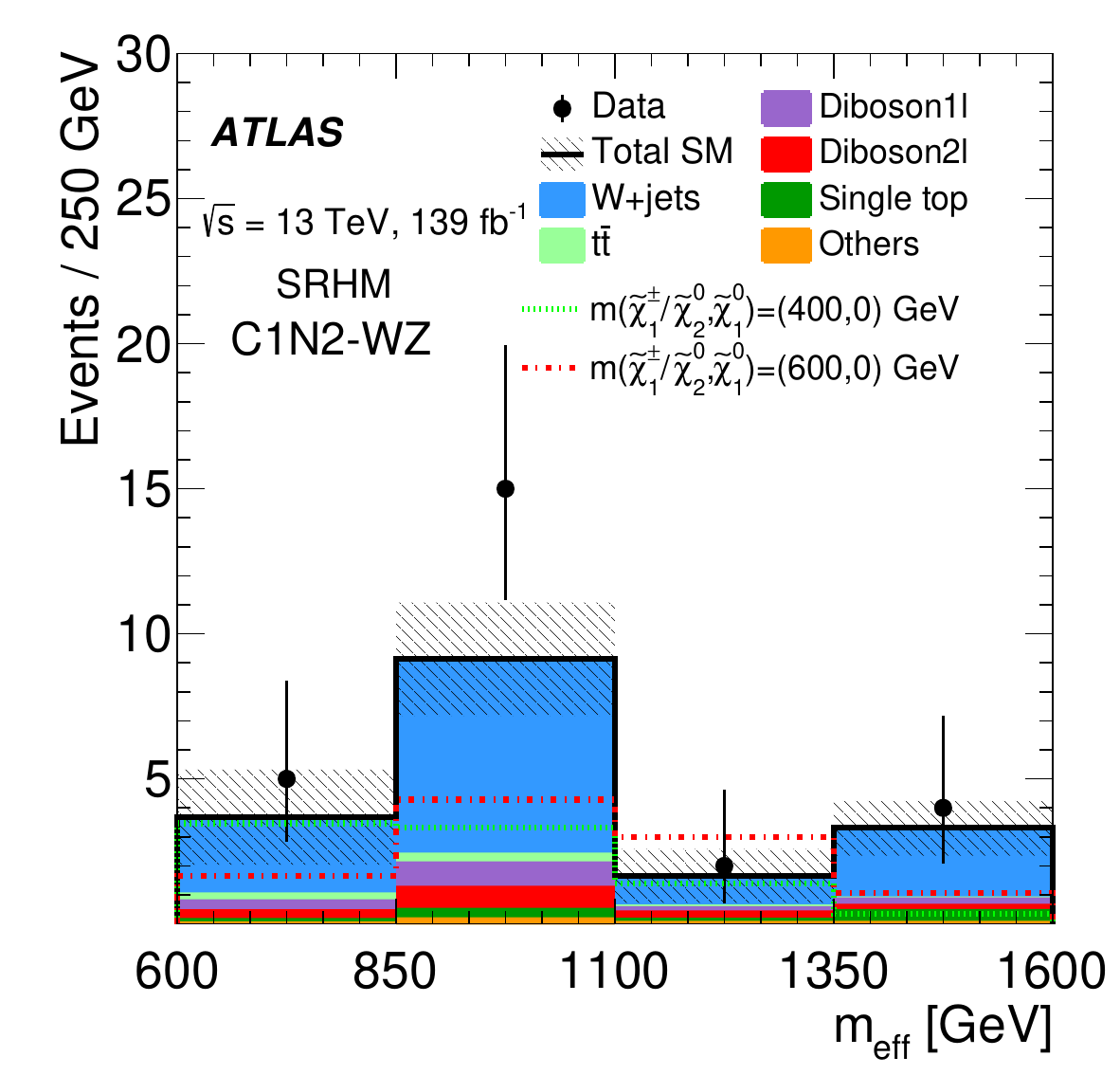}}
\end{center}
\caption{The post-fit \meff\ distributions in the exclusion signal regions (a, b) SRLM, (c, d) SRMM, and (e, f) SRHM for the C1C1-WW and C1N2-WZ models. The uncertainty bands plotted include all statistical and systematic uncertainties. The dashed lines represent the benchmark signal samples. The overflow events, where present,  are included in the last bin.}
\label{fig:SRs}
\end{figure}

Table~\ref{upperlimit_toys} summarises the observed ($S_{\mathrm{obs}}^{95}$) and expected ($S_{\mathrm{exp}}^{95}$) 95\% confidence level (CL) upper limits on the number of signal events and on the observed visible cross-section,
$\langle\epsilon{\mathrm{ \sigma}}\rangle_{\mathrm{ obs}}^{95}$,
for each SR(disc.). The discovery SRs are used to test for the presence of any beyond-the-Standard-Model (BSM) physics processes. Upper limits on contributions from new physics processes are estimated by using the `model-independent fit', where a generic BSM process is assumed to contribute only to the SR and not to the CRs, thus giving a conservative background estimate in the SR\@.
When normalised to the integrated luminosity of the data sample, the results can be interpreted as corresponding to observed upper limits
$\langle\epsilon{\mathrm{ \sigma}}\rangle_{\mathrm{ obs}}^{95}$,
defined as the product of the production cross-section, the acceptance, and the selection efficiency of a
BSM signal. The $p_{0}$ value and the $\textrm{CL}_{\textrm{B}}$ value are also provided.
The former represents the probability of the SM background alone to fluctuate to the
observed number of events or higher, and latter provides the confidence level observed for the background-only hypothesis.
The limits are validated by comparing pseudo-experiments and using asymptotic formulae, and found to be comparable.
All limits presented in this paper are calculated using asymptotic formulae.

 
\begin{table}
\small
\begin{center}
\setlength{\tabcolsep}{0.0pc}
\begin{tabular*}{\textwidth}{@{\extracolsep{\fill}}lccccccccc}
\noalign{\smallskip}\hline\noalign{\smallskip}{\textbf{Signal channel}}                 & Observed events &  Total SM background &     $\langle\epsilon{\mathrm{ \sigma}}\rangle_{\mathrm{ obs}}^{95}$[fb]  &  $S_{\mathrm{ obs}}^{95}$  & $S_{\mathrm{ exp}}^{95}$ & $\textrm{CL}_{\textrm{B}}$ & $p_{0}$ & Z \\
\noalign{\smallskip}\hline\noalign{\smallskip}{\textbf{ C1C1-WW model}} &&&&&&&& \\ 
SRLM (disc.)    &16 & $11.6 \pm 1.6$ & $0.09$ &  $13.0$ & $ { 8.8 }^{ +4.3 }_{ -1.5 }$ & $0.84$ & $ 0.14$ & $1.09$ \\
SRMM (disc.)    &9  & $~9.8 \pm 2.0$  & $0.06$ &  $7.9$  & $ { 9.0 }^{ +5.4 }_{ -1.4 }$ & $0.42$ & $ 0.50$ & $0.00$ \\
SRHM (disc.)    &12 & $10.8 \pm 2.5$ & $0.07$ &  $10.4$ & $ { 9.4 }^{ +4.1 }_{ -3.0 }$ & $0.60$ & $ 0.39$ & $0.29$  \\
 
{\textbf{ C1N2-WZ model}}   &&&&&&&& \\ 
SRLM (disc.)   &17 & $18.4 \pm 2.9$ & $0.08$ &  $11.5$ & $ { 13.7 }^{ +4.0 }_{ -4.5 }$ & $0.40$ & $ 0.50$ & $0.00$ \\
SRMM (disc.)   &9  & $~5.7 \pm 1.3$  & $0.07$ &  $10.2$ & $ { 6.8 }^{ +3.1 }_{ -0.9 }$ & $0.87$ & $ 0.13$ & $1.11$ \\
SRHM (disc.)   &21 & $13.7 \pm 2.3$ & $0.13$ &  $17.5$ & $ { 10.5 }^{ +4.4 }_{ -2.4 }$ & $0.92$ & $ 0.06$ & $1.54$ \\
 
{\textbf{ C1N2-Wh model}}   &&&&&&&& \\ 
SR (inclusive)  & $25$ & $21.4 \pm 4.2$ & $0.12$ &  $16.2$ & $ { 13.4 }^{ +5.5 }_{ -4.0 }$ & $0.70$ & $ 0.29$ & $0.57$ \\%
SR Bin 2--4 & $20$ & $12.0 \pm 3.0$ & $0.13$ &  $17.9$ & $ { 10.7 }^{ +4.7 }_{ -3.1 }$ & $0.92$ & $ 0.06$ & $1.58$ \\%
SR Bin 3--4 & $11$ & $6.3 \pm 1.6$ & $0.09$ &  $12.0$ & $ { 7.4 }^{ +3.5 }_{ -2.3 }$ & $0.89$ & $ 0.08$ & $1.42$ \\%
SR Bin 4  & $5$ & $2.2 \pm 1.0$  & $0.06$ &  $8.0$ & $ { 5.1 }^{ +2.7 }_{ -1.6 }$ & $0.86$ & $ 0.08$ & $1.37$ \\%
\noalign{\smallskip}\hline\noalign{\smallskip}
\end{tabular*}
\end{center}
\caption[Breakdown of upper limits.]{
The number of observed events, total SM background, 95\% CL upper limits on the visible cross-section
($\langle\epsilon\sigma\rangle_{\mathrm obs}^{95}$) and on the number of
signal events ($S_{\mathrm obs}^{95}$ ).  The fifth column
($S_{\mathrm exp}^{95}$) shows the 95\% CL upper limit on the number of
signal events, given the expected number (with $\pm 1$ standard deviation
excursions on the expectation) of background events.
The last three columns indicate the $\textrm{CL}_{\textrm{B}}$ value that provides a measure of compatibility of the observed data with the 95\% CL signal strength hypothesis relative to fluctuations of the background, the discovery $p$-value ($p_{0}$) that measures compatibility of the observed data with the background-only (zero signal strength) hypothesis relative to fluctuations of the background and the corresponding Gaussian significance (Z). Larger values indicate greater relative compatibility. The $p_{0}$ is not calculated in signal regions with a deficit relative to the nominal background prediction and here the $p_{0}$  value is capped at 0.50.
\label{upperlimit_toys}}
\end{table}


\begin{figure}
\begin{center}
\subfloat[]{
\includegraphics[width= 0.8\textwidth]{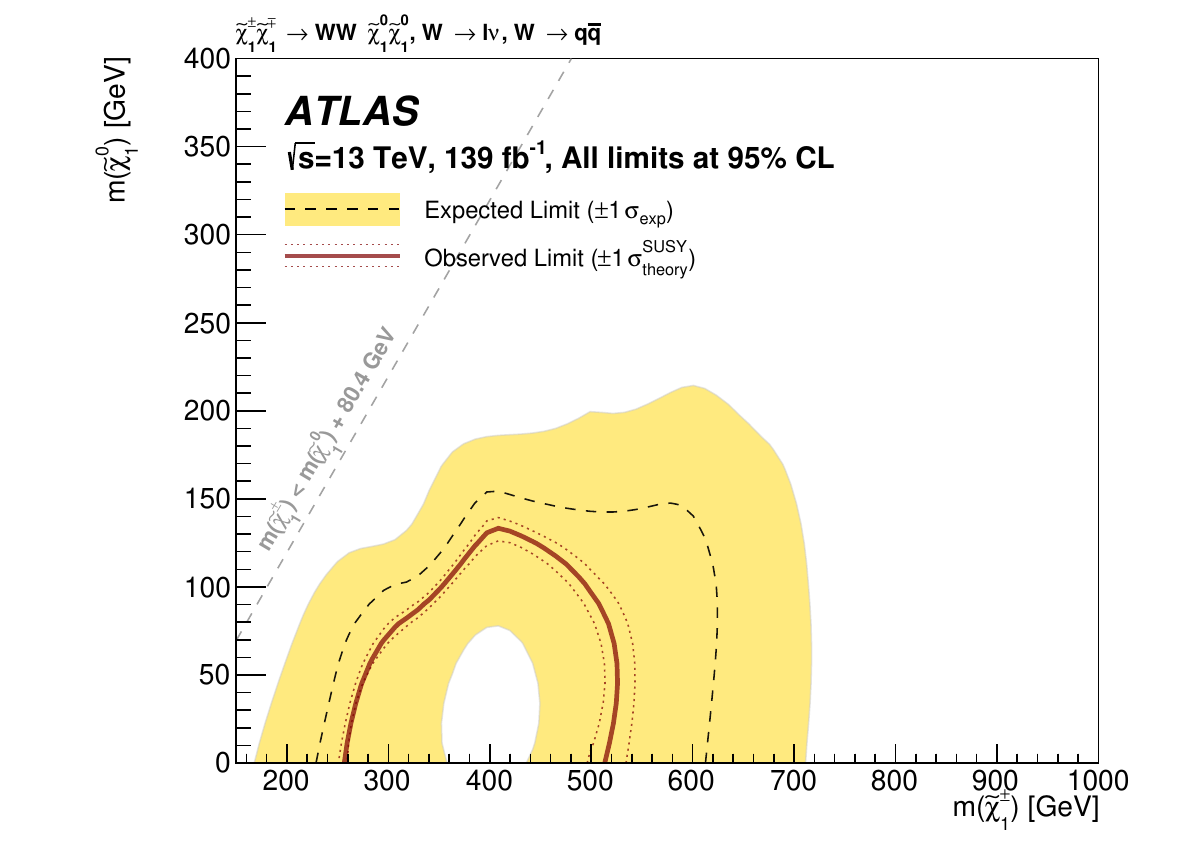}}
\quad
\subfloat[]{
\includegraphics[width= 0.8\textwidth]{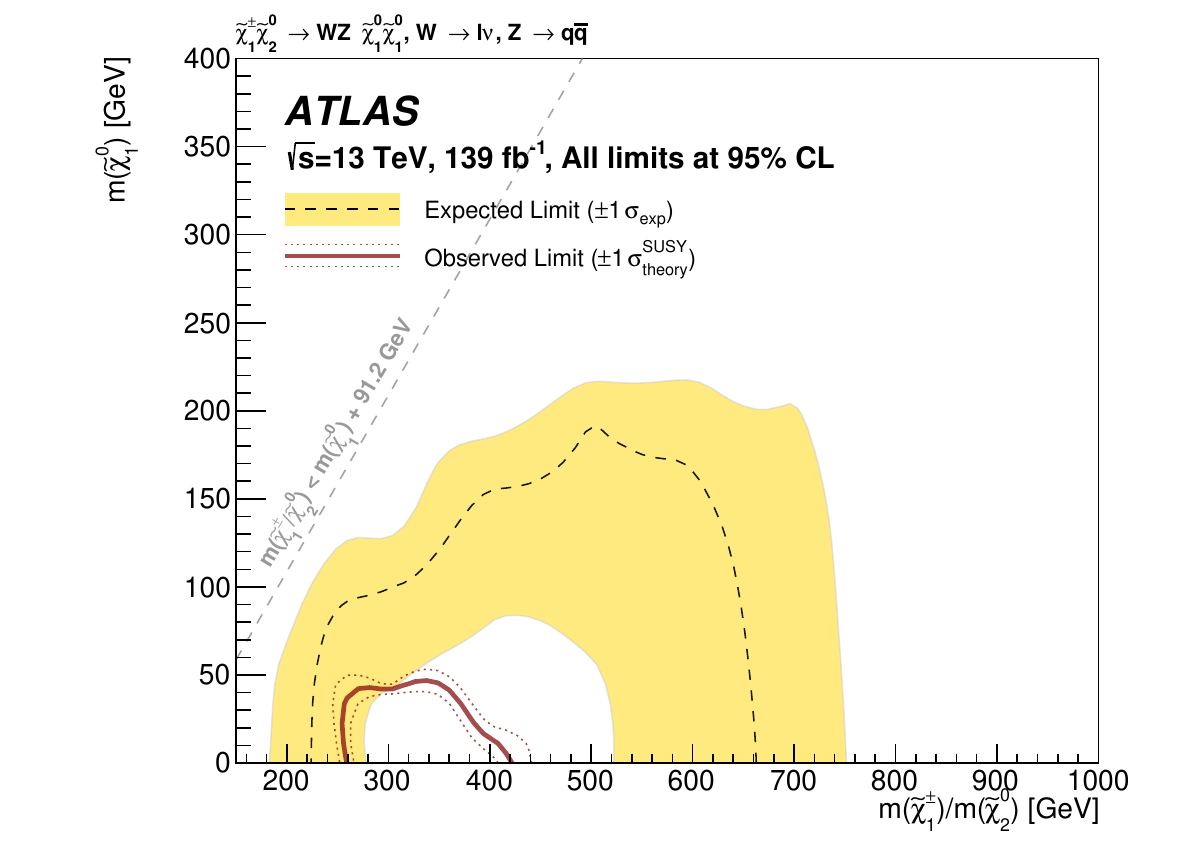}}
\end{center}
\caption{\label{sec:results:limit}Model-dependent exclusion contour at 95\% CL on (a) the chargino pair production and (b) the production of a chargino and a next-to-lightest neutralino. The observed limit is given by the solid line with the signal cross-section uncertainties shown by the dotted lines as indicated in the text. Expected limits are given by the dashed line with uncertainties shown by the shaded band.}
\end{figure}
 
\begin{figure}
\centering
\includegraphics[width= 0.8\textwidth]{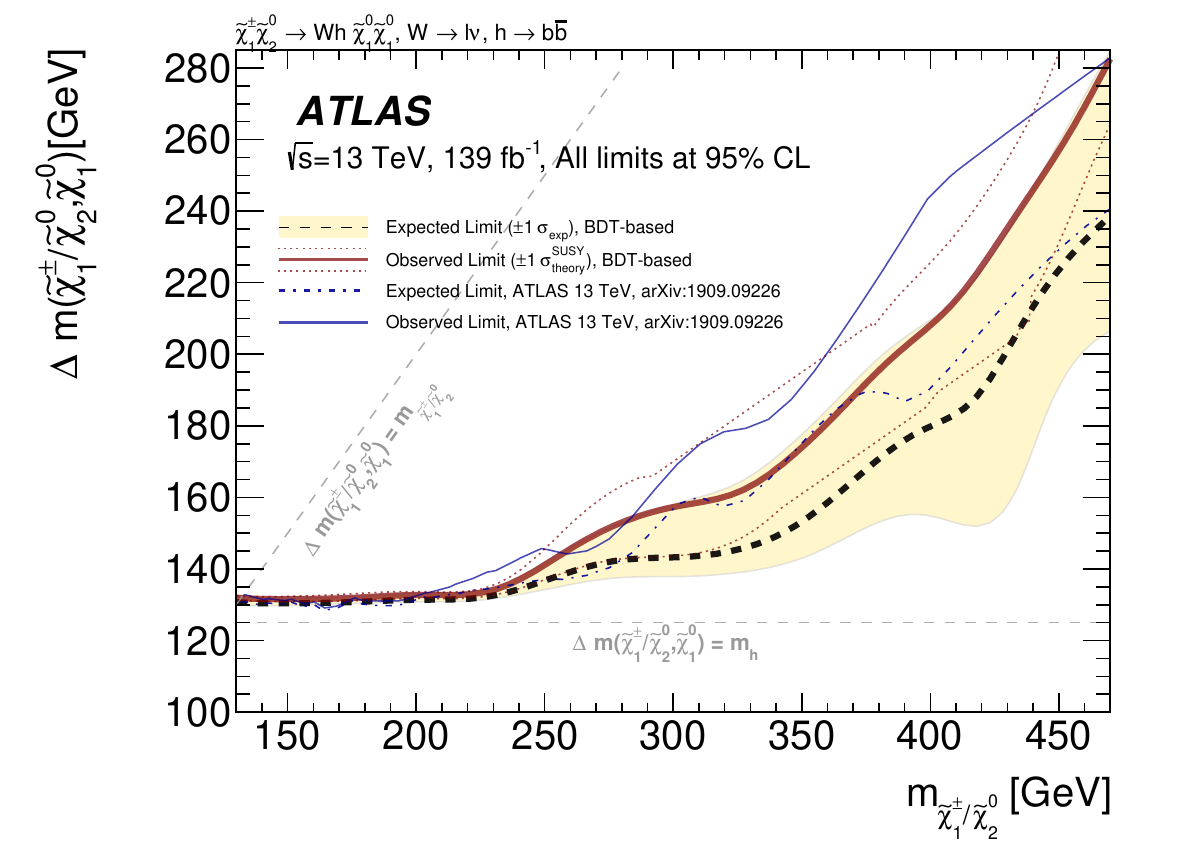}
\caption{\label{sec:results:limitWh}Model-dependent exclusion contour at 95\% CL on the chargino and a next-to-lightest neutralino mass versus the mass difference between chargino and LSP in the C1N2-Wh model. The observed limit is given by the solid line with the signal cross-section uncertainties shown by the dotted lines as indicated in the text. The area above the red observed limit line corresponds to the excluded region. Expected limits are given by the dashed line with uncertainties shown by the shaded band. The observed and expected limit contours from the previous ATLAS analysis of the same data sample using standard cut-and-count methods are also shown for comparison. }
\end{figure}

Model-dependent exclusion limits at 95\% CL are placed on the signal model. These limits are shown as a function of the masses of the SUSY particles in Figure~\ref{sec:results:limit} for C1C1-WW and C1N2-WZ models, and Figure~\ref{sec:results:limitWh} for C1N2-Wh models. 
A likelihood similar to the one used in the background-only fit, but with additional terms for the SRs, is used for the calculation. The exclusion SRs are included in the fit and are used to constrain normalisation and nuisance parameters. A signal is allowed in this likelihood in both the CRs and SRs. The VRs are not used in the fit. The $\mathrm{CL_s}$ method~\cite{Read:2002hq} is used to derive the CL of the exclusion for a particular signal model; signal models with a $\mathrm{CL_s}$ value below 0.05 are excluded at 95\% CL\@. The uncertainties in the observed limit are calculated by varying the cross-section for the signal up and down by its uncertainty.
For the C1C1-WW model, the $\chinoonepm$ mass of about 260--520~\GeV\ is excluded for a massless \ninoone, which complements the previous ATLAS limits in a 0-lepton channel~\cite{SUSY-2018-41} and a 2-lepton channel~\cite{SUSY-2018-32}.
For the C1N2-WZ model, the range of 260--420~\GeV\ in 
$m (\chinoonepm/\tilde{\chi}^{0}_{2})$
for a massless \ninoone\ is excluded. The limit in the high 
$m (\chinoonepm/\tilde{\chi}^{0}_{2})$
region is dominated by the high \meff\ bin of SR-HM\@. Combining the low and high \meff\ bins of SRMM for C1N2-WZ model leads to a signal significance of around 2.1~standard deviations. This differences between observed and expected events in bins with small numbers of events lead to an observed limit weaker than the expected one. Similar differences arise in the exclusion limit of C1N2-Wh models. In this case, limits are shown as a function of the mass of the chargino and next-to-lightest neutralino and the mass difference between that and the LSP, and are compared with previous ATLAS results on the same data sample. The presented mass range is chosen to illustrate the improved region only. While the low numbers of events and the large systematic uncertainties in the most constraining bins in $w_{\mathrm{sig}}$ reduce the expected sensitivity, the BDT approach exceed previous constraints at low $\Delta m (\chinoonepm/\ninotwo, \tilde{\chi}^{0}_{1})$ by up to 40~\GeV\ in the range of 200--260~\GeV\ and 280--470~\GeV\
in $m (\chinoonepm/\tilde{\chi}^{0}_{2})$.


\FloatBarrier
 
\section{Conclusion}
\label{sec:conclusion}
The results of three searches for electroweakino pair production $pp \rightarrow \tilde\chi^\pm_1 \tilde\chi^0_2 / $\chinoonep\chinoonem$  $  in which the chargino ($\tilde\chi^\pm_1$) decays into a $W$ boson and the lightest neutralino ($\tilde\chi^0_1$), while the heavier neutralino ($\tilde\chi^0_2$) decays into either a $Z$ or a Higgs boson and a second $\tilde\chi^0_1$ are presented. The searches are performed in events with one isolated lepton, jets and missing transverse momentum, using \(pp\) collisions provided by the LHC at a centre-of-mass energy of 13~\TeV.
Data collected with the ATLAS detector between 2015 and 2018 are used, corresponding to an integrated luminosity of $139~\mathrm{fb}^{-1}$.
No significant deviation from the expected Standard Model background is observed, and limits are set on the direct production of the electroweakinos in simplified models. Searches exploiting large radius jets to identify hadronically decaying $W$ and $Z$ bosons complement the previous ATLAS limits. In the $\chinoonep\chinoonem$ model, masses of $\tilde{\chi}^{\pm}_{1}$ ranging from 260 to 520~\GeV\ are excluded at 95\% confidence level for a massless $\tilde{\chi}^{0}_{1}$.  In the $\tilde\chi^\pm_1 \tilde\chi^0_2 $ model with $\tilde\chi^0_2$ decaying via a $Z$ boson, masses of $\tilde{\chi}^{\pm}_{1}/\tilde{\chi}^{0}_{2}$ ranging from 260 to 420~\GeV\ are excluded at 95\% CL for a massless $\tilde{\chi}^{0}_{1}$. Utilizing a BDT discriminant, the search targeting $\tilde\chi^\pm_1 \tilde\chi^0_2 $ models with $\tilde\chi^0_2$ decaying via a Higgs boson and mass-splitting between the mass-degenerate $\tilde\chi^\pm_1$ and $\tilde\chi^0_2 $ as small as the Higgs boson mass and the LSP exceed previous constraints by up to 40~\GeV\ in the range of 200--260~\GeV\ and 280--470~\GeV\
in $m (\chinoonepm/\tilde{\chi}^{0}_{2})$.
 
\section*{Acknowledgements}


We thank CERN for the very successful operation of the LHC, as well as the
support staff from our institutions without whom ATLAS could not be
operated efficiently.
 
We acknowledge the support of
ANPCyT, Argentina;
YerPhI, Armenia;
ARC, Australia;
BMWFW and FWF, Austria;
ANAS, Azerbaijan;
CNPq and FAPESP, Brazil;
NSERC, NRC and CFI, Canada;
CERN;
ANID, Chile;
CAS, MOST and NSFC, China;
Minciencias, Colombia;
MEYS CR, Czech Republic;
DNRF and DNSRC, Denmark;
IN2P3-CNRS and CEA-DRF/IRFU, France;
SRNSFG, Georgia;
BMBF, HGF and MPG, Germany;
GSRI, Greece;
RGC and Hong Kong SAR, China;
ISF and Benoziyo Center, Israel;
INFN, Italy;
MEXT and JSPS, Japan;
CNRST, Morocco;
NWO, Netherlands;
RCN, Norway;
MEiN, Poland;
FCT, Portugal;
MNE/IFA, Romania;
MESTD, Serbia;
MSSR, Slovakia;
ARRS and MIZ\v{S}, Slovenia;
DSI/NRF, South Africa;
MICINN, Spain;
SRC and Wallenberg Foundation, Sweden;
SERI, SNSF and Cantons of Bern and Geneva, Switzerland;
MOST, Taiwan;
TENMAK, T\"urkiye;
STFC, United Kingdom;
DOE and NSF, United States of America.
In addition, individual groups and members have received support from
BCKDF, CANARIE, Compute Canada and CRC, Canada;
PRIMUS 21/SCI/017 and UNCE SCI/013, Czech Republic;
COST, ERC, ERDF, Horizon 2020, ICSC-NextGenerationEU and Marie Sk{\l}odowska-Curie Actions, European Union;
Investissements d'Avenir Labex, Investissements d'Avenir Idex and ANR, France;
DFG and AvH Foundation, Germany;
Herakleitos, Thales and Aristeia programmes co-financed by EU-ESF and the Greek NSRF, Greece;
BSF-NSF and MINERVA, Israel;
Norwegian Financial Mechanism 2014-2021, Norway;
NCN and NAWA, Poland;
La Caixa Banking Foundation, CERCA Programme Generalitat de Catalunya and PROMETEO and GenT Programmes Generalitat Valenciana, Spain;
G\"{o}ran Gustafssons Stiftelse, Sweden;
The Royal Society and Leverhulme Trust, United Kingdom.
 
The crucial computing support from all WLCG partners is acknowledged gratefully, in particular from CERN, the ATLAS Tier-1 facilities at TRIUMF (Canada), NDGF (Denmark, Norway, Sweden), CC-IN2P3 (France), KIT/GridKA (Germany), INFN-CNAF (Italy), NL-T1 (Netherlands), PIC (Spain), ASGC (Taiwan), RAL (UK) and BNL (USA), the Tier-2 facilities worldwide and large non-WLCG resource providers. Major contributors of computing resources are listed in Ref.~\cite{ATL-SOFT-PUB-2023-001}.


\printbibliography
\clearpage
\addcontentsline{toc}{part}{Auxiliary Material}
\FloatBarrier

\clearpage
 
\begin{flushleft}
\hypersetup{urlcolor=black}
{\Large The ATLAS Collaboration}

\bigskip

\AtlasOrcid[0000-0002-6665-4934]{G.~Aad}$^\textrm{\scriptsize 102}$,
\AtlasOrcid[0000-0002-5888-2734]{B.~Abbott}$^\textrm{\scriptsize 120}$,
\AtlasOrcid[0000-0002-1002-1652]{K.~Abeling}$^\textrm{\scriptsize 55}$,
\AtlasOrcid[0000-0001-5763-2760]{N.J.~Abicht}$^\textrm{\scriptsize 49}$,
\AtlasOrcid[0000-0002-8496-9294]{S.H.~Abidi}$^\textrm{\scriptsize 29}$,
\AtlasOrcid[0000-0002-9987-2292]{A.~Aboulhorma}$^\textrm{\scriptsize 35e}$,
\AtlasOrcid[0000-0001-5329-6640]{H.~Abramowicz}$^\textrm{\scriptsize 151}$,
\AtlasOrcid[0000-0002-1599-2896]{H.~Abreu}$^\textrm{\scriptsize 150}$,
\AtlasOrcid[0000-0003-0403-3697]{Y.~Abulaiti}$^\textrm{\scriptsize 117}$,
\AtlasOrcid[0000-0002-8588-9157]{B.S.~Acharya}$^\textrm{\scriptsize 69a,69b,m}$
% ...
\bigskip
\\

% ...
$^{29}$Physics Department, Brookhaven National Laboratory, Upton NY; United States of America.\\
% ...
$^{35}$$^{(a)}$Facult\'e des Sciences Ain Chock, R\'eseau Universitaire de Physique des Hautes Energies - Universit\'e Hassan II, Casablanca;$^{(b)}$Facult\'{e} des Sciences, Universit\'{e} Ibn-Tofail, K\'{e}nitra;$^{(c)}$Facult\'e des Sciences Semlalia, Universit\'e Cadi Ayyad, LPHEA-Marrakech;$^{(d)}$LPMR, Facult\'e des Sciences, Universit\'e Mohamed Premier, Oujda;$^{(e)}$Facult\'e des sciences, Universit\'e Mohammed V, Rabat;$^{(f)}$Institute of Applied Physics, Mohammed VI Polytechnic University, Ben Guerir; Morocco.\\
% ...
$^{49}$Fakult\"{a}t Physik , Technische Universit{\"a}t Dortmund, Dortmund; Germany.\\
% ...
$^{55}$II. Physikalisches Institut, Georg-August-Universit\"{a}t G\"ottingen, G\"ottingen; Germany.\\
% ...
$^{69}$$^{(a)}$INFN Gruppo Collegato di Udine, Sezione di Trieste, Udine;$^{(b)}$ICTP, Trieste;$^{(c)}$Dipartimento Politecnico di Ingegneria e Architettura, Universit\`a di Udine, Udine; Italy.\\
% ...
$^{102}$CPPM, Aix-Marseille Universit\'e, CNRS/IN2P3, Marseille; France.\\
% ...
$^{117}$Department of Physics, New York University, New York NY; United States of America.\\
% ...
$^{120}$Homer L. Dodge Department of Physics and Astronomy, University of Oklahoma, Norman OK; United States of America.\\
% ...
$^{150}$Department of Physics, Technion, Israel Institute of Technology, Haifa; Israel.\\
$^{151}$Raymond and Beverly Sackler School of Physics and Astronomy, Tel Aviv University, Tel Aviv; Israel.\\
% ...
$^{a}$ Also Affiliated with an institute covered by a cooperation agreement with CERN.\\
$^{b}$ Also at An-Najah National University, Nablus; Palestine.\\
$^{c}$ Also at Borough of Manhattan Community College, City University of New York, New York NY; United States of America.\\
$^{d}$ Also at Center for High Energy Physics, Peking University; China.\\
$^{e}$ Also at Center for Interdisciplinary Research and Innovation (CIRI-AUTH), Thessaloniki; Greece.\\
$^{f}$ Also at Centro Studi e Ricerche Enrico Fermi; Italy.\\
% ...
$^{m}$ Also at Department of Physics, King's College London, London; United Kingdom.\\
% ...

\end{flushleft}


\FloatBarrier

\end{document}